\documentclass[prl,twocolumn,superscriptaddress,showpacs,amssymb,amsmath,amsfonts,aps]{revtex4}
\usepackage{graphicx}
\usepackage{dcolumn}
\usepackage{rotating}
\usepackage{epsfig}
\usepackage{layout}
\begin{document}
%
%
%
\title{{\large Single $\pi^{+}$ Electroproduction on the Proton in the First 
        and Second Resonance Regions at $0.25$~GeV$^2 < Q^{2} < 0.65$~GeV$^2$ Using CLAS}}
%
%
%

%
\newcommand*{\JLAB }{ Thomas Jefferson National Accelerator Facility, Newport News, Virginia 23606} 
\affiliation{\JLAB } 

\newcommand*{\WM }{ College of William and Mary, Williamsburg, Virginia 23187-8795} 
\affiliation{\WM } 

\newcommand*{\VIRGINIA }{ University of Virginia, Charlottesville, Virginia 22901} 
\affiliation{\VIRGINIA } 

\newcommand*{\YEREVAN }{ Yerevan Physics Institute, 375036 Yerevan, Armenia} 
\affiliation{\YEREVAN } 

\newcommand*{\ASU }{ Arizona State University, Tempe, Arizona 85287-1504} 
\affiliation{\ASU } 

\newcommand*{\SACLAY }{ CEA-Saclay, Service de Physique Nucl\'eaire, F91191 Gif-sur-Yvette, France} 
\affiliation{\SACLAY } 

\newcommand*{\UCLA }{ University of California at Los Angeles, Los Angeles, California  90095-1547} 
\affiliation{\UCLA } 

\newcommand*{\CMU }{ Carnegie Mellon University, Pittsburgh, Pennsylvania 15213} 
\affiliation{\CMU } 

\newcommand*{\CUA }{ Catholic University of America, Washington, D.C. 20064} 
\affiliation{\CUA } 

\newcommand*{\CNU }{ Christopher Newport University, Newport News, Virginia 23606} 
\affiliation{\CNU } 

\newcommand*{\UCONN }{ University of Connecticut, Storrs, Connecticut 06269} 
\affiliation{\UCONN } 

\newcommand*{\DUKE }{ Duke University, Durham, North Carolina 27708-0305} 
\affiliation{\DUKE } 

\newcommand*{\GBEDINBURGH }{ Edinburgh University, Edinburgh EH9 3JZ, United Kingdom} 
\affiliation{\GBEDINBURGH } 

\newcommand*{\FIU }{ Florida International University, Miami, Florida 33199} 
\affiliation{\FIU } 

\newcommand*{\FSU }{ Florida State University, Tallahassee, Florida 32306} 
\affiliation{\FSU } 

\newcommand*{\GWU }{ The George Washington University, Washington, DC 20052} 
\affiliation{\GWU } 

\newcommand*{\GBGLASGOW }{ University of Glasgow, Glasgow G12 8QQ, United Kingdom} 
\affiliation{\GBGLASGOW } 

\newcommand*{\INFNFR }{ INFN, Laboratori Nazionali di Frascati, Frascati, Italy} 
\affiliation{\INFNFR } 

\newcommand*{\INFNGE }{ INFN, Sezione di Genova, 16146 Genova, Italy} 
\affiliation{\INFNGE } 

\newcommand*{\ORSAY }{ Institut de Physique Nucleaire ORSAY, Orsay, France} 
\affiliation{\ORSAY } 

\newcommand*{\BONN }{ Institute f\"{u}r Strahlen und Kernphysik, Universit\"{a}t Bonn, Germany} 
\affiliation{\BONN } 

\newcommand*{\ITEP }{ Institute of Theoretical and Experimental Physics, Moscow, 117259, Russia} 
\affiliation{\ITEP } 

\newcommand*{\JMU }{ James Madison University, Harrisonburg, Virginia 22807} 
\affiliation{\JMU } 

\newcommand*{\KYUNGPOOK }{ Kungpook National University, Daegu 702-701, South Korea} 
\affiliation{\KYUNGPOOK } 

\newcommand*{\MIT }{ Massachusetts Institute of Technology, Cambridge, Massachusetts  02139-4307} 
\affiliation{\MIT } 

\newcommand*{\UMASS }{ University of Massachusetts, Amherst, Massachusetts  01003} 
\affiliation{\UMASS } 

\newcommand*{\MSU }{ Moscow State University, Skabeltsin Nuclear Physics Institute, 119899 Moscow, Russia}
\affiliation{\MSU}

\newcommand*{\UNH }{ University of New Hampshire, Durham, New Hampshire 03824-3568} 
\affiliation{\UNH } 

\newcommand*{\NSU }{ Norfolk State University, Norfolk, Virginia 23504} 
\affiliation{\NSU } 

\newcommand*{\OHIOU }{ Ohio University, Athens, Ohio  45701} 
\affiliation{\OHIOU } 

\newcommand*{\ODU }{ Old Dominion University, Norfolk, Virginia 23529} 
\affiliation{\ODU } 

\newcommand*{\PITT }{ University of Pittsburgh, Pittsburgh, Pennsylvania 15260} 
\affiliation{\PITT } 

\newcommand*{\ROMA }{ Universita' di ROMA III, 00146 Roma, Italy} 
\affiliation{\ROMA } 

\newcommand*{\RPI }{ Rensselaer Polytechnic Institute, Troy, New York 12180-3590} 
\affiliation{\RPI } 

\newcommand*{\RICE }{ Rice University, Houston, Texas 77005-1892} 
\affiliation{\RICE } 

\newcommand*{\URICH }{ University of Richmond, Richmond, Virginia 23173} 
\affiliation{\URICH } 

\newcommand*{\SCAROLINA }{ University of South Carolina, Columbia, South Carolina 29208} 
\affiliation{\SCAROLINA } 

\newcommand*{\UTEP }{ University of Texas at El Paso, El Paso, Texas 79968} 
\affiliation{\UTEP } 

\newcommand*{\UNIONC }{ Union College, Schenectady, NY 12308} 
\affiliation{\UNIONC } 

\newcommand*{\VT }{ Virginia Polytechnic Institute and State University, Blacksburg, Virginia   24061-0435} 
\affiliation{\VT } 


\newcommand*{\NOWNCATU }{ North Carolina Agricultural and Technical State University, Greensboro, NC 27411}

\newcommand*{\NOWGBGLASGOW }{ University of Glasgow, Glasgow G12 8QQ, United Kingdom}

\newcommand*{\NOWJLAB }{ Thomas Jefferson National Accelerator Facility, Newport News, Virginia 23606}

\newcommand*{\NOWSCAROLINA }{ University of South Carolina, Columbia, South Carolina 29208}

\newcommand*{\NOWFIU }{ Florida International University, Miami, Florida 33199}

\newcommand*{\NOWINFNFR }{ INFN, Laboratori Nazionali di Frascati, Frascati, Italy}

\newcommand*{\NOWOHIOU }{ Ohio University, Athens, Ohio  45701}

\newcommand*{\NOWCMU }{ Carnegie Mellon University, Pittsburgh, Pennsylvania 15213}

\newcommand*{\NOWINDSTRA }{ Systems Planning and Analysis, Alexandria, Virginia 22311}

\newcommand*{\NOWASU }{ Arizona State University, Tempe, Arizona 85287-1504}

\newcommand*{\NOWCISCO }{ Cisco, Washington, DC 20052}

\newcommand*{\NOWdeceased }{ Deceased}

\newcommand*{\NOWUK }{ University of Kentucky, LEXINGTON, KENTUCKY 40506}

\newcommand*{\NOWSACLAY }{ CEA-Saclay, Service de Physique Nucl\'eaire, F91191 Gif-sur-Yvette, Cedex, France}

\newcommand*{\NOWRPI }{ Rensselaer Polytechnic Institute, Troy, New York 12180-3590}

\newcommand*{\NOWUNCW }{ North Carolina}

\newcommand*{\NOWHAMPTON }{ Hampton University, Hampton, VA 23668}

\newcommand*{\NOWTulane }{ Tulane University, New Orleans, Lousiana  70118}

\newcommand*{\NOWKYUNGPOOK }{ Kungpook National University, Taegu 702-701, South Korea}

\newcommand*{\NOWCUA }{ Catholic University of America, Washington, D.C. 20064}

\newcommand*{\NOWGEORGETOWN }{ Georgetown University, Washington, DC 20057}

\newcommand*{\NOWJMU }{ James Madison University, Harrisonburg, Virginia 22807}

\newcommand*{\NOWURICH }{ University of Richmond, Richmond, Virginia 23173}

\newcommand*{\NOWCALTECH }{ California Institute of Technology, Pasadena, California 91125}

\newcommand*{\NOWMOSCOW }{ Moscow State University, General Nuclear Physics Institute, 119899 Moscow, Russia}

\newcommand*{\NOWVIRGINIA }{ University of Virginia, Charlottesville, Virginia 22901}

\newcommand*{\NOWYEREVAN }{ Yerevan Physics Institute, 375036 Yerevan, Armenia}

\newcommand*{\NOWRICE }{ Rice University, Houston, Texas 77005-1892}

\newcommand*{\NOWINFNGE }{ INFN, Sezione di Genova, 16146 Genova, Italy}

\newcommand*{\NOWBATES }{ MIT-Bates Linear Accelerator Center, Middleton, MA 01949}

\newcommand*{\NOWODU }{ Old Dominion University, Norfolk, Virginia 23529}

\newcommand*{\NOWVSU }{ Virginia State University, Petersburg,Virginia 23806}

\newcommand*{\NOWORST }{ Oregon State University, Corvallis, Oregon 97331-6507}

\newcommand*{\NOWGWU }{ The George Washington University, Washington, DC 20052}

\newcommand*{\NOWMIT }{ Massachusetts Institute of Technology, Cambridge, Massachusetts  02139-4307}

\newcommand*{\NOWCNU }{ Christopher Newport University, Newport News, Virginia 23606}

\newcommand*{\ALTSKRY} {Current Address: Sakarya University, Sakarya, Turkey}
\author{H.~Egiyan}
     \affiliation{\JLAB} \affiliation{\WM}
\author{I.G.~Aznauryan}
     \affiliation{\YEREVAN}
\author{V.D.~Burkert}
     \affiliation{\JLAB}
\author{K.A.~Griffioen}
     \affiliation{\WM}
\author{K.~Joo}
     \affiliation{\UCONN} \affiliation{\JLAB}
\author{R.~Minehart}
     \affiliation{\VIRGINIA}
\author{L.C.~Smith}
     \affiliation{\VIRGINIA}
  
\author{G.~Adams}
     \affiliation{\RPI}  
\author{P.~Ambrozewicz}
     \affiliation{\FIU}
\author{E.~Anciant}
     \affiliation{\SACLAY}
% ...

\collaboration{The CLAS Collaboration}
     \noaffiliation
%

%
%
%
\date{\today}
%
%
\begin{abstract}
The $ep \rightarrow e^{\prime} \pi^{+} n$ reaction was studied in the 
first and second nucleon resonance regions in the $0.25$~GeV$^2 < Q^{2} < 0.65$~GeV$^2$ 
range using the CLAS detector at Thomas Jefferson National Accelerator Facility. 
For the first time the absolute cross sections were measured covering nearly the full angular 
range in the hadronic center-of-mass frame. The  structure functions $\sigma_{TL}$, 
$\sigma_{TT}$ and the linear combination $\sigma_{T}+\epsilon \sigma_{L}$ were 
extracted by fitting the $\phi$-dependence of the measured cross sections, and 
were compared to the MAID and Sato-Lee  models. 
\end{abstract}
%
%
%
\maketitle
%
%
%
%
\section{Introduction}
\begin{figure}
\begin{center}
  \epsfig{file=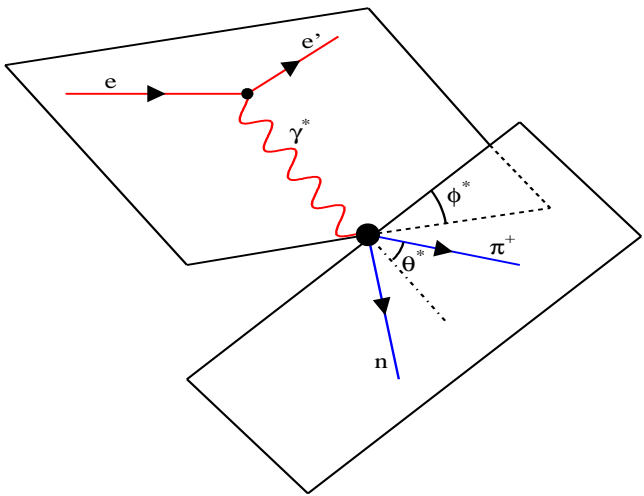,  totalheight=6.5cm, width=8.5cm, angle=0}
  \caption[Kinematic diagram for single pion production]
   {(Color online) Kinematic diagram of  single-pion electroproduction.}
 \label{FigKin}
\end{center}
\end{figure}

The structure of the nucleon and its  excited  states 
has been one of the most extensively studied subjects in nuclear 
and particle physics for many years. It allows us to 
understand important aspects of the underlying theory of the 
strong interaction, QCD, in the confinement regime where solutions 
are very difficult to obtain. Elastic electron scattering experiments provide 
information on the ground state of the nucleon, while   
studying the $Q^2$ evolution of the transition amplitudes from the nucleon 
ground state into the excited states provides insight into the internal 
structure of the excited nucleon. Single-pion electroproduction 
is one of the most suitable processes for studying the 
transitions to states  with masses below $1.7$~GeV 
because of the large $\pi N$ coupling for these 
states \cite{Burkert:04}.  The detection of  two out of three outgoing 
particles is sufficient to achieve a complete measurement of the 
differential cross sections in order to attempt the extraction of 
the amplitudes for the individual resonances. 
The kinematic quantities of the  $ep \rightarrow e^{\prime} \pi^{+} n$
 reaction is shown in Fig.~\ref{FigKin}.
The virtual photon is described by the  
four-momentum transfer $Q^{2}$, energy transfer $\nu$ and the polarization 
parameter $\epsilon$:
%
%
\begin{eqnarray}
   && Q^{2} =  4 E_{i} E_{f} \sin^{2}{\frac{\theta_{e}} {2} } , 
\label{Eq_Q2_Def}   \\
   && \epsilon  =  \left[ 1 + 2 \left( 1 + \frac {\nu^2} {Q^2} \right) 
        \tan^{2}{\frac {\theta_e} {2}} \right] ^{-1} ,   
\label{Eq_Eps_Def}  \\
   && \nu = E_{i} - E_{f},
\label{Eq_Nu_Def} 	
\end{eqnarray}	
%
%
where $E_{i}$ and $E_{f}$ are the initial and final energies of the 
electron and $\theta_{e}$ is the electron scattering angle. The mass of 
the hadronic system is given by:
%
%
\begin{eqnarray}
    W &=& \sqrt{ M^{2} + 2 M \nu - Q^{2} } , 
\label{Eq_W_Def} 	
\end{eqnarray}	
%
%
where  $M$ is the 
proton mass. The two hadron production angles $\theta$ and $\phi$ are 
defined in the center-of-mass (c.m.) reference frame, with $\theta$ being the 
angle between the outgoing pion and the direction of the three-momentum transfer, 
and $\phi$ being the angle between the electron scattering plane 
and the hadron production plane.
The unpolarized cross section 
for  single-pion electroproduction can be written as \cite{Foster:83}:
%
%
\begin{eqnarray}
   &&\frac {\partial^{5}\sigma} {\partial E_{f} \partial \Omega_{e} \partial \Omega_{\pi}^{*}}  = 
     \Gamma \cdot \frac {d \sigma} {d \Omega_{\pi}^{*}} , 
\label{Eq_SPP_CS}   \\ 
   &&\Gamma  =  \frac{\alpha} {2\pi^{2}Q^{2}} \frac{(W^{2} - M^{2} )E_{f} }{M E_{i}} 
   \frac{1}{1-\epsilon} ,   
\label{Eq_GammaFactor} \\
   &&\frac {d \sigma} {d \Omega_{\pi}^{*}}  =  
     \sigma_{T} + \epsilon \sigma_{L} +                    \nonumber \\ 
	&& \epsilon \sigma_{TT} \cos{2 \phi} + 
     \sqrt{2 \epsilon (1+\epsilon)} \sigma_{TL} \cos{\phi}  ,   
\label{Eq_SSP_CS_PHT}
\end{eqnarray}
%
%
where $\Gamma$ is the 
virtual photon flux, and $ \frac {d \sigma} {d \Omega_{\pi}^{*}}$ is the virtual
photoproduction cross section. The $\sigma_{T}$, $\sigma_{L}$, $\sigma_{TT}$ 
and $\sigma_{TL}$  structure functions are bilinear combinations 
of the  helicity amplitudes, depending only on the variables $Q^2$, $W$ and $\theta$. 
The analysis of the angular distributions  
provides information for extracting the electroproduction 
amplitudes for different resonances.
%
%
%
%
%
%

\begin{figure}
\begin{center}
  \epsfig{file=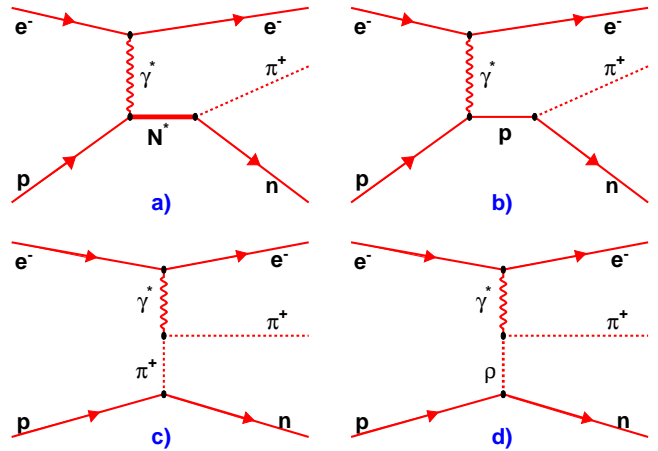,  totalheight=8.5cm, angle=270}
  \caption[Main Feynman diagrams for $\pi^{+}$ electroproduction ]
   {(Color online) Some of the main diagrams contributing to single $\pi^{+}$  
    electroproduction.}
 \label{FigFeynDiag}
\end{center}
\end{figure}

The main tree-level Feynman diagrams contributing to the $ep \rightarrow e^{\prime} \pi^{+} n$ 
process are shown in Fig.~\ref{FigFeynDiag}. The $s$-channel resonance 
excitation process is represented by the diagram in Fig.~\ref{FigFeynDiag}a. 
The hadronic vertex of this process is  known from  $\pi N$ elastic 
scattering experiments \cite{Arndt:04}. Therefore studies of  pion electroproduction 
can yield the $Q^2$ evolution  of the photocoupling amplitudes describing the 
$\gamma^{*} N N^{*}$ vertex. 
For the purpose of studying the excitation of nucleon resonances, 
the other diagrams are considered as physical background. 
The largest non-resonant contribution to the cross section comes from the 
$t$-channel pion exchange diagram, shown in Fig.~\ref{FigFeynDiag}c. 
Although this process mainly contributes in the forward region  due 
to the pion  propagator pole, it still accounts for a significant 
part of the cross section even at large angles. The diagrams in 
Fig.~\ref{FigFeynDiag}b and Fig.~\ref{FigFeynDiag}d correspond   
to the $s$-channel nucleon pole and $t$-channel $\rho$-meson 
exchange amplitudes. 
Sophisticated analysis procedures are necessary to separate the 
resonant contributions from the non-resonant background, 
and to extract the resonant amplitudes for different overlapping 
excited states. The extraction of resonance multipoles is beyond 
the scope of this paper. In this contribution we describe the experiment 
and data analysis, and the extraction of fully exclusive and differential 
cross sections, and determination of response functions.
%
%
%
%
%
%
%
%

Electroexcitation  of a nucleon resonance  can be described in terms of 
three photocoupling amplitudes $A_{1/2}$, $A_{3/2}$ and  $S_{1/2}$.
The first two are due to the coupling of transverse photons with the proton 
resulting in a combined helicity $h=\frac{1}{2}$ or $h=\frac{3}{2}$ 
respectively. The $S_{1/2}$ amplitude is present due to the possibility 
of a longitudinal polarization for virtual photons. Alternatively, pion  
electroproduction  can be described using multipole amplitudes 
$E_{l\pm}$, $M_{l\pm}$ and $S_{l\pm}$. The 
$l$-index represents the orbital angular momentum of the $\pi N$ 
system, and the $\pm$ sign indicates how the nucleon spin is coupled to the 
orbital momentum. For each excited state the helicity amplitudes 
can be expressed in terms of multipole amplitudes and vice versa \cite{Foster:83}.

Quark models predict that the $\frac {E_{1+}}{M_{1+}}$ 
and $\frac {S_{1+}}{M_{1+}}$ ratios for the $P_{33}(1232)$  are small 
 at low $Q^{2}$ \cite{Isgur:82,Capstick:90}, while perturbative QCD predicts  
$\frac {E_{1+}}{M_{1+}} = 1$ and$\frac {S_{1+}}{M_{1+}}$ is   
independent of $Q^2$ as $Q^2 \rightarrow \infty$ \cite{Carlson:86}. A transition between 
these two regimes is expected at some finite $Q^2$.  At low $Q^2$ the deviations of these 
ratios from zero can be interpreted as non-spherical deformation of the nucleon or the 
$\Delta(1232)$ \cite{Buchmann:01}.  Usually these ratios for $\Delta(1232)$ 
are obtained through measurements in the $\pi^{0} p$ decay channel with an 
assumption that the uncertainty  due to the isospin $I=\frac{1}{2}$ background 
is negligible. High quality 
data in the $\pi^{+} n$ channel  will enable us to separate the isospin 
$I=\frac{1}{2}$ and $I=\frac{3}{2}$ components of the transition form-factors
for the $P_{33}(1232)$ and to determine these ratios with smaller uncertainties 
coming from non-resonant contributions.

The second resonance region is dominated by the three known isospin 
$I=\frac{1}{2}$ states, $P_{11}(1440)$, $D_{13}(1520)$ and $S_{11}(1535)$.
These resonances, produced in electron-proton  scattering, are twice as 
likely to decay through  the $\pi^+ n$ channel than through $\pi^0 p$. Therefore, 
cross section measurements of  the $ep \rightarrow e^{\prime} \pi^{+} n$ 
process are crucial for understanding the properties of these states.  
The nature of the $P_{11}(1440)$ resonance is not understood in the framework of the 
constituent quark model (CQM) \cite{Isgur:85}, and there are suggestions 
that the Roper resonance may be a hybrid state \cite{Burkert:92}  
or a small quark core with a large vector meson cloud \cite{Gonzalez:98}.
The $Q^2$ evolution of the $A_{1/2}$ photocoupling amplitude for the Roper is 
predicted to be different for $3$-quark  and hybrid states. Previous analyses 
\cite{Arndt:02,Gerhard:80} indicate a rapid fall-off of $A_{1/2}$ between $Q^2=0$ and 
$Q^2=0.5$~GeV$^2$, therefore high quality data in this region will be very 
valuable in understanding the nature of the $P_{11}(1440)$.

The experimental data for the $A_{1/2}$ transition amplitude for $S_{11}(1535)$ 
show a significantly slower $Q^2$ fall-off than predicted by constituent 
quark models. Most of these results are obtained through analysis 
of  $\eta$-meson electroproduction data, where there can be no  
$I=\frac{3}{2}$ background. The proximity of the $S_{11}(1535)$ mass 
to the $\eta$-production threshold complicates the analysis of the data.
High quality  single $\pi^{+}$ data currently 
exist only at the photoproduction point, and there is very little data for 
non-zero $Q^2$. The results from analyses of pion and $\eta$ photoproduction 
data are significantly different \cite{Hagiwara:02}, and the source of these discrepancies is 
still not understood. New electroproduction data will allow for a similar comparison 
between the results from the two channels from CLAS to check the consistency 
of the analysis frameworks. These data will also allow for a future  combined 
analysis of pion and $\eta$ production data, which will provide more stringent 
constraints on the fit.

Until now there have only been three experiments  
\cite{Evangelides:74,Breuker:78,Breuker:82} 
measuring  single $\pi^{+}$ electroproduction cross section in the resonance 
regions in this range of $Q^{2}$. In all of these experiments the lack of  
angular coverage in the center-of-mass reference frame 
significantly reduced the sensitivity 
 to the resonant amplitudes.  The aim of the present experiment 
is to provide differential cross sections for the $\pi^{+} n$ channel  over 
a large kinematic region and  with  high statistical accuracy, that can be 
used together with other channels to obtain more reliable results on the 
resonance photocoupling amplitudes.
\section{Experiment}
The measurement was carried out  using the CEBAF Large Acceptance Spectrometer 
(CLAS) \cite{Mecking:03} at the Thomas Jefferson National Accelerator 
Facility (Jefferson Lab),
located in Newport News, Virginia. CLAS is a nearly 
$4\pi$ detector, providing almost complete angular coverage 
for the $e p \rightarrow e^{\prime} \pi^{+} n$ 
reaction in the center-of-mass frame. It is well suited 
for conducting  experiments which require detection of two or more 
particles in the final state. Such a detector and the continuous 
beam produced by CEBAF provide excellent conditions for measuring the 
$ep \rightarrow e^{\prime} \pi^{+} n$ electroproduction cross section by 
detecting the outgoing electron and pion in coincidence.
 \subsection{Apparatus}
\begin{figure}
\begin{center}
  \epsfig{file=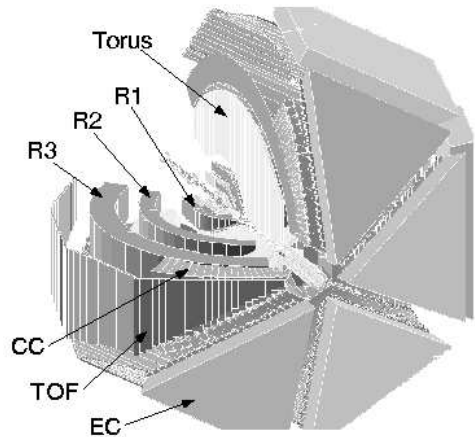,  totalheight=6cm, angle=0}
  \caption[Three dimensional view of CLAS]
   {Three dimensional view of CLAS.}
 \label{FigCLAS}
\end{center}
\end{figure}

   The main magnetic field of CLAS is provided by  six 
superconducting coils, which produce an approximately toroidal field  in the 
azimuthal direction around the beam axis. The gaps between the cryostats 
are instrumented with  six identical 
detector packages, also referred to here as  ``sectors'', as shown in 
Fig~\ref{FigCLAS}. Each sector consists of three regions (R1, R2, R3) of 
Drift Chambers (DC) \cite{Mestayer:00}  to determine the trajectories 
of the charged particles, {\v C}erenkov Counters (CC)  \cite{Adams:01} 
for electron identification, Scintillator Counters (SC) \cite{Smith:99} 
for charged particle identification using the Time-Of-Flight  
(TOF) method, and Electromagnetic Calorimeters (EC) \cite{Amarian:01} used 
for electron identification and detection of neutral particles. The liquid-hydrogen 
target was located in the center of the detector on the beam axis. To reduce the 
electromagnetic background resulting  from M{\o}ller scattering off atomic electrons,
 a second smaller normal-conducting toroidal magnet 
(mini-torus) was placed symmetrically around the target. 
This additional magnetic field  prevented  the  
 M{\o}ller electrons from reaching the detector volume. 
A totally absorbing Faraday cup, located at the very end of the beam line, 
was used to determine the integrated beam charge passing through the target. 
%
%
%
%
%
%
%
The CLAS detector can provide $\frac {\delta p} {p} <0.5\%$ momentum resolution, 
and  $\approx 80\%$ of $4 \pi$  solid-angle coverage. 
The efficiency of  detection and  reconstruction 
for stable charged particles in  fiducial regions of CLAS is $\epsilon > 95\%$. 
The combined 
information from the tracking in  the DC and the TOF systems allows us to 
reliably separate protons from positive pions for momenta up to $3$~GeV. 
 \subsection{Data taking and data reduction}
    The data were taken in the spring of $1999$ as  part of the  experimental 
program of the CLAS collaboration.  The CEBAF  $1.5$~GeV electron beam  was incident on 
a $5$-cm long  liquid hydrogen target at $20.5$~K temperature. 
The data were taken at $3$~nA nominal beam current, with $\pm 0.04$~nA current fluctuations, 
at luminosities of $\sim 4 \times 10^{33}$~cm$^{-2}$s$^{-1}$. The size 
of the beam spot at the target was  $\sim 0.2$~mm, with position 
fluctuations of $\pm 0.04$~mm.  
The  main torus current was set at 
$1500 $~A, which created a magnetic field of about $0.8$~Tesla in the forward direction. 
The magnetic field of the spectrometer is significantly lower at large  angles. 
The CLAS event readout was triggered by a coincidence of  signals from the 
electromagnetic calorimeter  and the {\v C}erenkov counters in 
a single sector, generating an event rate of $\sim 2$~kHz.  The total number 
of accumulated triggers at these detector  settings was about $4.5 \times 10^{8}$. 
The raw data were written onto a tape silo of the  Jefferson Lab Computer 
Center. During the off-line processing each file was retrieved 
from the tape silo and analyzed 
to produce files for  general use containing 4-vectors of the reconstructed particles. 
ROOT \cite{Brun:97} files, containing the specific  information 
relevant for single $\pi^{+}$ electroproduction, were created and stored on a disk. 
These files were further analyzed  to extract the differential  
cross sections for the  $e p \rightarrow e^{\prime} \pi^{+} n$ reaction.
 \subsection{Particle identification}
%
%
%
%
\begin{figure}
\begin{center}
  \epsfig{file=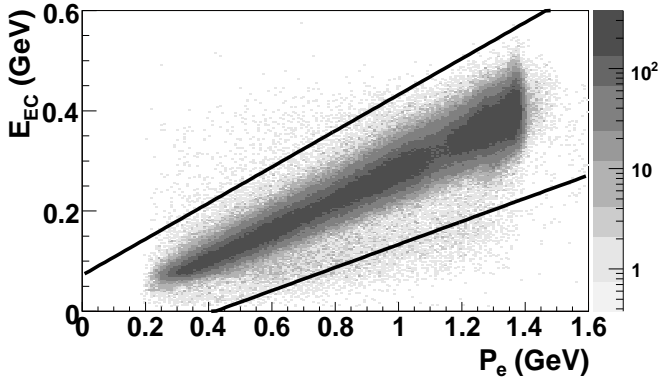,  totalheight=5.5cm, width=8.8cm, angle=0}
  \caption[Electron ID cut]
	{Energy deposited by the electron candidates in the 
	electromagnetic calorimeter versus their momenta. The 
	black lines show the cut applied for the electron identification.}
 \label{FigEID}
\end{center}
\end{figure}
%
%
%
One of the key issues in electron scattering experiments is the ability of 
the detector to reliably identify  electrons. Electron identification 
at the trigger level was accomplished  by requiring a minimum amount of deposited energy in 
the electromagnetic calorimeter in coincidence with a signal in the {\v C}erenkov 
counter in the same sector. 
Additional requirements were applied in the off-line 
analysis to select events containing an electron. 
First, a geometrical matching was required between the EC and CC hits and the 
associated negatively charged tracks in the drift chambers. 
The values of the geometrical cuts in the software are given in 
Table~\ref{TableGeomCuts}. 
A sampling fraction cut was imposed on the dependence of the EC visible energy
on the momentum to reject the background coming from negative pions (see 
Fig.~\ref{FigEID}).  The electron identification in the off-line analysis can 
be summarized by:
%
\begin{equation}
    EID  = TRK \otimes CC \otimes EC \otimes SF ,   
\label{EqElectronID}
\end{equation}
%
where $TRK$ stands for track reconstruction in the drift chambers, 
$CC$ and $EC$ are the {\v C}erenkov 
counter and the calorimeter geometrically matched hits and  
$SF$ is the sampling fraction cut described above. In order to avoid  
inefficiencies due to the trigger threshold in the electromagnetic 
calorimeter, only events containing an electron with at least $500$~MeV 
momentum were used in the analysis.
%
%
%
\begin{table}
\begin{center}
\begin{tabular}{|c|c|}   \hline
        Matching               &  Tolerance   \\  \hline
\hline
        $TRK \otimes EC$       &   $30$~cm    \\  \hline
        $TRK \otimes CC$       &   $5^o$      \\  \hline 
        $EC  \otimes CC$       &   $5^o$       \\  \hline 
\end{tabular}
\end{center}
\caption[Cuts for geometrical matching]
{Cuts for the geometrical matching in the offline analysis software.}
\label{TableGeomCuts}
\end{table}
%
%
%
In addition,  fiducial cuts, discussed later, were applied to select only electrons 
in the regions where the {\v C}erenkov counter efficiency was greater than $92\%$. The 
final cross  sections were corrected for the remaining inefficiency 
of the {\v C}erenkov counters \cite{Adams:01}. 

%
%
%
\begin{figure}[b]
\begin{center}
  \vspace{3mm}
  \epsfig{file=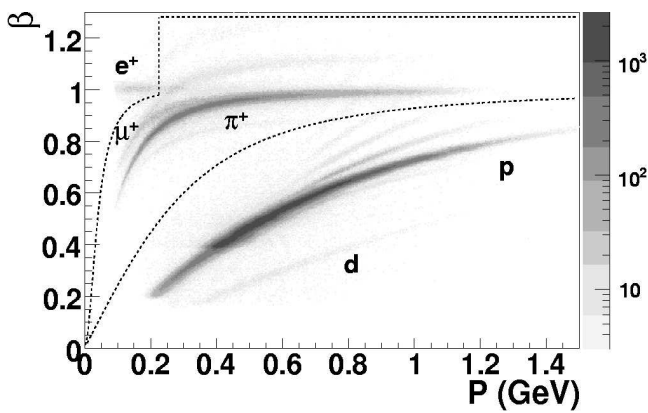,  totalheight=8.5cm, width=5.3cm, angle=270}
  \caption[Charged particle ID]
   {Distribution of the number of positively charged particles versus 
	$\beta$ and $p$. The visible bands are due to  positrons, muons, 
	pions, protons and deuterons. All positive particles within the area  
	outlined by the dashed lines are considered as $\pi^{+}$'s in this 
	analysis. }
 \label{FigPID}
\end{center}
\end{figure}
%
%
%
%
Charged hadron identification in the CLAS detector is accomplished using the momentum 
determined from the tracking and the timing information from the scintillation counters. 
Fig.~\ref{FigPID} shows the distribution of  positively charged particles   
 at $1.515$~GeV electron beam energy plotted versus velocity 
$\beta$ and momentum $P$.  Bands due to positrons, pions, 
protons and deuterons can be easily identified. 
At low  momentum  the muon band is visible as 
well. The deuterons are produced from electron scattering on the aluminum 
windows of the target cell.
All positive particles in the region outlined by the dashed lines 
were considered as $\pi^{+}$. Positrons can be separated from pions at low momenta, 
but at higher momenta the pion and positron bands merge. Background 
due to muons and positrons is significantly reduced by the missing mass 
and  vertex cuts described below. The remaining contamination 
is evaluated as a systematic uncertainty.
 \subsection{Momentum corrections}
%
%
%
\begin{figure}
\begin{center}
  \epsfig{file=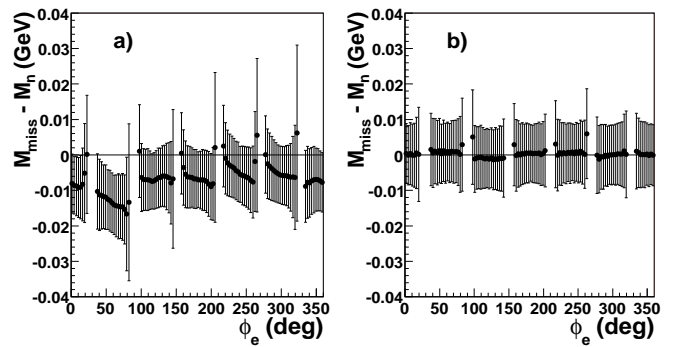,  totalheight=4.8cm, width=8.8cm, angle=0}
  \caption[Missing mass showing momentum corrections]
   {Position of the missing mass peak in the $e p \rightarrow e^{\prime} \pi^{+} X$ 
	reaction versus $\phi_{e}$ in the lab frame: a) - before, 
	b) - after momentum corrections are applied. The error bars 
	represent the width of the distribution around the neutron mass.  }
 \label{FigMomCor}
\end{center}
\end{figure}
%
%
%
%
%
When extracting the resonant parameters for  excited states, it is 
important to have the correct value for the invariant mass of the hadronic 
state. Therefore, it is necessary to measure the electron momentum with  high accuracy. 
For this reason additional  corrections were applied to the electron momentum 
reconstructed by the standard CLAS software package. 
These corrections were determined using  elastic scattering events from the 
same runs that were used in the single pion  analysis. 
It was found that the missing mass determined from the elastically 
scattered  electrons is typically $\sim 5$~MeV below  the proton mass 
$M_{p} = 0.938$~GeV. Assuming that the scattering angle of the electron is measured 
correctly, and using:
%
%
\begin{equation} 
       E_{f} = \frac { 2 M E_{i} - ( W^{2} - M^{2} )}  
	{2 M + 4 E_{i}  \sin^{2}{\frac{\theta_{e}} {2} }}
\label{EqEvsW}
\end{equation}
the momentum correction factor can be found as:
%
%
%
\begin{equation}
	C_{p} \equiv \frac {\delta E_{f}} {E_{f}} = - \frac{ W^2 - M^2 } { 2 M E_{i} },
\label{EqMomCor}
\end{equation}
where $E_{i}$ is the electron beam energy and $W$ is the measured recoil mass. 
This quantity was calculated for 
different bins in $\theta_{e} \in [15^{\circ}, 55^{\circ}]$ and 
$\phi^{*}_{e} \in [-30^{\circ}, +30^{\circ}]$ in the laboratory frame for 
each sector and stored in a look-up table. The azimuthal angle $\phi^{*}_{e}$ 
is defined within a sector, with 
$\phi^{*}_{e}= 0$  corresponding to the mid-plane of the sector. 
The momenta of the electrons from single $\pi^{+}$ 
production data were corrected using this table on an event-by-event basis.
This procedure relies on the fact that the relative momentum offset is independent 
of $W$ for a fixed value of $\theta_{e}$. 
It was found that after these  corrections were applied, 
the neutron peak in the missing mass of the $ep \rightarrow e^{\prime} \pi^{+} X$ 
reaction was within $\pm 2.0$~MeV of the neutron mass for $1.1$~GeV$< W < 1.6$~GeV range. 
Fig.~\ref{FigMomCor}  shows the difference between the missing mass in 
$e p \rightarrow e^{\prime} \pi^{+} X$ reaction and the neutron mass with and 
without momentum corrections. The six gaps between the points are due to the six coils 
of the magnet. The dependence of the peak position on $\phi_{e}$ is practically eliminated 
by this procedure, and the peak is located much closer to the neutron mass. 
 \subsection{Fiducial cuts}
%
%
%
%
\begin{figure}[b]
\begin{center}
  \epsfig{file=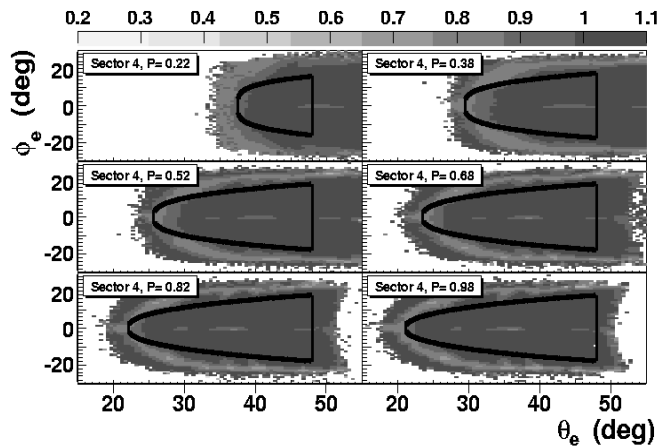,  totalheight=6.2cm, width=9.3cm, angle=0}
  \caption[Electron fiducial cuts]
   {{\v C}erenkov counter efficiency versus the $\theta_{e}$ and $\phi_{e}$  
	electron angles in the laboratory frame for Sector 4. The black curves indicate 
	the outer edges of the electron fiducial regions. Each momentum bin 
	is $200$~MeV wide.}
 \label{FigElFidReg}
\end{center}
\end{figure}
%
%
%
%
Although CLAS is a nearly $4\pi$ detector, it still contains significant 
inactive volumes without particle detectors. 
In addition, some of the detection inefficiencies in the active volumes 
are not adequately reproduced in the detector simulation software. These 
areas are near the edges of the electromagnetic calorimeter, {\v C}erenkov 
counter mirrors, the main torus and mini-torus coils, regions with broken  
wires in the drift chambers, and malfunctioning phototubes in the TOF system.
To eliminate  events with particles 
traveling through these regions, a set of fiducial cuts was developed. For electrons 
the main boundary of the  fiducial region was  defined by the efficiency of 
the {\v C}erenkov counters and the edges of the electromagnetic calorimeter. 
Fig.~\ref{FigElFidReg} shows the dependence of the  {\v C}erenkov counter 
efficiency versus the $\theta_{e}$ and $\phi_{e}$  electron angles in the laboratory 
frame for six $200$~MeV wide momentum bins in Sector 4. Due to the optics design of the 
CLAS {\v C}erenkov counters \cite{Adams:01} 
there are areas with relatively lower efficiency shown with 
the lighter shade. These features are difficult to implement in the detector simulation.
 Only  events in the regions within the black curves and with the  
{\v C}erenkov counter efficiency above $92\%$ were used in the analysis. 
An additional  set of geometrical cuts was applied 
to reject  electrons  hitting  malfunctioning scintillator counters 
or traversing  regions with missing or inefficient wires in the drift chambers. 

Two sets of fiducial cuts were used to define the outer boundary of the 
fiducial regions for the positive pions. The first set, similar to the 
electron cuts,  was defined in such a way that the $\phi^{lab}_{\pi}$ 
distributions of the number of events  be uniform within the 
fiducial region. The second set of cuts was applied to ensure equivalent solid 
angle coverage for pions in the Monte-Carlo simulation and the real data. 
This mismatch was due to distortions of the minitorus coils which were not 
implemented in the detector simulation package. As in the 
case with the electrons, tracks in the regions with malfunctioning 
scintillator counters or broken  wires were rejected by another set of fiducial 
cuts.

 \subsection{Kinematic cuts}

\begin{figure}[h]
\begin{center}
  \vspace{2mm}
  \epsfig{file=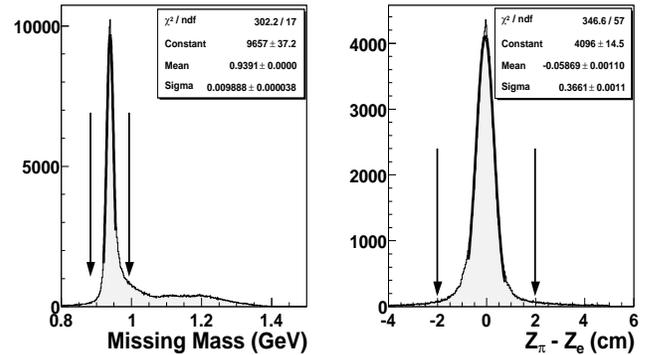,  totalheight=5.2cm, width=8.7cm, angle=0}
  \caption[Missing mass and vertex cuts]
   {Missing mass spectrum (a) and the distribution of the events 
     versus $Z_{\pi}-Z_{e}$ (b). The arrows represent the cuts 
     applied in the data analysis. The solid lines are gaussian fits to 
     the data.}
 \label{FigMissMass}
\end{center}
\end{figure}

The exclusive final state  was selected by detecting the outgoing 
electron and the $\pi^{+}$, and by requiring that the missing particle be a neutron.
The missing-mass spectrum  in Fig.~\ref{FigMissMass}a shows 
 a prominent neutron peak as well as  events from the 
radiative tail and from  multi-pion production channels. The arrows 
indicate the cuts used in the analysis. 
The number of rejected events in the tails is recovered by 
imposing the same cuts on the simulated events in the acceptance 
calculations.
%
%
%
%
%
%

GEANT-based Monte-Carlo studies showed  that about $18\%$ of the positive 
pions decay in-flight into $\mu^{+} \nu_{\mu}$. Most of the 
momentum of the original pion is carried by the $\mu^{+}$, which  is, 
therefore, often detected 
and reconstructed as  a $\pi^{+}$ with a significantly different momentum vector.
In order to reduce the number of the events with decaying pions, a vertex cut 
$\mid Z_{\pi} - Z_{e} \mid < 2$~cm was applied on the difference of the $Z$-coordinates 
along the beam-line for the electron and the pion tracks in the same event  
(see Fig.~\ref{FigMissMass}b).
This led to a reduction in the number of events with decaying $\pi^{+}$ to 
$4\%$ with less than $1\%$ losses in the number of events when the pion did not 
decay. 

\begin{figure}
\begin{center}
  \epsfig{file=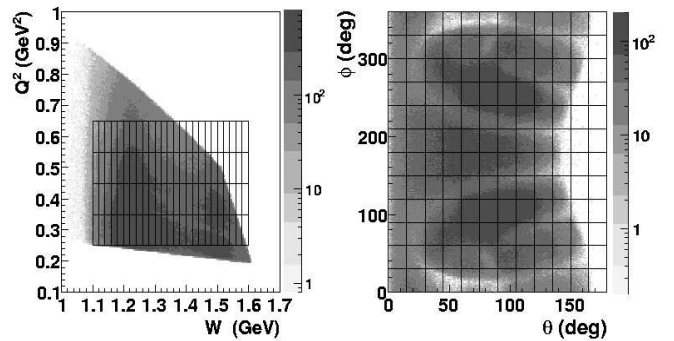,  totalheight=5cm, width=8.7cm, angle=0}
  \caption[Coverage and the binning]
   {Distribution of the number of single $\pi^{+}$ events  versus: 
        $Q^2$ and  $W$ - left; $\phi$ and  $\theta$ center-of-mass 
	angles - right.}
 \label{FigBinning}
\end{center}
\end{figure}

The kinematic coverage of this experiment is shown in Fig.~\ref{FigBinning}. The 
grating on the figures shows the binning of the data. In the first resonance 
region this experiment covers a $Q^2$ range from $0.25$~GeV$^2$ to  $0.65$~GeV$^2$, 
while in the second resonance region the upper boundary of the $Q^2$ coverage is 
reduced to $0.45$~GeV$^2$. 
The angular coverage in the hadronic center-of-mass frame is  nearly complete, 
with the exception of the region $\theta > 140^{\circ}$. This limitation at 
larger angles  is related to the fact that the CLAS coverage for charged particles 
is limited to $140^{\circ}$ in laboratory frame. The number and the sizes of the 
cross section bins are given in Table~\ref{TableDataBinning}. 
%
%
%
%
%
%
%
%
%
%
%
%
%
%
\begin{table} 
\begin{center}
\begin{tabular}{|c|c|c|c|c|}   \hline
   Variable     &   $\#$ of bins &   Lower limit       &    Upper limit       &    Width            \\  \hline
                         							                        \hline 
  $Q^2$         &   $4$          & $0.25$~GeV$^2$      &  $0.65$~GeV$^2$      & $0.10$~GeV$^2$      \\  \hline
  $W$	        &   $25$         & $1.1$~GeV           &  $1.6$~GeV           & $20$~MeV            \\  \hline
  $\theta$      &   $12$         & $0^{\circ}$         &  $180^{\circ}$       & $15^{\circ}$        \\  \hline
  $\phi$        &   $12$         & $0^{\circ}$         &  $360^{\circ}$       & $30^{\circ}$        \\  \hline       
\end{tabular}
\end{center}
\caption[The number and sizes of data bins]
{The number and the sizes of the data bins. Values for the  
limits indicate the upper and lower edges of the bins, rather than
the bin centers.}
\label{TableDataBinning}
\end{table}
%
%
%
%
%
%
%
%
%
 \subsection{Acceptance corrections}
In order to relate the experimental yields to cross sections,  acceptance correction 
factors were calculated using the Monte-Carlo method. The GEANT-based detector simulation 
package GSIM incorporated the survey geometry of  CLAS, realistic drift chamber and timing 
resolutions along with  missing wires and malfunctioning photomultiplier tubes. Because CLAS is a 
complicated detector covering almost $4\pi$ of solid angle, it is virtually impossible 
to separate the efficiency 
calculations from the geometrical acceptance calculations. In this work the term acceptance 
correction refers to a combined correction factor due to the geometry of the detector and 
the inefficiencies of the detection and reconstruction. It is defined as the ratio of the 
number of reconstructed Monte-Carlo events to the number of  simulated  events in a given bin:
%
%
\begin{equation}
 	A = \frac {N_{rec}} {N_{sim}}. 
\label{EqAccDef}
\end{equation}
With this  definition of the acceptance it is desirable to have a realistic 
physics model in the event generator because of the finite bin sizes and  
bin migration effects, which  are described later. 
In this work the MAID2000 model \cite{Drechsel:99}, 
which reasonably reproduces both $p \pi^{0}$ \cite{Joo:02} and the 
current $n \pi^{+}$  CLAS data, was used 
as an input to the Monte-Carlo event generator.  The simulated $200$ million events   
were processed using the same software package and analyzed  with the same 
cuts that were applied 
to the real data. An acceptance table with  $8 \times 25 \times 24 \times 48$ bins 
in $Q^2$, $W$, $\theta$ and $\phi$, respectively, was calculated 
using the definition in Eq.~(\ref{EqAccDef}). The fine binning of the acceptance 
look-up table reduces the model dependence of the cross sections. 
The statistical errors for the 
acceptance corrections were estimated using the binomial distribution:
%
\begin{eqnarray}
  \delta A_{bin} & = & \sqrt{ \frac{ A_{bin} ( 1 - A_{bin} ) } 
        { N_{gen} - 1  } } , 
\label{EqBinomErr}
\end{eqnarray}
%
where $N_{gen}$ is the number of the Monte-Carlo events generated in the bin.
These errors are included in the statistical error of the final cross sections.
%
%
%
\begin{figure}
 \begin{center}
  \epsfig{file=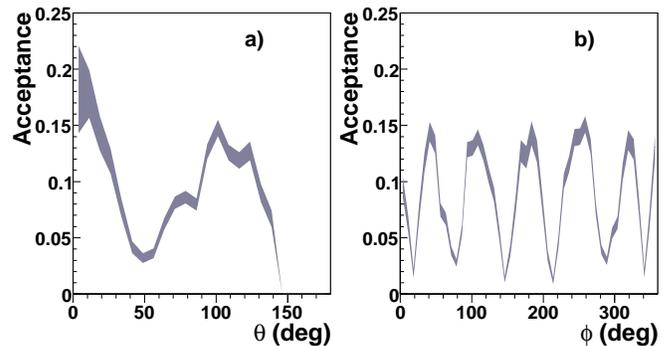, height=5.0cm, width=8.7cm, angle=0}
  \caption[Sample plot of CLAS acceptance in the $Q^2=0.3$~GeV$^2$ bin.]
        {Sample plots of acceptance corrections versus $\theta$   
	and $\phi$ pion angles in the center-of-mass frame in the  
	$Q^2=0.3$~GeV$^2$ and $W=1.23$~GeV bin. The $\theta$-dependence 
	(a) is shown at $\phi=116.25^{\circ}$ and the $\phi$-dependence is 
	at $\theta=108.75^{\circ}$. The width of the curves represent 
	the statistical uncertainty for the acceptance.}
    \label{FigAccepCor}
 \end{center}
\end{figure}
%
%
%

The acceptance of CLAS for single-pion electroproduction at $Q^2=0.3$~GeV$^2$ 
and $W=1.23$~GeV is shown in Fig.~\ref{FigAccepCor}. The $\theta$-dependence 
of the acceptance (see Fig.~\ref{FigAccepCor}a) 
exhibits a dip at $\sim 45^{\circ}$, which is due to the 
forward beam pipe. Six sectors of CLAS can be clearly identified in the 
plot showing acceptance versus $\phi$-angle  (see Fig.~\ref{FigAccepCor}b). 
The width of the curves in these graphs represent the statistical error bands. 
Since a single cross section bin contains sixteen acceptance bins, 
the contribution of the acceptance statistical error to the total  
uncertainty of cross sections is on average approximately four times 
smaller than the errors seen in these plots. 
 \subsection{Radiative corrections}
In addition to processes which result in the exclusive $e^{\prime} \pi^{+} n$ final state, 
there are radiative processes represented by Feynman diagrams similar 
to the original single photon exchange diagrams,  but with an additional photon leg, 
that  also contribute to  the cross sections. The experimentally 
measured cross sections must be corrected for such  processes, also known as internal radiation. 
The radiative cross section for an exclusive process can be written as \cite{Afanasev:02} :
%
%
%
%
\begin{eqnarray}
d\sigma_{r} & =  & \frac { (4 \pi \alpha )^{3} dQ^{2} dW^{2} d\Omega_{\pi}^{*} }
	            { 2 (4 \pi )^{7} S^{2} W^{2} } \times \nonumber \\ 
	            & & \int{ d\Omega_{k} dv \frac{v \sqrt{\lambda_{W}}} { f_{W}^{2} Q^{4} } 
	                                          L_{\mu \nu}^{(r)} W_{\mu \nu} }, 
\label{EqRadCSDef}
\end{eqnarray}
where $S \equiv 2 E_{i}M_{p}$, $d\Omega_{\pi}^{*}$ is the differential center-of-mass 
solid angle of the $\pi^+$, $v \equiv M_{X}^2-M_{n}^2$,   
$L_{\mu \nu}^{(r)}$ and $W_{\mu \nu}$ are the leptonic and the hadronic tensors respectively, 
and
\begin{eqnarray} 
\lambda_{W} & \equiv & ( W^2-m_{\pi^+}^{2}-M_{miss}^{2} )^{2} -4m_{\pi^{+}}^{2}W^2  
                                                                \label{EqLamdaWDef} \\
f_{W} & \equiv & W - E_{\pi} + p_{\pi}  ( \cos{\theta_{\pi}} \cos{\theta_{k}} 
	                                    \nonumber \\
             & & + \sin{\theta_{\pi}} \sin{\theta_{k}} \cos{(\phi_{\pi}-\phi_{k} )} ) .	    	
 \label{EqFWDef}
\end{eqnarray}
%
%
Here, $\theta_{\pi}$,  $\phi_{\pi}$, $\theta_{k}$ and  $\phi_{k}$ are the pion and
radiated photon's angles in the hadronic center-of-mass reference frame. The integral 
in Eq.~(\ref{EqFWDef}) is taken over the photon angles and variable $v$. 

In addition,  there is also  a nonzero probability 
that in the presence of the electromagnetic field of the atoms of the target the electron
will emit one or more photons before or after interacting with the nucleus of the target  
(external radiation). The  probability of emitting a real photon of a particular energy 
is proportional to the path length of the electron in the target material. 
The size of the external radiative  corrections for these measurements was significantly 
smaller than for internal because of the small amount of the target material ($t = 0.5\%$ of 
radiation length).

\begin{figure}
 \begin{center}
  \epsfig{file=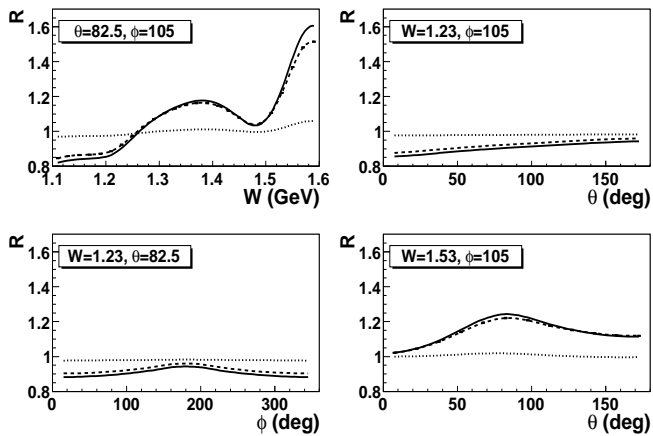, height=6.0cm, width=8.8cm, angle=0}
  \caption[Sample radiative corrections at $Q^2=0.3$~GeV$^2$.]
        {Sample plots of the radiative correction factor $R$ at 
	$Q^2=0.3$~GeV$^2$. The dotted lines are the external corrections, 
	the dashed lines are internal corrections, and the solid lines show 
	the combined radiative correction. }
    \label{FigRadCor}
 \end{center}
\end{figure}

The internal radiative corrections for the cross sections were calculated using the \textit{ExcluRad} 
program \cite{Afanasev:02} as multiplicative correction factors for each data bin. 
The external radiative corrections were done using the Mo and Tsai formalism \cite{Mo:69}.
The unradiated structure functions, needed as an input for the correction procedure, 
were calculated using a parameterization of the multipole amplitudes using a fit of these 
CLAS data based on the unitary isobar model \cite{Aznauryan:03}. 
The size of the required corrections varied up to $55\%$, 
depending on the kinematics. Fig. \ref{FigRadCor} shows dependences of the radiative 
corrections $R \equiv \frac {\sigma_{rad}} {\sigma_{Born}}$ on the variables $W$, $\theta$ and 
$\phi$  at $Q^2=0.3$~GeV$^2$. The dotted line shows the  
correction due to external radiation, and the dashed line is the correction factor obtained 
using the \textit{ExcluRad} program. The solid line is the combined correction 
factor calculated as the product of the two. Because of the short length of the target, 
the external radiative corrections are much smaller than the internal corrections.
%
%
%
%
%
%
%
%
 \subsection{Corrections for binning effects}
Because of the finite detector resolution and finite bin size, the measured 
values of the cross section in the center of the data bin can be distorted by 
up to $10\%$. The 
experimentally measured quantity is the cross section averaged over a full data 
bin, while usually it is more desirable to determine the value of the cross section 
at the center of the bin.
To account for such distortions, multiplicative corrections were introduced 
as the ratio of the cross section in the center of a bin to the average cross section 
in that bin:
\begin{eqnarray}
    B &=& \frac{ \frac {d\sigma} {d\Omega_{\pi}^{*}}  \mid_{ctr} }
	       { \frac {d\sigma} {d\Omega_{\pi}^{*}}  \mid_{avg} }.
\label{EqBinCorr}
\end{eqnarray} 
The averaged cross sections were evaluated using two models: the $Q^2$ dependence of 
the cross sections was taken from the MAID2000 model \cite{Drechsel:99}, while the 
cross sections at fixed values of $Q^2$ were obtained using a unitary isobar
\cite{Aznauryan:03} fit to these data in the first iteration.  
 \subsection{Normalization}
\begin{figure}
 \begin{center}
  \vspace{4mm}
  \epsfig{file=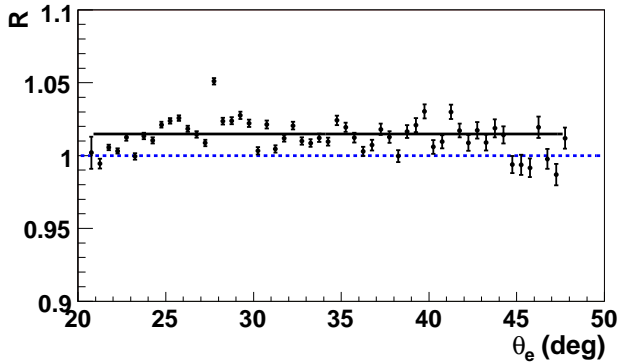, totalheight=5.4cm, width = 9.2cm, angle=0}
  \caption[Elastic cross section from CLAS]
        {The ratio of the measured elastic  cross section to  
	the  parameterization of the  
	world data \cite{Bosted:95}. The error bars represent the statistical 
	uncertainty only. The solid line is from the fit of the data points 
	to a constant.}
    \label{FigElastic}
 \end{center}
\end{figure}

The integrated charge of the electron beam passing through the target was measured 
using the Faraday cup located at the end of the Hall B beam line.
It generated pulses with a frequency proportional to the beam current with $10$~Hz per $1$~nA 
linear slope parameter. The calibration parameters of this device are known with less  
than $0.5\%$ uncertainty. The measured charge was 
corrected for the data acquisition  live-time, calculated as the ratio of the counts 
from two scalers. These scalers  were connected to a single $100$~kHz pulse 
generator. One of them was ungated, while the other one was gated by the data acquisition 
``live'' signal. To ensure the quality of the analyzed data sample,  
software cuts were imposed on the live-time, elastic scattering 
and single $\pi^{+}$ electroproduction rates. The portions of the runs for which these  
quantities were outside of the imposed limits were excluded from the analysis, 
with the corresponding beam charge being subtracted from the total charge. 
As was mentioned above, the {\v C}erenkov counter efficiencies were parameterized during 
the calibration procedure \cite{Adams:01}, and the appropriate corrections were 
applied to the cross sections. The comparison of the elastic 
scattering cross sections versus $\theta_{e}$ from CLAS and the model calculation 
using a parameterization \cite{Bosted:95} for the elastic form factors 
is shown in Fig.~\ref{FigElastic}. The model cross section includes  
radiative effects, according to the Mo and Tsai formalism \cite{Mo:69}. 
The error bars on the data points represent statistical uncertainties only. 
The solid line at $R=1.015$ shows the result of  fitting a constant to the 
ratio of the measured cross sections to the parameterization \cite{Bosted:95}. 
The fluctuations around this line can be used to estimate the systematic 
uncertainty of electron detection and reconstruction. 

The contributions from  scattering off the target cell walls was estimated 
to be $1.5\%$ using empty target runs. This correction factor was applied to 
the cross section in every data bin. 

%
%
%
%
%
%
%
%
\section{Systematic Errors Studies}
 A number of studies were carried out to estimate the systematic uncertainties 
on the measured cross sections. The primary  method used in these studies 
 was to vary different 
independent parameters of the analysis to determine the corresponding 
change in the resulting cross sections and the structure functions. 

Because of the finite bin size, the result of averaging 
the acceptance over an acceptance bin depends on the distribution of  
events in that bin. If the physics model used in the Monte-Carlo 
simulation differs from the real  data, then the averaging over a bin may result 
in an incorrect cross section. The  introduced error depends
on the shape of the acceptance function and the cross sections as well as on 
the acceptance bin size. In addition, because of the finite detector resolution, 
some of the events produced in one acceptance bin will be 
reconstructed in a different bin. This may cause significant distortions 
in the final cross section distributions. In order to correctly  account 
for these  effects, a  realistic physics generator and detector simulation 
are required. 
To estimate the errors of the final results due to the model used in the 
acceptance calculations, we calculated the acceptance table with two different models. 
The comparison of the results with the two  acceptance corrections  allowed us to 
estimate  the systematic errors due to the physics model in the acceptance 
calculations. 

As was mentioned above, we use missing mass and  vertex cuts 
to select the single-pion production events and to reduce the number 
of events with decaying pions. These cuts  cause losses of some single-pion events as well. 
The true number of events is expected to be recovered by applying 
the acceptance corrections by using exactly the same cuts 
on the Monte-Carlo data. The remaining systematic errors associated with 
these cuts were estimated by varying the sizes of the windows. 
The absolute value  of the cross section variations calculated 
with different cut windows, averaged over $\phi$ at fixed 
$Q^2$, $W$ and $\theta$, was considered as the systematic uncertainty for all $\phi$ 
for that fixed $Q^2$, $W$ and $\theta$.  

One of the possible sources of systematic errors in this experiment is the uncertainty in  
the normalization. This can arise from miscalibrations of the Faraday cup, target
density variations,  errors in determining  the target length and its   
temperature along with data acquisition live-time and other factors. 
However, the presence of the elastic 
events in the data set allows us to account for  the normalization uncertainties 
of the cross sections by comparing the elastic cross sections 
to the parameterization of the world data \cite{Bosted:95}. 
This way we were able to combine the normalization, 
electron detection, electron tracking and electron identification errors into one 
global uncertainty factor. A comparison of the measured elastic cross sections  
for different $\theta_{e}$ and $\phi_{e}$ electron angles  allowed us to 
assign a $5.2\%$ global uncertainty due to the normalization and electron 
efficiency uncertainties.

The systematic uncertainty due to the model used in the radiative corrections 
was estimated by performing a second iteration. The radiatively corrected 
experimental cross sections from CLAS were fitted once more, and using the fit the 
new correction factors were calculated and compared with the previous 
iteration. The comparison indicated an uncertainty on the order of $~2\%$ 
due to the model dependence of the radiative corrections. 

Using the kinematically over-determined reaction 
$ep \rightarrow e^{\prime} \pi^{+} \pi^{-} p$ allows us to determine 
the $\pi^{+}$ efficiency by detecting the outgoing electron, $\pi^{-}$ and 
proton. The efficiency of the $\pi^{+}$ detection can be found as the ratio 
of the number of events where the $\pi^{+}$ was detected to the number of 
events where the $\pi^{+}$ was expected to be detected. 
 A comparison of the pion efficiency calculated from the real data with the 
efficiency from GEANT-based Monte-Carlo  simulation lead to a systematic error 
estimate of $2.5\%$. 

In order to estimate the background coming from  two-pion production, a  
sample of two-pion Monte-Carlo events  was processed as if it were the actual data sample. 
The analysis of these events showed that this background would contribute less than $1\%$ 
uncertainty to the differential cross sections. The systematic error due to 
the $\pi^{+}$ misidentification was estimated to be about $\sim 0.5\%$ by 
varying the cut in the proton-pion separation in the analysis. 

The total systematic error in each bin was calculated as the square root of 
the sum of the squares of these different contributions. The size of the systematic 
errors is typically slightly larger than the size of the statistical uncertainties 
and is shown in Fig.~\ref{FigSampleCS} as the shaded bands.  
\section{Results} 
 \subsection{Cross sections}
\begin{figure}
 \begin{center}
  \epsfig{file=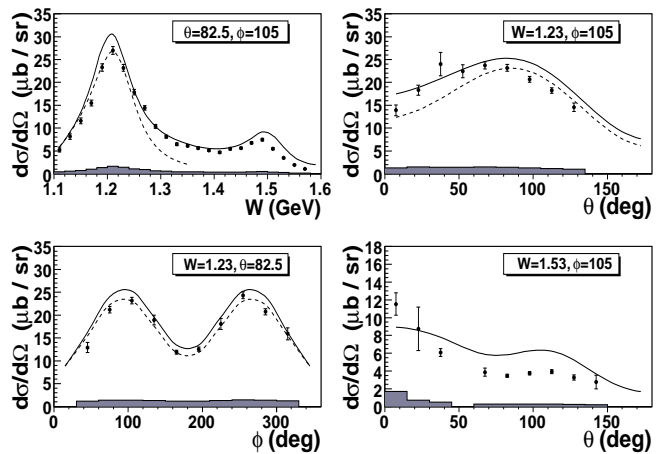, height=6.2cm, width=8.8cm, angle=0}
  \caption[Sample cross sections at $Q^2=0.3$~GeV$^2$.]
        {Sample plots of $\pi^{+}$ virtual photoproduction 
	cross sections at $Q^2=0.3$~GeV$^2$ for different 
	kinematics. The shaded bands represent the systematic 
        uncertainties. The solid curve is  MAID2003  \cite{Tiator:04} , 
	and the dashed curve is a calculation from Sato-Lee \cite{Sato:96}.}
    \label{FigSampleCS}
 \end{center}
\end{figure}

The experimental differential cross section for each data bin  was 
determined  using the following formula:
%
%
%
\begin{eqnarray}   
  && \frac {\partial^{5}\sigma} 
        {\partial E_{f} \partial \Omega_{e} \partial \Omega_{\pi}^{*}}  = 
  {R \cdot B}  \sum_{events} \frac { 1 }  { A L \epsilon_{cc}}  \cdot \nonumber \\
 && \cdot \frac {1} {\Delta Q^2 \Delta W \sin{\theta} \Delta \theta \Delta \phi } 
 \cdot \frac {\partial (W, Q^2)} {\partial (E_{f}, \cos{\theta_{e}})} ,  
\label{Eq_EXP_CS} \\
 &&L  =  \frac { Q_{tot} } { e } N_{A} \rho L_{T} ,   
\label{Eq_Luminocity} \\  
 && \frac {\partial (W, Q^2)} {\partial (E_{f}, \cos{\theta_{e}})}  =  
   \frac { 2 M E_{i} E_{f} } {W} ,  
\label{Eq_Jacobian}
\end{eqnarray}  
%
%
where the sum in Eq.~\ref{Eq_EXP_CS} runs over the $e \pi^{+} (n)$ events reconstructed 
in the fiducial region of CLAS. Here $A$ is the acceptance correction 
factor for an event, 
$L$ is the integrated luminosity, $N_{A}$ is Avogadro's number, $\rho$
is the target density, $L_{T}$ is the target length, $Q_{tot}$ is the 
integrated charge corrected for the data acquisition live-time, 
$e$ is the electron charge,   
$\epsilon_{cc}$ is the {\v C}erenkov efficiency correction factor, $\Delta Q^2$, 
$\Delta W$, $\Delta \theta$, $\Delta \phi$ are the  bin sizes, 
$ \frac {\partial (W, Q^2)} {\partial (E_{f}, \cos{\theta_{e}})}$ is
the Jacobian  between the $(W, Q^2)$ and $(E_{f}, \cos{\theta_{e}})$ 
sets of variables, and $R$ and $B$ are the radiative and binning correction factors, respectively. 
The values of all  kinematic variables are calculated for each 
particular event, as opposed to being taken at the center of the bin.
The virtual photoproduction cross section can be obtained, according to 
Eq.~(\ref{Eq_SPP_CS}), by dividing 
the left-hand side of Eq.~(\ref{Eq_EXP_CS})  
by the virtual photon flux $\Gamma$ factor defined in Eq.~(\ref{Eq_GammaFactor}). 
Sample plots of the differential  cross sections 
compared with models are shown in 
Fig.~\ref{FigSampleCS}. The solid line shows  the cross sections 
calculated using MAID2003 \cite{Tiator:04} multipoles with $l \le 5$
(here simply  referred to as MAID2003 model). The MAID model uses effective 
Langrangian approach to calculate the Born background, including $\omega$ and 
$\rho$ meson calculations. The background is unitarized in the K-matrix approximation. 
The resonant amplitudes are determined by fitting the world pion production data.  
The dashed line in Fig.~\ref{FigSampleCS} corresponds to the model 
by Sato and Lee \cite{Sato:96}. This model obtains an effective Hamiltonian 
from the interaction Lagrangian using the method of unitary transformations. 
Due to complicated calculations Sato-Lee model only includes the $P_{33}(1232)$ 
resonance and, therefore, it's validity domain  is limited 
to the first resonance region.
The shaded areas at the bottom represent the estimated systematic 
uncertainties. Typically the systematic uncertainties are slightly larger 
than the statistical errors.
 The data and the models are globally in qualitative 
agreement, while in certain regions there are quantitative discrepancies. 
Due to the large number of  data points, it is more 
convenient in this paper to discuss the structure functions rather than the cross sections 
themselves. The values of the measured cross sections will be available from 
the CLAS physics database \cite{CLAS_DB}  or upon request~\footnote{e-mail: 
Hovanes.Egiyan@jlab.org} .  
%
%
%
%
%
%
%
%
%
 \subsection{Structure functions}
The structure functions $\sigma_{TT}$, $\sigma_{TL}$ and the linear combination 
$\sigma_{T}+\epsilon \sigma_{L}$ were obtained by fitting  the 
$\phi$-dependence of the cross section to a function of the form:
%
%
\begin{equation}
F(\phi) = A + B \cos{\phi} + C \cos{2 \phi} .
\label{EqFitFunc}
\end{equation} 

\begin{figure}[b]
 \begin{center}
  \epsfig{file=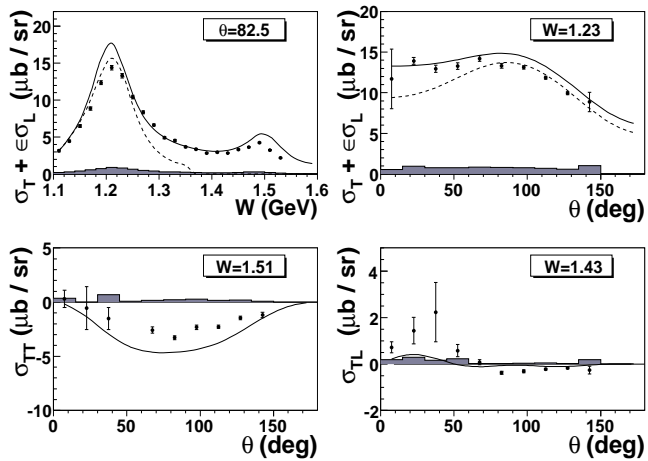, height=6.3cm, width=8.7cm, angle=0}
  \caption[Sample plots of structure functions]
        {Sample plots of structure functions from CLAS at $Q^2=0.4$~GeV$^2$. 
	The solid curves are from MAID2003 \cite{Tiator:04} and the dashed curves are 
	from the Sato-Lee calculations \cite{Sato:96}. The shadowed 
	areas show the systematic uncertainty. }
    \label{FigSampleSF}
 \end{center}
\end{figure}

The large angular  coverage of CLAS in the center-of-mass reference 
frame allowed us to extract the structure functions up to $145^{\circ}$ 
in the center-of-mass $\theta$ angle.  
The $W$ and $\theta$ dependencies of the structure functions are 
shown in Fig.~\ref{FigSampleSF} and Appendix~A. The solid curves 
in Fig.~\ref{FigSampleSF} and Appendix~A are from MAID2003 
\cite{Tiator:04} calculations, while the dashed curves are from 
the Sato-Lee model \cite{Sato:96}. The table of the structure functions 
are presented in Appendix~B. The error column in the table shows the statistical 
and systematic uncertainties added in quadrature. 

A typical $W$-distribution of the 
$\sigma_{T}+\epsilon \sigma_{L}$ and $\sigma_{TT}$ terms features a 
distinct $\Delta(1232)$ peak, followed by the less prominent second resonance region. 
The $\theta$-dependence of $\sigma_{T}+\epsilon \sigma_{L}$ is 
mostly flat at low values of $W$ near the single-pion production threshold. 
This is consistent with $E_{0+}$ dominance at low energies. 
With increasing invariant mass, structures are  
developing at $\theta \approx 90^{\circ}$, 
which is characteristic of  resonance production. 
Above the first resonance region the $\theta$-dependence of 
$\sigma_{T}+\epsilon \sigma_{L}$   becomes 
monotonically falling, consistent with the $t$-channel pion exchange 
mechanism dominance. 
%
%
%
%
%
%
%
%

From the plots in Figs.~19,21,23,25  
in Appendix A, one can see that in the first resonance region  the measurements of $\sigma_{TT}$ 
agree very well with the  Sato-Lee \cite{Sato:96} model. For these values of hadronic 
center-of-mass energy this structure function is dominated 
by the $\Delta(1232)$ resonance contributions. But the model predictions for the 
$\sigma_{T}+\epsilon \sigma_{L}$ linear combination disagree with the measured values, 
although the discrepancy is within the systematic errors. 
The present experiment does not separate the longitudinal and transverse structure functions. 
Given the successful description 
of the $\sigma_{TT}$ term, which can be expressed in terms of helicity amplitudes 
as $\sigma_{TT} \sim Re(H_{3}H_{2}^{*}-H_{4}H_{1}^{*})$, 
one may assume that the  structure function 
$\sigma_{T} \sim |H_1|^2 + |H_2|^2 + |H_3|^2 + |H_4|^2$, which depends on the 
same helicity amplitudes,  may also be  described reasonably  well. 
Then the discrepancies in the sum  $\sigma_{T} + \epsilon \sigma_{L}$ could   
be due to incomplete knowledge of the non-resonant physical background 
contributing directly to $\sigma_{L}$. Therefore, these data can be used 
to improve our understanding of the non-resonant background in the first 
resonance region. The measured $\sigma_{TL}$ structure function 
is the smallest, and the relative systematic uncertainties are large.
The predictions for $\sigma_{TL}$ from  Sato-Lee  \cite{Sato:96} 
are in agreement with the measured values within the error bars. 

The MAID2003 model, which is a fit to predominantly $\pi^{0} p$ channel, describes 
our data surprisingly well, with the curves in the plots following most 
of the features of the experimental data. But the absolute values for the 
 $\sigma_{TT}$ and $\sigma_{T} + \epsilon \sigma_{L}$ structure functions 
are typically  overestimated, especially in the second resonance region. 
This may be indicative of our relatively poor knowledge of the $D_{13}(1520)$ and 
$S_{11}(1535)$ strength in the $ep \rightarrow e^{\prime} \pi^{+} n$ 
channel. We also observe a distinct 
structure in the $\theta$ dependence of the $\sigma_{TL}$ 
amplitude for $W>1.32$~GeV, which is not reproduced by MAID2003, where 
the sign is in fact opposite (see Fig.~\ref{FigSampleSF}). Inclusion of these 
data in the MAID fit can improve our knowledge of the resonance parameters, 
background terms and the branching ratios for the states in the second resonance 
region. 

 \section{Summary}
In conclusion, for the first time we have measured the unpolarized electroproduction 
cross sections for the $ep \rightarrow  e^{\prime} \pi^{+} n$ process 
covering a large angular range in the center-of-mass frame,  and we 
have extracted the $\sigma_{TT}$, 
$\sigma_{TL}$ and $\sigma_{T} + \epsilon \sigma_{L}$ linear combinations 
of the structure functions. The combined statistical and 
systematic errors are of the order of $10\%$ in  most of the measured 
kinematic region. In the first resonance region the measured  cross 
sections and the structure functions  are in qualitative agreement 
with the MAID2003 \cite{Tiator:04} and the Sato-Lee \cite{Sato:96} models, 
with a  quantitative discrepancy with the  MAID2003 model  \cite{Tiator:04}. 
In the second resonance region MAID2003  overestimates the height of the resonance peak. 
Together with $p \pi^{0}$ channel these data will provide the basis for 
the analysis of resonance transition form-factors in a coupled-channel analysis.
\section*{Acknowledgments}
\begin{acknowledgments}
 We would like to thank the staff of the Accelerator and Physics 
Divisions at the Jefferson Laboratory for their outstanding efforts 
to provide us with the high quality  beam and the facilities for the 
data analysis.
This work was supported by the U.S. Department of Energy and the  
National Science Foundation, the French Commissariat \`{a} l'Energie 
Atomique, the Italian Istituto Nazionale di Fisica Nucleare, and the 
Korean Science and Engineering Foundation.  The Southeastern Universities 
Research Association (SURA) operates the Thomas Jefferson National 
Accelerator Facility for the United States Department of Energy under 
Contract No. DE-AC05-84ER40150.  
\end{acknowledgments}

%
%
%

%
%

%
%
%
%
\appendix
 \section{Appendix A: Plots of the structure functions}\label{AppxPlts}

%
%

\begin{figure*}
 \begin{center}
  \epsfig{file=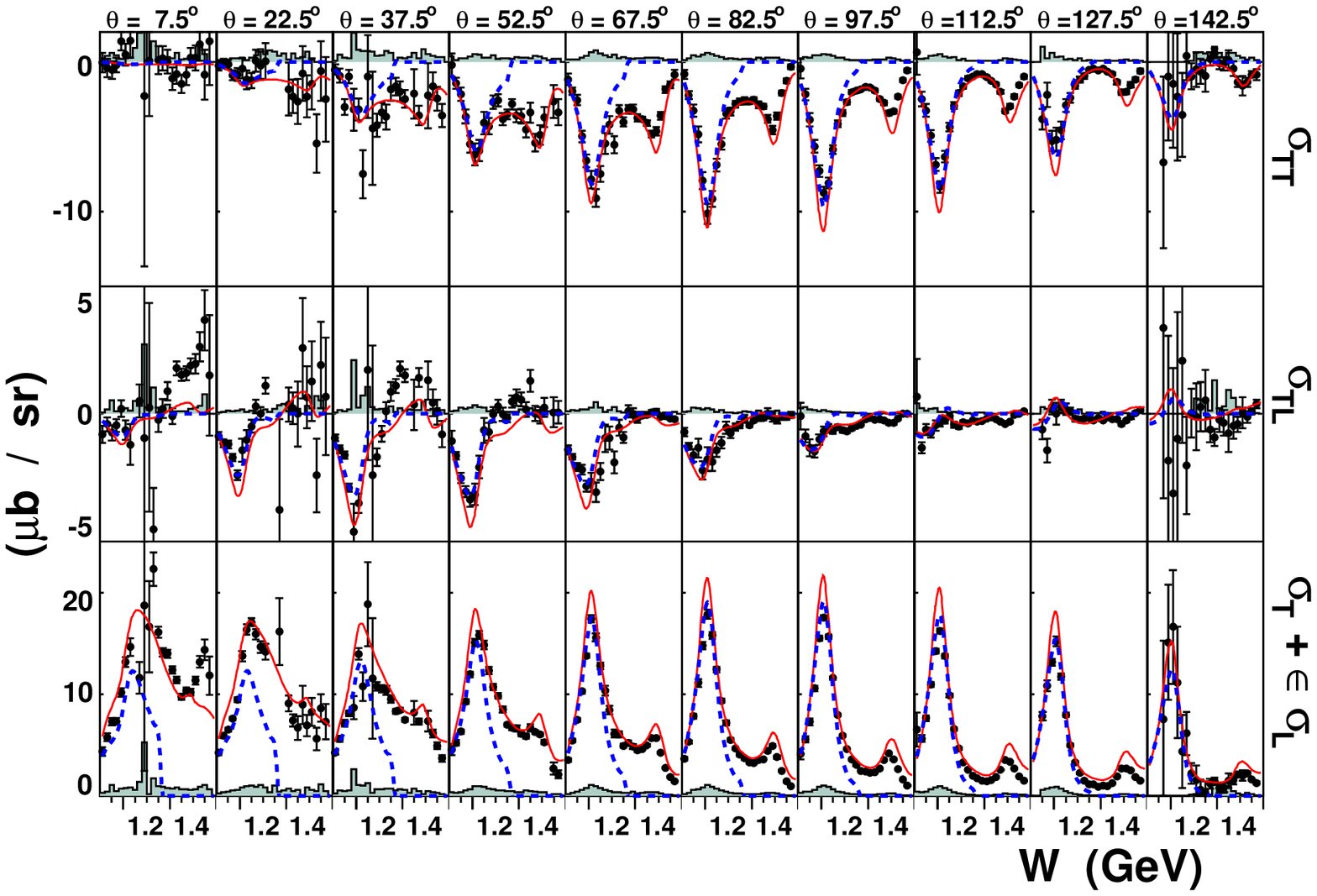,  width=17cm,  totalheight=10.0cm, angle=0}
          {\caption[Structure Functions at Q^{2}=$0.3$~GeV$^2$] 
	{(Color online) Structure functions versus $W$ at $Q^2 = 0.3$~GeV$^2$. 
		Solid curves represent MAID2003 calculations, while the 
	 	dashed curves show the predictions of the Sato-Lee model.
		Shaded areas represent the systematic uncertainties. } }
    \label{FigStrFun5}
 \end{center}
\end{figure*}

\begin{figure*}
 \begin{center}
  \epsfig{file=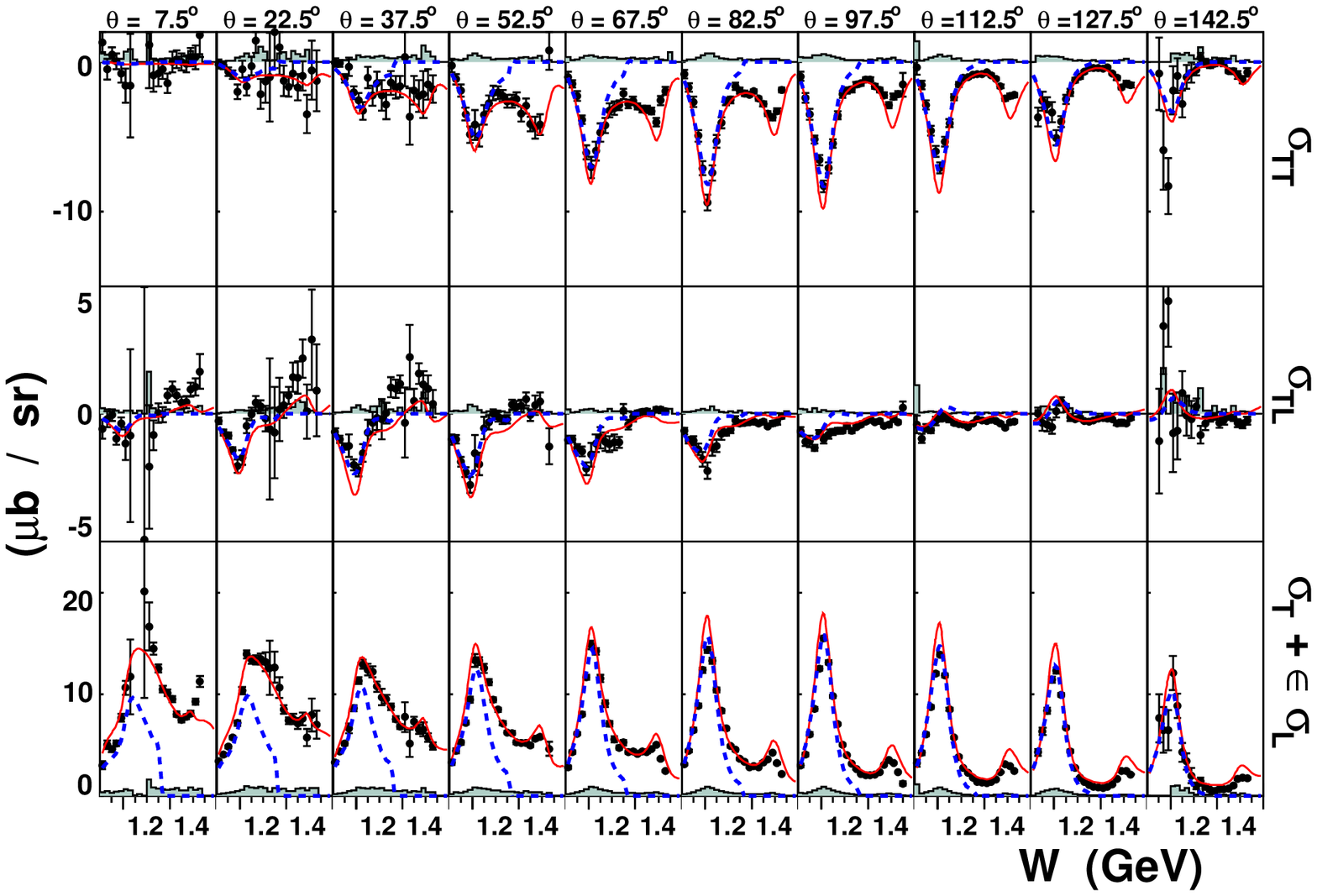,  width=17cm,  totalheight=10.0cm, angle=0}
          {\caption[Structure Functions at Q^{2}=$0.4$~GeV$^2$] 
	{(Color online) Structure functions versus $W$ at $Q^2 = 0.4$~GeV$^2$. 
		Solid curves represent MAID2003 calculations, while the 
	 	dashed curves show the predictions of the Sato-Lee model.
		Shaded areas represent the systematic uncertainties. } }
    \label{FigStrFun6}
 \end{center}
\end{figure*}

\begin{figure*}
 \begin{center}
  \epsfig{file=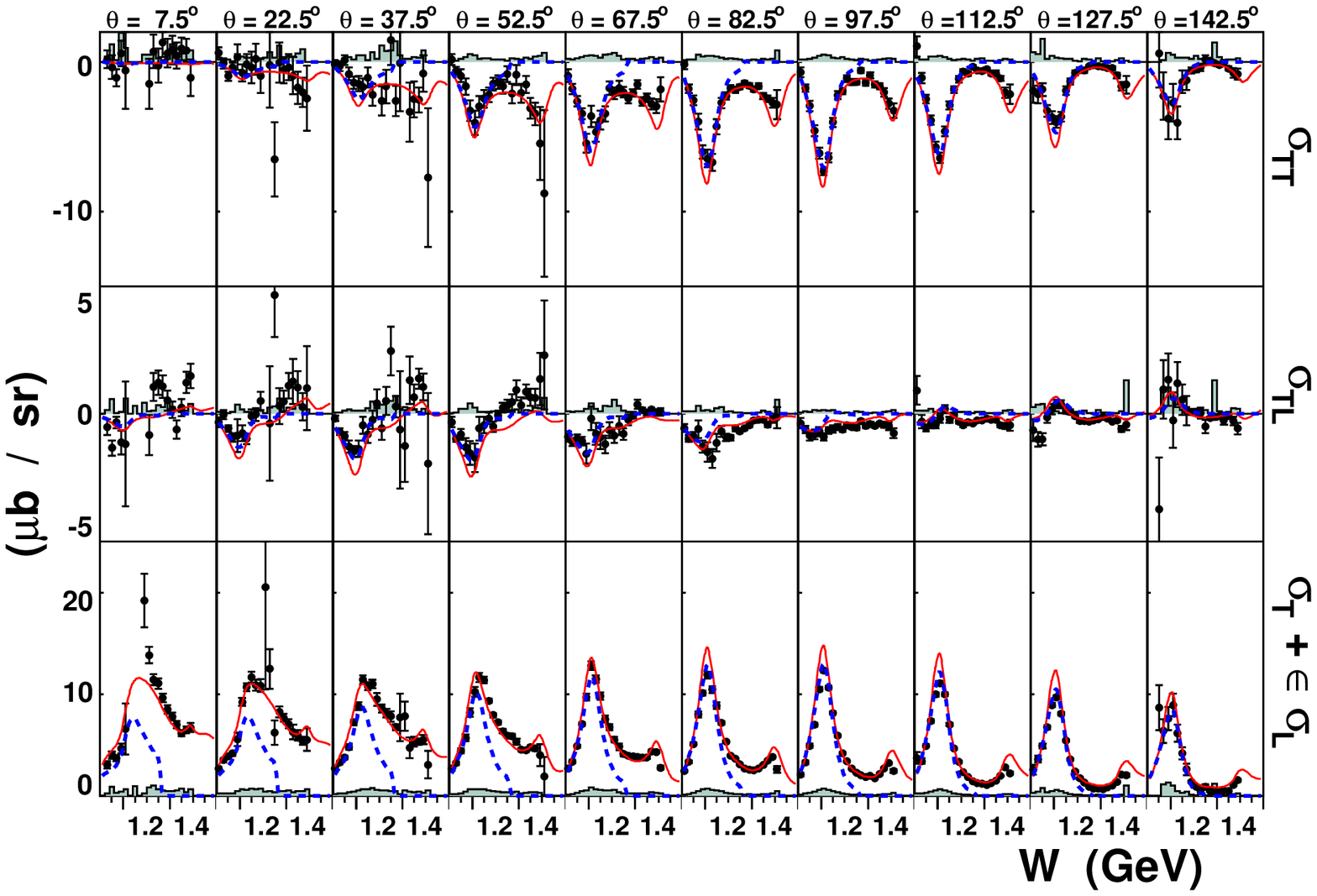,  width=17cm,  totalheight=10.0cm, angle=0}
          {\caption[Structure Functions at Q^{2}=$0.5$~GeV$^2$] 
	{(Color online) Structure functions versus $W$ at $Q^2 = 0.5$~GeV$^2$. 
		Solid curves represent MAID2003 calculations, while the 
	 	dashed curves show the predictions of the Sato-Lee model.
		Shaded areas represent the systematic uncertainties. } }
    \label{FigStrFun7}
 \end{center}
\end{figure*}

\begin{figure*}
 \begin{center}
  \epsfig{file=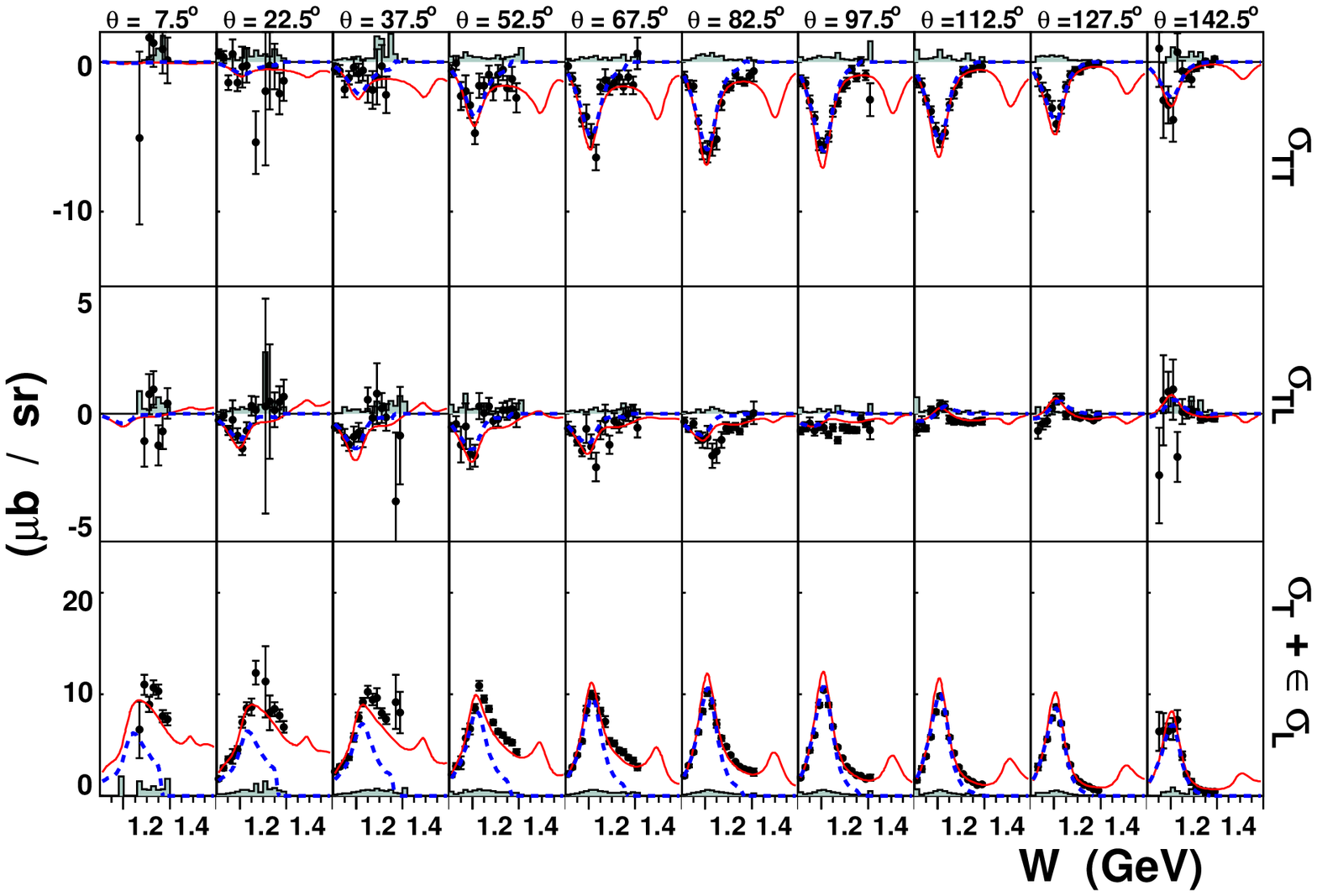,  width=17cm,  totalheight=10.0cm, angle=0}
          {\caption[Structure Functions at Q^{2}=$0.6$~GeV$^2$] 
	{(Color online) Structure functions versus $W$ at $Q^2 = 0.6$~GeV$^2$. 
		Solid curves represent MAID2003 calculations, while the 
	 	dashed curves show the predictions of the Sato-Lee model.
		Shaded areas represent the systematic uncertainties. } }
    \label{FigStrFun8}
 \end{center}
\end{figure*}

%
%

\begin{figure*}
 \begin{center}
  \epsfig{file=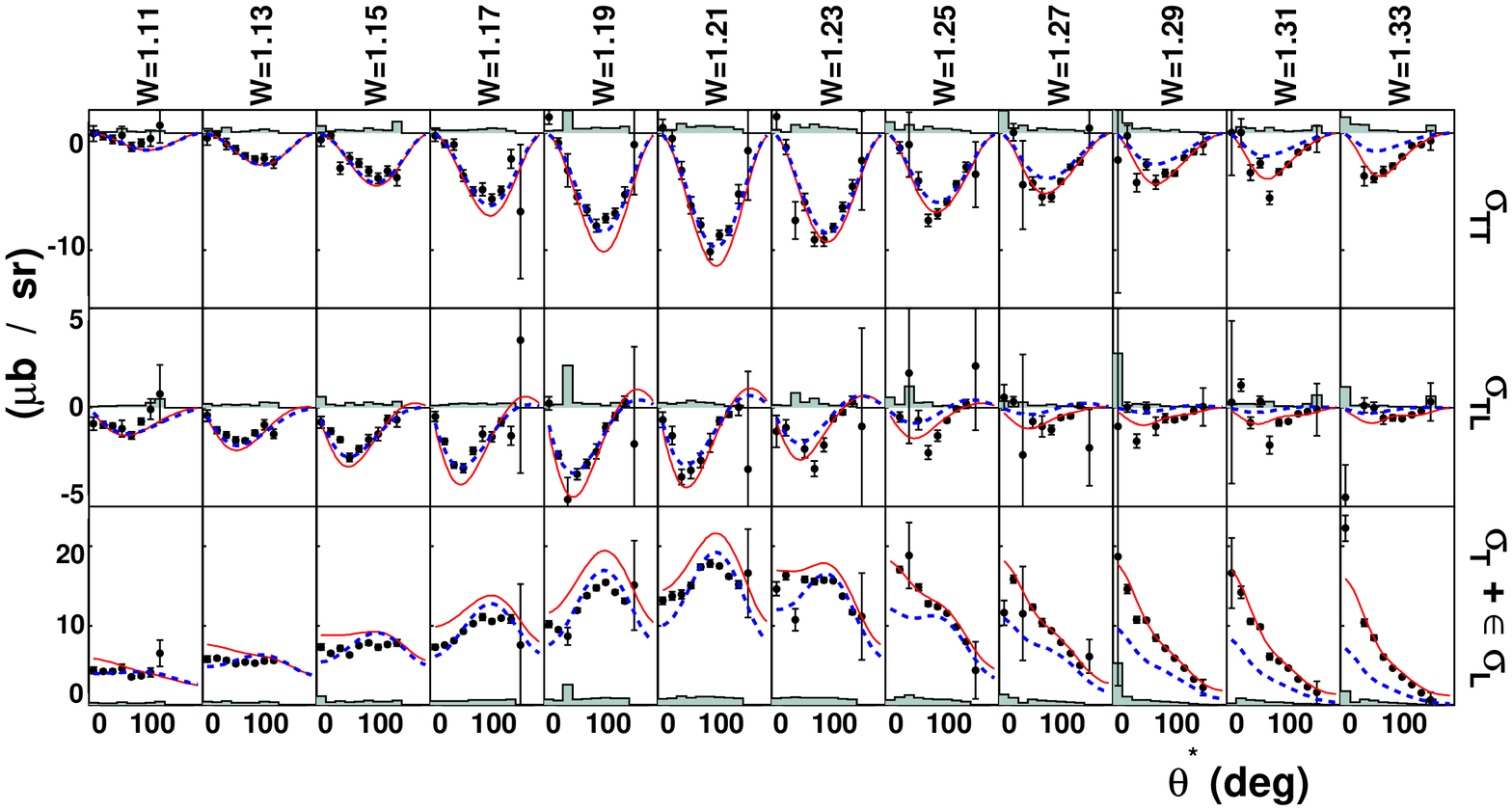,  width=9.2cm,   angle=270}
          {\caption[Structure Functions at Q^{2}=$0.3$~GeV$^2$] 
	{(Color online) Structure functions versus c.m. $\theta$  at $Q^2 = 0.3$~GeV$^2$. 
		Solid curves represent MAID2003 calculations, while the 
	 	dashed curves show the predictions of the Sato-Lee model.
		Shaded areas represent the systematic uncertainties. } }
    \label{FigStrFun1_0}
 \end{center}
\end{figure*}

\begin{figure*}
 \begin{center}
  \epsfig{file=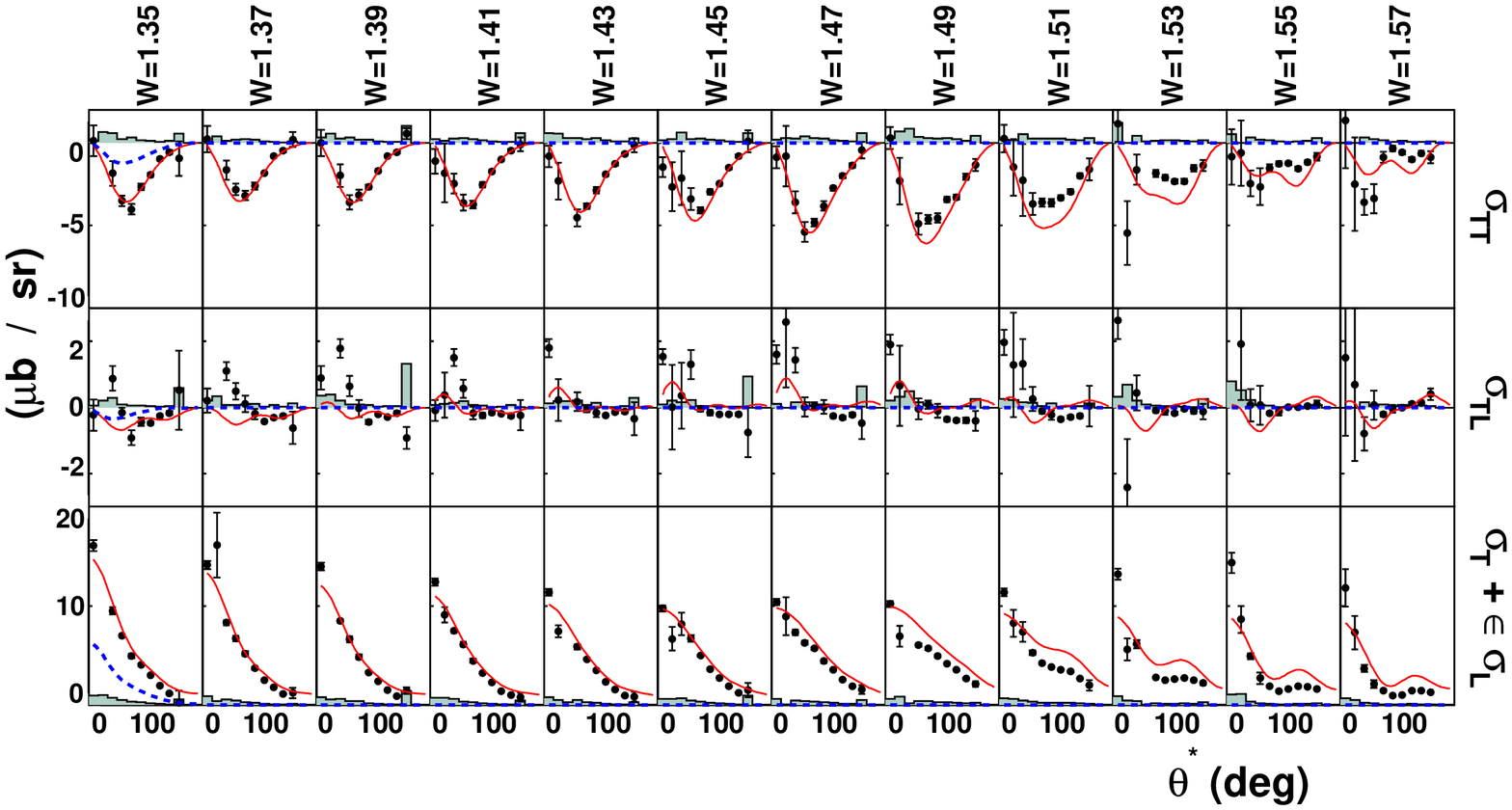,  width=9.2cm,   angle=270}
          {\caption[Structure Functions at Q^{2}=$0.3$~GeV$^2$] 
	{(Color online) Structure functions versus c.m. $\theta$  at $Q^2 = 0.3$~GeV$^2$. 
		Solid curves represent MAID2003 calculations, while the 
	 	dashed curves show the predictions of the Sato-Lee model.
		Shaded areas represent the systematic uncertainties. } }
    \label{FigStrFun1_1}
 \end{center}
\end{figure*}

\begin{figure*}
 \begin{center}
  \epsfig{file=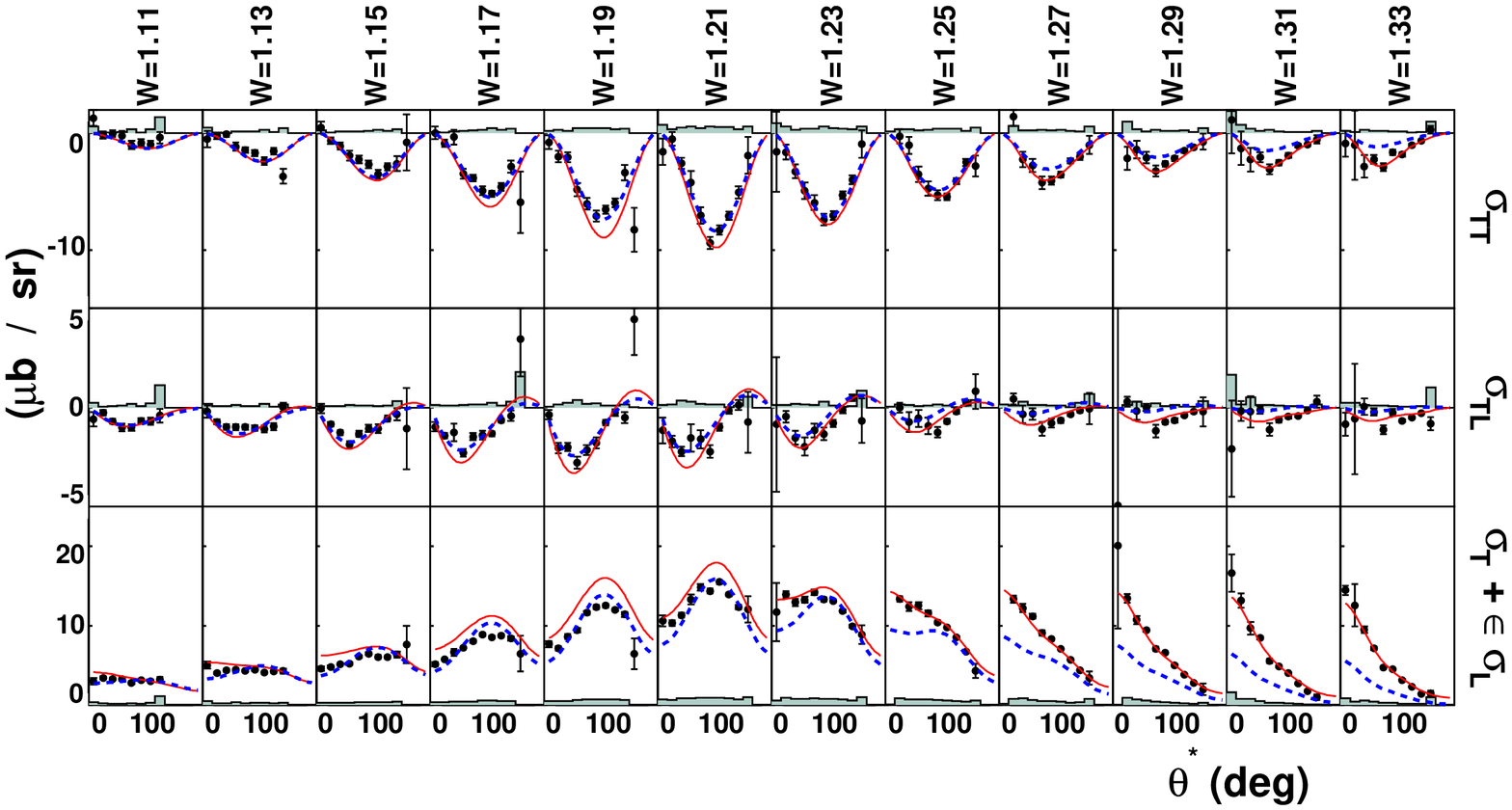,  width=9.2cm,   angle=270}
          {\caption[Structure Functions at Q^{2}=$0.4$~GeV$^2$] 
	{(Color online) Structure functions versus c.m. $\theta$  at $Q^2 = 0.4$~GeV$^2$. 
		Solid curves represent MAID2003 calculations, while the 
	 	dashed curves show the predictions of the Sato-Lee model.
		Shaded areas represent the systematic uncertainties. } }
    \label{FigStrFun2_0}
 \end{center}
\end{figure*}

\begin{figure*}
 \begin{center}
  \epsfig{file=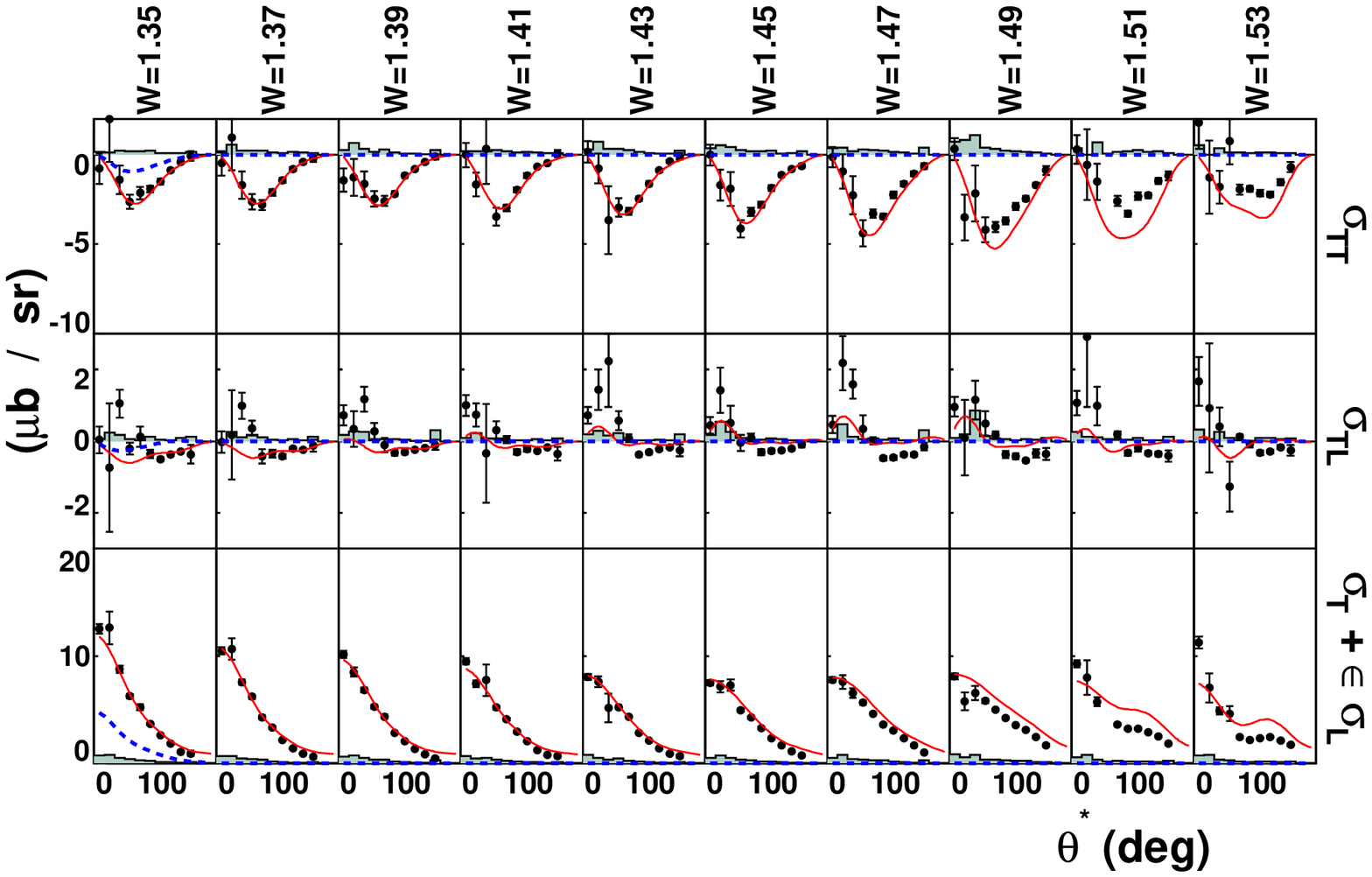,  width=9.2cm,   angle=270}
          {\caption[Structure Functions at Q^{2}=$0.4$~GeV$^2$] 
	{(Color online) Structure functions versus c.m. $\theta$  at $Q^2 = 0.4$~GeV$^2$. 
		Solid curves represent MAID2003 calculations, while the 
	 	dashed curves show the predictions of the Sato-Lee model.
		Shaded areas represent the systematic uncertainties. } }
    \label{FigStrFun2_1}
 \end{center}
\end{figure*}

\begin{figure*}
 \begin{center}
  \epsfig{file=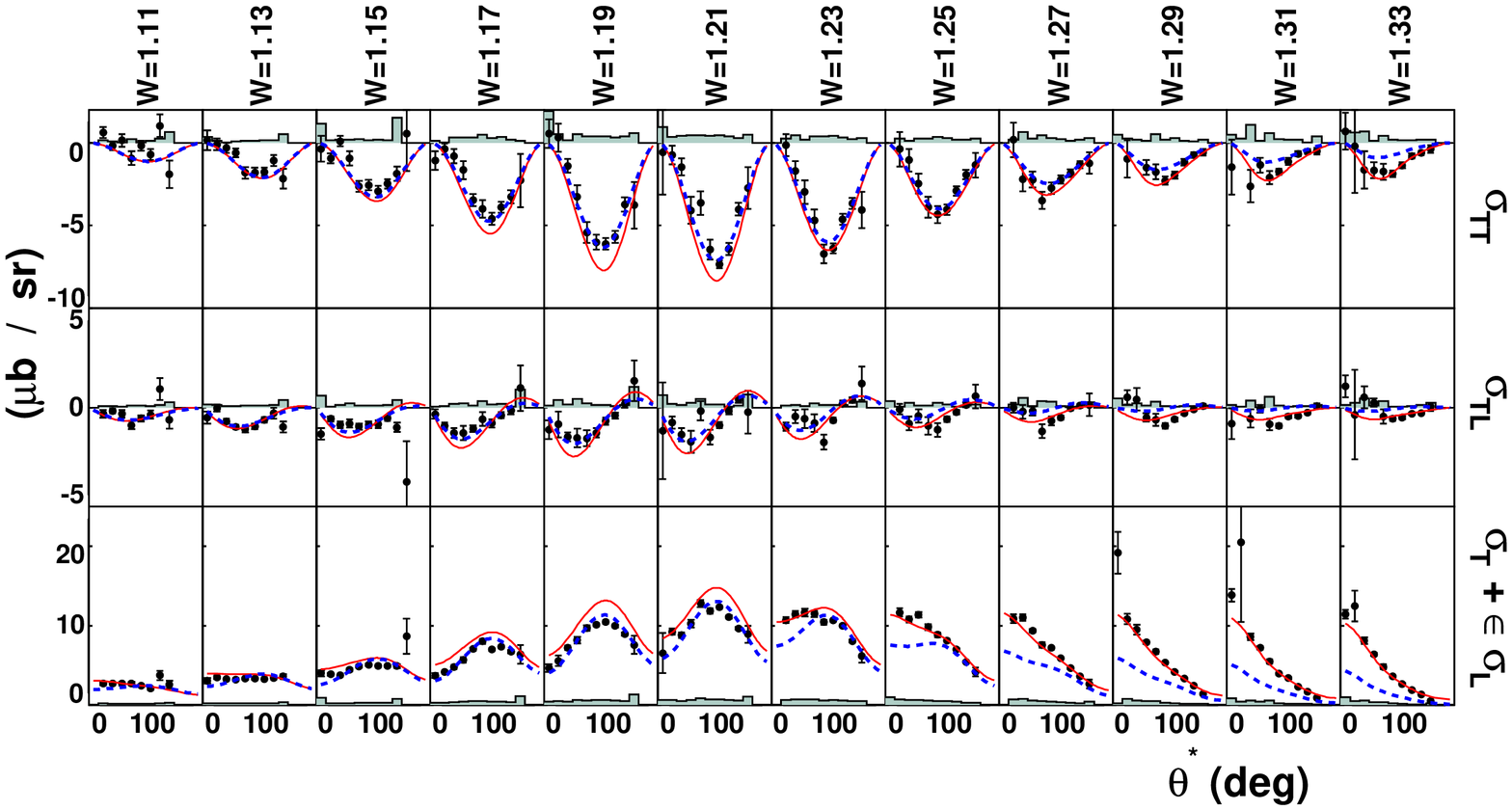,  width=9.2cm,   angle=270}
          {\caption[Structure Functions at Q^{2}=$0.5$~GeV$^2$] 
	{(Color online) Structure functions versus c.m. $\theta$  at $Q^2 = 0.5$~GeV$^2$. 
		Solid curves represent MAID2003 calculations, while the 
	 	dashed curves show the predictions of the Sato-Lee model.
		Shaded areas represent the systematic uncertainties. } }
    \label{FigStrFun3_0}
 \end{center}
\end{figure*}

\begin{figure*}
 \begin{center}
  \epsfig{file=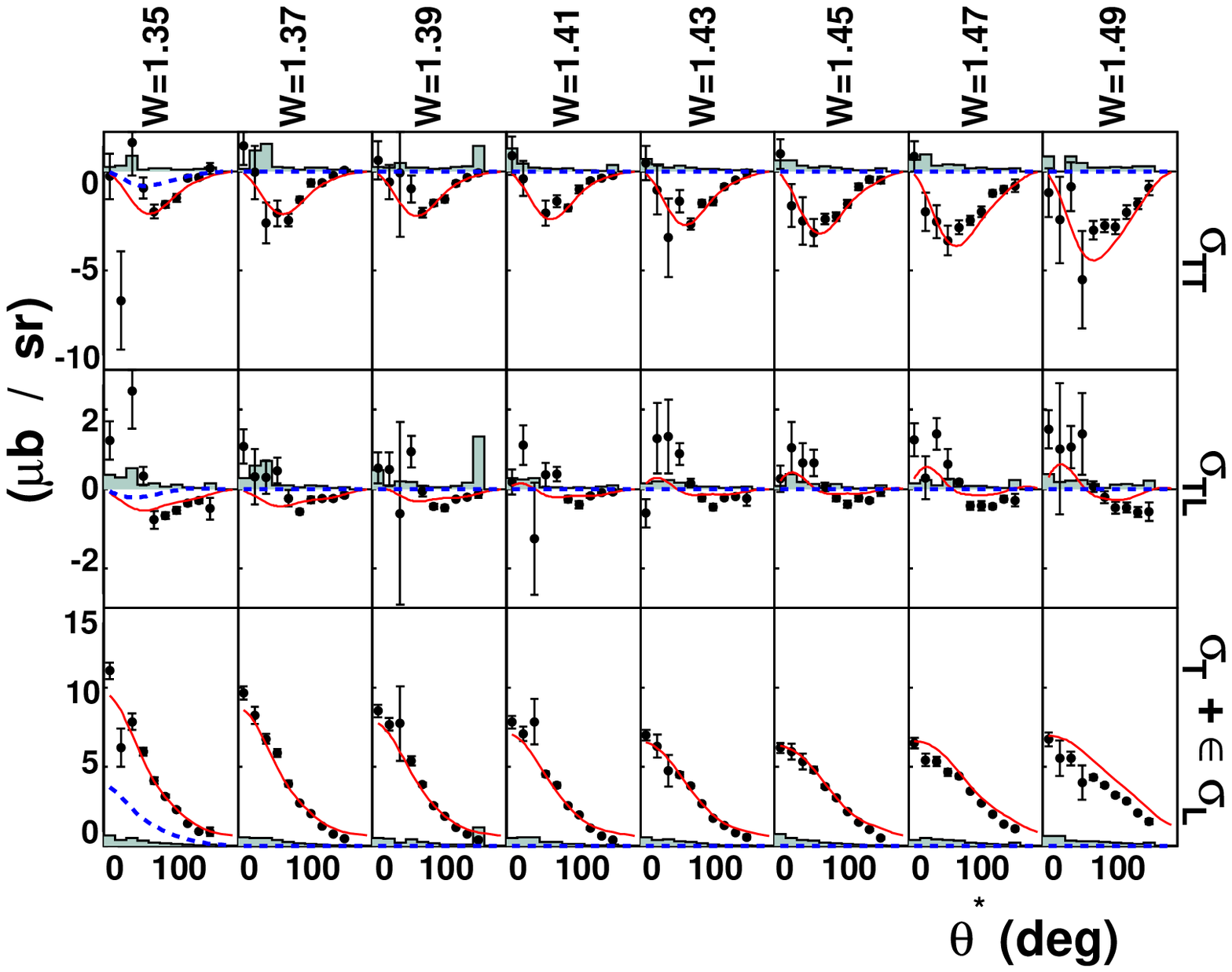,  width=9.2cm,  angle=270}
          {\caption[Structure Functions at Q^{2}=$0.5$~GeV$^2$] 
	{(Color online) Structure functions versus c.m. $\theta$  at $Q^2 = 0.5$~GeV$^2$. 
		Solid curves represent MAID2003 calculations, while the 
	 	dashed curves show the predictions of the Sato-Lee model.
		Shaded areas represent the systematic uncertainties. } }
    \label{FigStrFun3_1}
 \end{center}
\end{figure*}

\begin{figure*}
 \begin{center}
  \epsfig{file=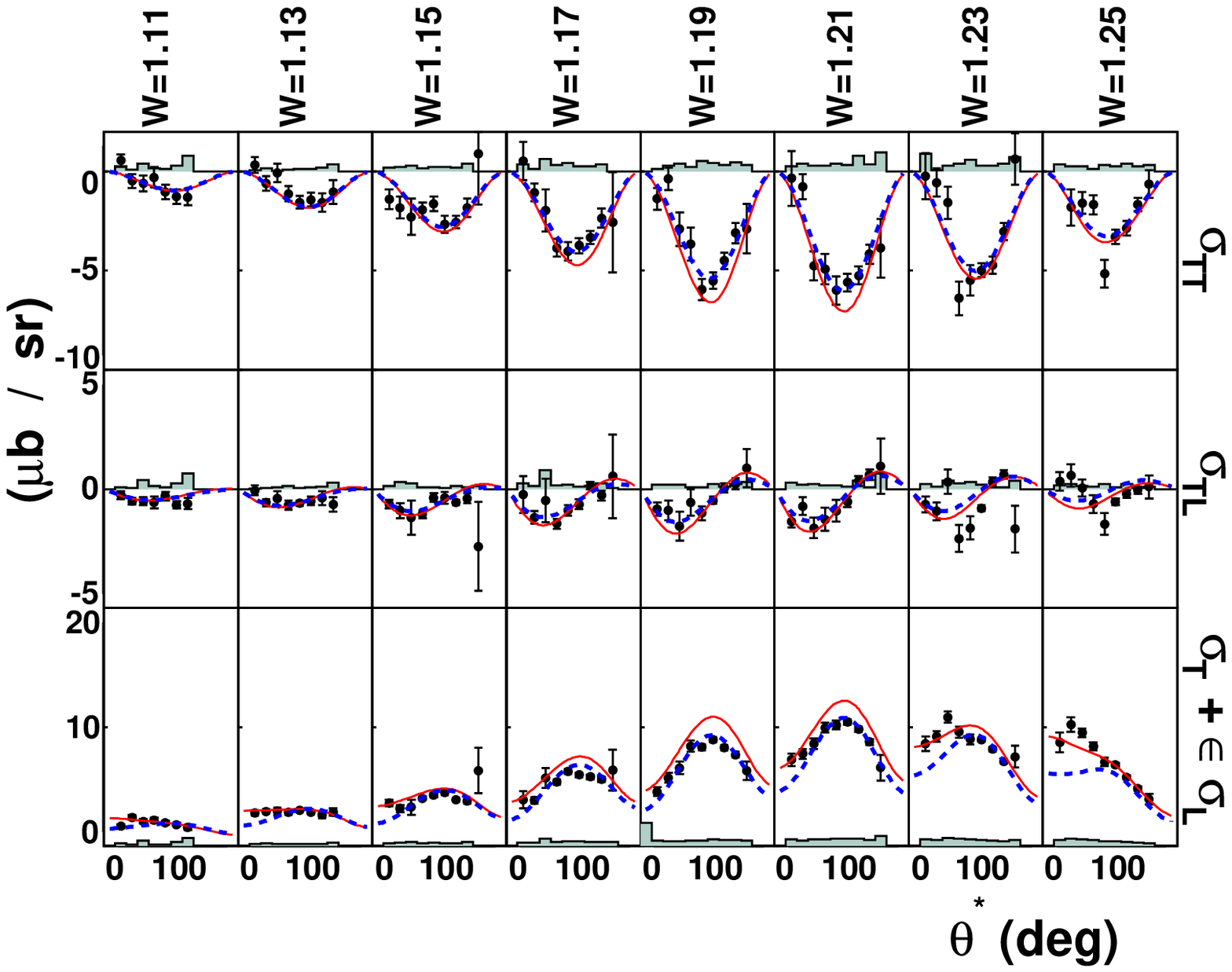,  width=9.2cm,   angle=270}
          {\caption[Structure Functions at Q^{2}=$0.6$~GeV$^2$] 
	{(Color online) Structure functions versus c.m. $\theta$  at $Q^2 = 0.6$~GeV$^2$. 
		Solid curves represent MAID2003 calculations, while the 
	 	dashed curves show the predictions of the Sato-Lee model.
		Shaded areas represent the systematic uncertainties. } }
    \label{FigStrFun4_0}
 \end{center}
\end{figure*}

\begin{figure*}
 \begin{center}
  \epsfig{file=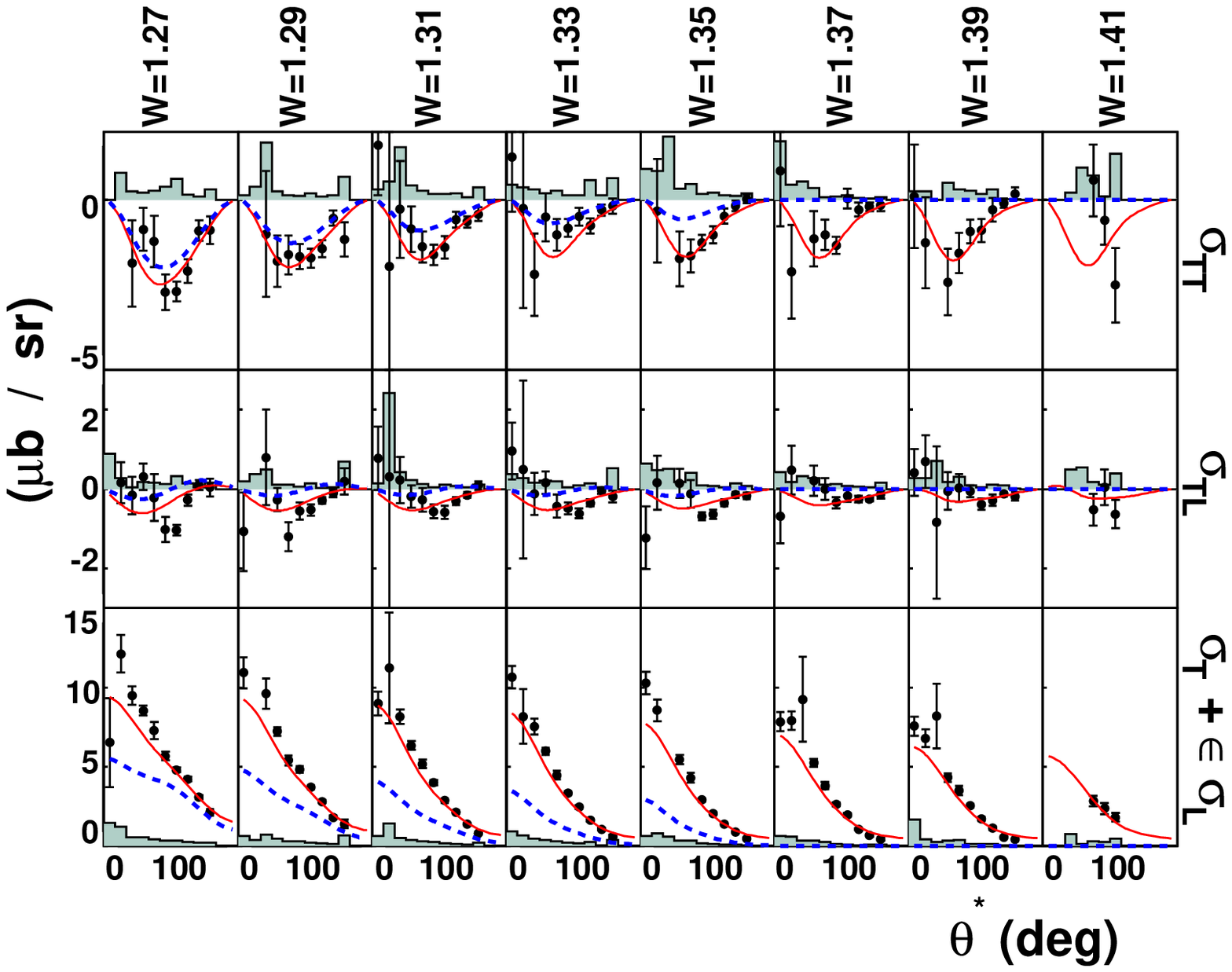,  width=9.2cm,  angle=270}
          {\caption[Structure Functions at Q^{2}=$0.6$~GeV$^2$] 
	{(Color online) Structure functions versus c.m. $\theta$  at $Q^2 = 0.6$~GeV$^2$. 
		Solid curves represent MAID2003 calculations, while the 
	 	dashed curves show the predictions of the Sato-Lee model.
		Shaded areas represent the systematic uncertainties. } }
    \label{FigStrFun4_1}
 \end{center}
\end{figure*}

\clearpage

 \section{Appendix B: Table of the structure functions}\label{AppxTbls}
   { \tiny \ttfamily 
\begin{verse}
  {$Q^2$~~$W$~~~$\epsilon$~~~~~$\theta$~~~~~$\sigma_{T}+\epsilon\sigma_{L}$~~~~~~$\sigma_{TL}$~~~~~~~~~$\sigma_{TT}$ } \\ 
 \underline{\hspace{6.15cm}} \\ \vspace{0.1cm}
{\verb|0.3 1.11 0.890   7.5   4.396|}$\pm${\verb| 0.52  -0.806|}$\pm${\verb| 0.34  -0.042|}$\pm${\verb| 0.20|}\\
{\verb|0.3 1.11 0.890  22.5   4.243|}$\pm${\verb| 0.32  -0.864|}$\pm${\verb| 0.18  -0.241|}$\pm${\verb| 0.07|}\\
{\verb|0.3 1.11 0.890  37.5   4.285|}$\pm${\verb| 0.34  -0.935|}$\pm${\verb| 0.15  -0.526|}$\pm${\verb| 0.05|}\\
{\verb|0.3 1.11 0.890  52.5   4.580|}$\pm${\verb| 0.65  -1.066|}$\pm${\verb| 0.43  -0.199|}$\pm${\verb| 0.27|}\\
{\verb|0.3 1.11 0.890  67.5   3.556|}$\pm${\verb| 0.28  -1.410|}$\pm${\verb| 0.20  -1.171|}$\pm${\verb| 0.08|}\\
{\verb|0.3 1.11 0.890  82.5   3.673|}$\pm${\verb| 0.29  -0.702|}$\pm${\verb| 0.19  -0.833|}$\pm${\verb| 0.08|}\\
{\verb|0.3 1.11 0.890  97.5   4.121|}$\pm${\verb| 0.62  -0.082|}$\pm${\verb| 0.57  -0.466|}$\pm${\verb| 0.39|}\\
{\verb|0.3 1.11 0.890 112.5   6.547|}$\pm${\verb| 1.72   0.693|}$\pm${\verb| 1.52   0.658|}$\pm${\verb| 1.79|}\\
{\verb|0.3 1.13 0.883   7.5   5.784|}$\pm${\verb| 0.56  -0.406|}$\pm${\verb| 0.34  -0.404|}$\pm${\verb| 0.17|}\\
{\verb|0.3 1.13 0.883  22.5   5.917|}$\pm${\verb| 0.41  -1.149|}$\pm${\verb| 0.19  -0.110|}$\pm${\verb| 0.07|}\\
{\verb|0.3 1.13 0.883  37.5   5.645|}$\pm${\verb| 0.61  -1.384|}$\pm${\verb| 0.22  -0.922|}$\pm${\verb| 0.09|}\\
{\verb|0.3 1.13 0.883  52.5   5.241|}$\pm${\verb| 0.40  -1.641|}$\pm${\verb| 0.19  -1.427|}$\pm${\verb| 0.07|}\\
{\verb|0.3 1.13 0.883  67.5   5.460|}$\pm${\verb| 0.37  -1.674|}$\pm${\verb| 0.20  -1.951|}$\pm${\verb| 0.07|}\\
{\verb|0.3 1.13 0.883  82.5   5.316|}$\pm${\verb| 0.38  -1.276|}$\pm${\verb| 0.17  -2.225|}$\pm${\verb| 0.06|}\\
{\verb|0.3 1.13 0.883  97.5   5.600|}$\pm${\verb| 0.55  -0.870|}$\pm${\verb| 0.37  -2.175|}$\pm${\verb| 0.19|}\\
{\verb|0.3 1.13 0.883 112.5   5.657|}$\pm${\verb| 0.52  -1.327|}$\pm${\verb| 0.32  -2.496|}$\pm${\verb| 0.15|}\\
{\verb|0.3 1.15 0.877   7.5   7.311|}$\pm${\verb| 1.18  -0.739|}$\pm${\verb| 0.61  -0.551|}$\pm${\verb| 0.42|}\\
{\verb|0.3 1.15 0.877  22.5   6.537|}$\pm${\verb| 0.45  -1.194|}$\pm${\verb| 0.20  -0.158|}$\pm${\verb| 0.08|}\\
{\verb|0.3 1.15 0.877  37.5   7.158|}$\pm${\verb| 0.63  -1.611|}$\pm${\verb| 0.17  -2.997|}$\pm${\verb| 0.06|}\\
{\verb|0.3 1.15 0.877  52.5   6.338|}$\pm${\verb| 0.44  -2.519|}$\pm${\verb| 0.32  -2.156|}$\pm${\verb| 0.15|}\\
{\verb|0.3 1.15 0.877  67.5   7.509|}$\pm${\verb| 0.48  -2.084|}$\pm${\verb| 0.21  -2.558|}$\pm${\verb| 0.08|}\\
{\verb|0.3 1.15 0.877  82.5   7.848|}$\pm${\verb| 0.66  -1.623|}$\pm${\verb| 0.36  -3.256|}$\pm${\verb| 0.18|}\\
{\verb|0.3 1.15 0.877  97.5   7.300|}$\pm${\verb| 0.56  -1.332|}$\pm${\verb| 0.37  -3.871|}$\pm${\verb| 0.20|}\\
{\verb|0.3 1.15 0.877 112.5   7.640|}$\pm${\verb| 0.53  -0.607|}$\pm${\verb| 0.25  -3.243|}$\pm${\verb| 0.11|}\\
{\verb|0.3 1.15 0.877 127.5   7.845|}$\pm${\verb| 0.73  -0.612|}$\pm${\verb| 0.48  -3.815|}$\pm${\verb| 0.28|}\\
{\verb|0.3 1.17 0.869   7.5   7.330|}$\pm${\verb| 0.58  -0.454|}$\pm${\verb| 0.28  -0.207|}$\pm${\verb| 0.13|}\\
{\verb|0.3 1.17 0.869  22.5   7.603|}$\pm${\verb| 0.57  -1.715|}$\pm${\verb| 0.19  -0.908|}$\pm${\verb| 0.07|}\\
{\verb|0.3 1.17 0.869  37.5   8.122|}$\pm${\verb| 0.62  -2.896|}$\pm${\verb| 0.24  -0.985|}$\pm${\verb| 0.10|}\\
{\verb|0.3 1.17 0.869  52.5   9.284|}$\pm${\verb| 0.59  -3.055|}$\pm${\verb| 0.29  -3.644|}$\pm${\verb| 0.13|}\\
{\verb|0.3 1.17 0.869  67.5  10.270|}$\pm${\verb| 0.65  -2.184|}$\pm${\verb| 0.26  -4.977|}$\pm${\verb| 0.11|}\\
{\verb|0.3 1.17 0.869  82.5  11.112|}$\pm${\verb| 0.79  -1.331|}$\pm${\verb| 0.43  -4.833|}$\pm${\verb| 0.25|}\\
{\verb|0.3 1.17 0.869  97.5  10.570|}$\pm${\verb| 0.69  -1.480|}$\pm${\verb| 0.29  -5.635|}$\pm${\verb| 0.13|}\\
{\verb|0.3 1.17 0.869 112.5  10.971|}$\pm${\verb| 0.71  -0.730|}$\pm${\verb| 0.27  -4.924|}$\pm${\verb| 0.12|}\\
{\verb|0.3 1.17 0.869 127.5  10.856|}$\pm${\verb| 0.91  -1.431|}$\pm${\verb| 0.51  -2.195|}$\pm${\verb| 0.33|}\\
{\verb|0.3 1.17 0.869 142.5   7.540|}$\pm${\verb| 7.69   3.381|}$\pm${\verb| 6.69  -6.707|}$\pm${\verb|17.31|}\\
{\verb|0.3 1.19 0.861   7.5  10.190|}$\pm${\verb| 0.83   0.214|}$\pm${\verb| 0.36   1.385|}$\pm${\verb| 0.20|}\\
{\verb|0.3 1.19 0.861  22.5   9.489|}$\pm${\verb| 0.65  -2.395|}$\pm${\verb| 0.24  -0.853|}$\pm${\verb| 0.10|}\\
{\verb|0.3 1.19 0.861  37.5   8.671|}$\pm${\verb| 2.86  -4.625|}$\pm${\verb| 2.39  -3.194|}$\pm${\verb| 3.29|}\\
{\verb|0.3 1.19 0.861  52.5  11.957|}$\pm${\verb| 0.75  -3.364|}$\pm${\verb| 0.37  -5.475|}$\pm${\verb| 0.19|}\\
{\verb|0.3 1.19 0.861  67.5  13.780|}$\pm${\verb| 0.85  -2.834|}$\pm${\verb| 0.28  -6.597|}$\pm${\verb| 0.12|}\\
{\verb|0.3 1.19 0.861  82.5  14.775|}$\pm${\verb| 0.94  -2.213|}$\pm${\verb| 0.40  -7.931|}$\pm${\verb| 0.23|}\\
{\verb|0.3 1.19 0.861  97.5  15.489|}$\pm${\verb| 0.95  -1.002|}$\pm${\verb| 0.23  -7.290|}$\pm${\verb| 0.10|}\\
{\verb|0.3 1.19 0.861 112.5  14.238|}$\pm${\verb| 0.89  -0.452|}$\pm${\verb| 0.30  -6.858|}$\pm${\verb| 0.14|}\\
{\verb|0.3 1.19 0.861 127.5  13.144|}$\pm${\verb| 0.92   0.281|}$\pm${\verb| 0.30  -5.281|}$\pm${\verb| 0.15|}\\
{\verb|0.3 1.19 0.861 142.5  15.094|}$\pm${\verb| 5.65  -1.826|}$\pm${\verb| 4.89  -0.976|}$\pm${\verb|10.83|}\\
{\verb|0.3 1.21 0.852   7.5  13.173|}$\pm${\verb| 0.90  -0.630|}$\pm${\verb| 0.45   0.481|}$\pm${\verb| 0.26|}\\
{\verb|0.3 1.21 0.852  22.5  13.751|}$\pm${\verb| 1.01  -1.414|}$\pm${\verb| 0.49  -0.441|}$\pm${\verb| 0.32|}\\
{\verb|0.3 1.21 0.852  37.5  13.970|}$\pm${\verb| 1.25  -3.500|}$\pm${\verb| 0.45  -3.181|}$\pm${\verb| 0.26|}\\
{\verb|0.3 1.21 0.852  52.5  15.069|}$\pm${\verb| 1.03  -3.163|}$\pm${\verb| 0.54  -6.181|}$\pm${\verb| 0.34|}\\
{\verb|0.3 1.21 0.852  67.5  17.425|}$\pm${\verb| 1.13  -2.677|}$\pm${\verb| 0.45  -7.868|}$\pm${\verb| 0.26|}\\
{\verb|0.3 1.21 0.852  82.5  17.842|}$\pm${\verb| 1.18  -1.774|}$\pm${\verb| 0.52 -10.137|}$\pm${\verb| 0.34|}\\
{\verb|0.3 1.21 0.852  97.5  17.510|}$\pm${\verb| 1.05  -0.677|}$\pm${\verb| 0.25  -8.730|}$\pm${\verb| 0.10|}\\
{\verb|0.3 1.21 0.852 112.5  16.208|}$\pm${\verb| 0.98  -0.341|}$\pm${\verb| 0.17  -8.342|}$\pm${\verb| 0.06|}\\
{\verb|0.3 1.21 0.852 127.5  15.215|}$\pm${\verb| 1.01   0.015|}$\pm${\verb| 0.44  -5.204|}$\pm${\verb| 0.28|}\\
{\verb|0.3 1.21 0.852 142.5  16.611|}$\pm${\verb| 5.58  -3.116|}$\pm${\verb| 4.93  -1.479|}$\pm${\verb|10.93|}\\
{\verb|0.3 1.23 0.843   7.5  14.642|}$\pm${\verb| 1.22  -1.197|}$\pm${\verb| 0.82   1.423|}$\pm${\verb| 0.73|}\\
{\verb|0.3 1.23 0.843  22.5  16.339|}$\pm${\verb| 1.09  -1.020|}$\pm${\verb| 0.41  -1.202|}$\pm${\verb| 0.25|}\\
{\verb|0.3 1.23 0.843  37.5  10.757|}$\pm${\verb| 1.73  -6.482|}$\pm${\verb| 1.53  -7.487|}$\pm${\verb| 1.69|}\\
{\verb|0.3 1.23 0.843  52.5  15.822|}$\pm${\verb| 1.02  -2.081|}$\pm${\verb| 0.53  -5.919|}$\pm${\verb| 0.34|}\\
{\verb|0.3 1.23 0.843  67.5  15.578|}$\pm${\verb| 1.03  -3.074|}$\pm${\verb| 0.60  -9.102|}$\pm${\verb| 0.39|}\\
{\verb|0.3 1.23 0.843  82.5  15.771|}$\pm${\verb| 0.98  -1.882|}$\pm${\verb| 0.40  -9.143|}$\pm${\verb| 0.23|}\\
{\verb|0.3 1.23 0.843  97.5  15.657|}$\pm${\verb| 0.94  -0.554|}$\pm${\verb| 0.19  -8.102|}$\pm${\verb| 0.07|}\\
{\verb|0.3 1.23 0.843 112.5  13.738|}$\pm${\verb| 0.83  -0.247|}$\pm${\verb| 0.16  -6.324|}$\pm${\verb| 0.05|}\\
{\verb|0.3 1.23 0.843 127.5  11.743|}$\pm${\verb| 0.78   0.303|}$\pm${\verb| 0.31  -4.575|}$\pm${\verb| 0.15|}\\
{\verb|0.3 1.23 0.843 142.5  11.163|}$\pm${\verb| 5.48  -0.964|}$\pm${\verb| 4.96  -2.334|}$\pm${\verb|11.03|}\\
{\verb|0.3 1.25 0.833   7.5  26.183|}$\pm${\verb| 9.95   7.984|}$\pm${\verb| 8.65   6.982|}$\pm${\verb|25.36|}\\
{\verb|0.3 1.25 0.833  22.5  17.060|}$\pm${\verb| 1.08  -0.482|}$\pm${\verb| 0.29  -1.241|}$\pm${\verb| 0.14|}\\
{\verb|0.3 1.25 0.833  37.5  18.852|}$\pm${\verb| 4.32   1.724|}$\pm${\verb| 3.68  -0.984|}$\pm${\verb| 6.71|}\\
{\verb|0.3 1.25 0.833  52.5  14.817|}$\pm${\verb| 1.00  -0.636|}$\pm${\verb| 0.55  -4.077|}$\pm${\verb| 0.36|}\\
{\verb|0.3 1.25 0.833  67.5  12.789|}$\pm${\verb| 0.85  -2.270|}$\pm${\verb| 0.42  -7.480|}$\pm${\verb| 0.23|}\\
{\verb|0.3 1.25 0.833  82.5  12.440|}$\pm${\verb| 0.79  -1.427|}$\pm${\verb| 0.28  -6.889|}$\pm${\verb| 0.14|}\\
{\verb|0.3 1.25 0.833  97.5  11.586|}$\pm${\verb| 0.71  -0.653|}$\pm${\verb| 0.15  -5.865|}$\pm${\verb| 0.05|}\\
{\verb|0.3 1.25 0.833 112.5   9.878|}$\pm${\verb| 0.61  -0.075|}$\pm${\verb| 0.14  -4.361|}$\pm${\verb| 0.05|}\\
{\verb|0.3 1.25 0.833 127.5   7.996|}$\pm${\verb| 0.52   0.190|}$\pm${\verb| 0.23  -2.902|}$\pm${\verb| 0.11|}\\
{\verb|0.3 1.25 0.833 142.5   4.385|}$\pm${\verb| 3.57   2.094|}$\pm${\verb| 3.23  -3.533|}$\pm${\verb| 5.81|}\\
{\verb|0.3 1.27 0.821   7.5  11.628|}$\pm${\verb| 2.07   0.516|}$\pm${\verb| 0.72   7.546|}$\pm${\verb| 0.55|}\\
{\verb|0.3 1.27 0.821  22.5  15.891|}$\pm${\verb| 1.06   0.292|}$\pm${\verb| 0.27   0.074|}$\pm${\verb| 0.13|}\\
{\verb|0.3 1.27 0.821  37.5  11.540|}$\pm${\verb| 5.94  -2.392|}$\pm${\verb| 5.08  -4.437|}$\pm${\verb|11.43|}\\
{\verb|0.3 1.27 0.821  52.5  12.337|}$\pm${\verb| 0.81  -0.700|}$\pm${\verb| 0.36  -4.296|}$\pm${\verb| 0.21|}\\
{\verb|0.3 1.27 0.821  67.5  10.379|}$\pm${\verb| 0.78  -0.954|}$\pm${\verb| 0.54  -5.463|}$\pm${\verb| 0.39|}\\
{\verb|0.3 1.27 0.821  82.5   9.371|}$\pm${\verb| 0.58  -1.127|}$\pm${\verb| 0.28  -5.473|}$\pm${\verb| 0.13|}\\
{\verb|0.3 1.27 0.821  97.5   7.954|}$\pm${\verb| 0.48  -0.509|}$\pm${\verb| 0.12  -4.138|}$\pm${\verb| 0.04|}\\
{\verb|0.3 1.27 0.821 112.5   6.566|}$\pm${\verb| 0.42  -0.418|}$\pm${\verb| 0.14  -2.914|}$\pm${\verb| 0.04|}\\
{\verb|0.3 1.27 0.821 127.5   4.991|}$\pm${\verb| 0.36   0.106|}$\pm${\verb| 0.14  -2.405|}$\pm${\verb| 0.05|}\\
{\verb|0.3 1.27 0.821 142.5   6.153|}$\pm${\verb| 2.12  -2.024|}$\pm${\verb| 1.92   0.413|}$\pm${\verb| 2.66|}\\
{\verb|0.3 1.29 0.809   7.5  18.693|}$\pm${\verb|17.07  -0.955|}$\pm${\verb|14.64  -2.284|}$\pm${\verb|54.72|}\\
{\verb|0.3 1.29 0.809  22.5  14.630|}$\pm${\verb| 1.22   0.020|}$\pm${\verb| 0.28  -0.232|}$\pm${\verb| 0.13|}\\
{\verb|0.3 1.29 0.809  37.5  10.761|}$\pm${\verb| 0.79  -1.691|}$\pm${\verb| 0.39  -4.241|}$\pm${\verb| 0.22|}\\
{\verb|0.3 1.29 0.809  52.5  10.713|}$\pm${\verb| 0.74  -0.022|}$\pm${\verb| 0.34  -2.690|}$\pm${\verb| 0.17|}\\
{\verb|0.3 1.29 0.809  67.5   8.456|}$\pm${\verb| 0.67  -0.940|}$\pm${\verb| 0.48  -4.205|}$\pm${\verb| 0.31|}\\
{\verb|0.3 1.29 0.809  82.5   7.188|}$\pm${\verb| 0.48  -0.537|}$\pm${\verb| 0.24  -3.444|}$\pm${\verb| 0.11|}\\
{\verb|0.3 1.29 0.809  97.5   5.947|}$\pm${\verb| 0.37  -0.632|}$\pm${\verb| 0.10  -3.431|}$\pm${\verb| 0.03|}\\
{\verb|0.3 1.29 0.809 112.5   4.667|}$\pm${\verb| 0.30  -0.481|}$\pm${\verb| 0.11  -2.093|}$\pm${\verb| 0.03|}\\
{\verb|0.3 1.29 0.809 127.5   3.277|}$\pm${\verb| 0.23  -0.185|}$\pm${\verb| 0.12  -1.574|}$\pm${\verb| 0.04|}\\
{\verb|0.3 1.29 0.809 142.5   2.237|}$\pm${\verb| 1.02   0.032|}$\pm${\verb| 0.96  -0.966|}$\pm${\verb| 0.94|}\\
{\verb|0.3 1.31 0.796   7.5  16.633|}$\pm${\verb| 4.46   0.257|}$\pm${\verb| 4.10   0.073|}$\pm${\verb| 8.28|}\\
{\verb|0.3 1.31 0.796  22.5  14.194|}$\pm${\verb| 1.15   1.115|}$\pm${\verb| 0.35   0.040|}$\pm${\verb| 0.19|}\\
{\verb|0.3 1.31 0.796  37.5  10.545|}$\pm${\verb| 0.75  -0.773|}$\pm${\verb| 0.36  -3.395|}$\pm${\verb| 0.19|}\\
{\verb|0.3 1.31 0.796  52.5   9.837|}$\pm${\verb| 0.64   0.306|}$\pm${\verb| 0.30  -2.568|}$\pm${\verb| 0.14|}\\
{\verb|0.3 1.31 0.796  67.5   6.128|}$\pm${\verb| 0.56  -1.904|}$\pm${\verb| 0.52  -5.526|}$\pm${\verb| 0.33|}\\
{\verb|0.3 1.31 0.796  82.5   5.535|}$\pm${\verb| 0.38  -0.760|}$\pm${\verb| 0.19  -3.315|}$\pm${\verb| 0.08|}\\
{\verb|0.3 1.31 0.796  97.5   4.649|}$\pm${\verb| 0.30  -0.693|}$\pm${\verb| 0.10  -2.687|}$\pm${\verb| 0.02|}\\
{\verb|0.3 1.31 0.796 112.5   3.312|}$\pm${\verb| 0.21  -0.323|}$\pm${\verb| 0.08  -1.649|}$\pm${\verb| 0.02|}\\
 {$Q^2$~~$W$~~~$\epsilon$~~~~~$\theta$~~~~~$\sigma_{T}+\epsilon\sigma_{L}$~~~~~~$\sigma_{TL}$~~~~~~~~~$\sigma_{TT}$ } \\ 
 \underline{\hspace{6.15cm}} \\ \vspace{0.1cm}
{\verb|0.3 1.31 0.796 127.5   2.281|}$\pm${\verb| 0.19  -0.220|}$\pm${\verb| 0.20  -1.197|}$\pm${\verb| 0.08|}\\
{\verb|0.3 1.31 0.796 142.5   1.589|}$\pm${\verb| 1.41  -0.109|}$\pm${\verb| 1.45  -0.505|}$\pm${\verb| 1.58|}\\
{\verb|0.3 1.33 0.782   7.5  22.290|}$\pm${\verb| 2.40  -4.533|}$\pm${\verb| 1.94   5.550|}$\pm${\verb| 2.35|}\\
{\verb|0.3 1.33 0.782  37.5  10.384|}$\pm${\verb| 0.83   0.104|}$\pm${\verb| 0.41  -3.681|}$\pm${\verb| 0.23|}\\
{\verb|0.3 1.33 0.782  52.5   8.508|}$\pm${\verb| 0.64  -0.010|}$\pm${\verb| 0.27  -3.841|}$\pm${\verb| 0.14|}\\
{\verb|0.3 1.33 0.782  67.5   6.063|}$\pm${\verb| 0.49  -0.543|}$\pm${\verb| 0.33  -3.271|}$\pm${\verb| 0.18|}\\
{\verb|0.3 1.33 0.782  82.5   4.680|}$\pm${\verb| 0.31  -0.519|}$\pm${\verb| 0.15  -2.827|}$\pm${\verb| 0.06|}\\
{\verb|0.3 1.33 0.782  97.5   3.584|}$\pm${\verb| 0.23  -0.563|}$\pm${\verb| 0.08  -2.019|}$\pm${\verb| 0.02|}\\
{\verb|0.3 1.33 0.782 112.5   2.611|}$\pm${\verb| 0.18  -0.366|}$\pm${\verb| 0.08  -1.086|}$\pm${\verb| 0.02|}\\
{\verb|0.3 1.33 0.782 127.5   1.536|}$\pm${\verb| 0.12  -0.185|}$\pm${\verb| 0.07  -0.959|}$\pm${\verb| 0.02|}\\
{\verb|0.3 1.33 0.782 142.5   0.625|}$\pm${\verb| 1.23   0.286|}$\pm${\verb| 1.11  -0.636|}$\pm${\verb| 1.03|}\\
{\verb|0.3 1.35 0.767   7.5  16.085|}$\pm${\verb| 1.11  -0.229|}$\pm${\verb| 0.48   0.125|}$\pm${\verb| 0.33|}\\
{\verb|0.3 1.35 0.767  37.5   9.509|}$\pm${\verb| 0.81   0.877|}$\pm${\verb| 0.47  -1.840|}$\pm${\verb| 0.28|}\\
{\verb|0.3 1.35 0.767  52.5   6.997|}$\pm${\verb| 0.56  -0.163|}$\pm${\verb| 0.18  -3.507|}$\pm${\verb| 0.07|}\\
{\verb|0.3 1.35 0.767  67.5   4.947|}$\pm${\verb| 0.40  -0.918|}$\pm${\verb| 0.23  -4.008|}$\pm${\verb| 0.11|}\\
{\verb|0.3 1.35 0.767  82.5   4.056|}$\pm${\verb| 0.31  -0.434|}$\pm${\verb| 0.14  -2.658|}$\pm${\verb| 0.05|}\\
{\verb|0.3 1.35 0.767  97.5   3.038|}$\pm${\verb| 0.19  -0.465|}$\pm${\verb| 0.06  -1.934|}$\pm${\verb| 0.01|}\\
{\verb|0.3 1.35 0.767 112.5   1.982|}$\pm${\verb| 0.13  -0.262|}$\pm${\verb| 0.05  -0.966|}$\pm${\verb| 0.01|}\\
{\verb|0.3 1.35 0.767 127.5   1.163|}$\pm${\verb| 0.10  -0.165|}$\pm${\verb| 0.07  -0.553|}$\pm${\verb| 0.02|}\\
{\verb|0.3 1.35 0.767 142.5   0.238|}$\pm${\verb| 1.36   0.527|}$\pm${\verb| 1.33  -0.940|}$\pm${\verb| 1.38|}\\
{\verb|0.3 1.37 0.751   7.5  14.138|}$\pm${\verb| 1.00   0.214|}$\pm${\verb| 0.42   0.237|}$\pm${\verb| 0.24|}\\
{\verb|0.3 1.37 0.751  22.5  16.143|}$\pm${\verb| 3.30  -3.769|}$\pm${\verb| 3.59   4.075|}$\pm${\verb| 6.77|}\\
{\verb|0.3 1.37 0.751  37.5   8.308|}$\pm${\verb| 0.62   1.104|}$\pm${\verb| 0.28  -1.647|}$\pm${\verb| 0.14|}\\
{\verb|0.3 1.37 0.751  52.5   6.724|}$\pm${\verb| 0.56   0.491|}$\pm${\verb| 0.26  -2.826|}$\pm${\verb| 0.13|}\\
{\verb|0.3 1.37 0.751  67.5   5.199|}$\pm${\verb| 0.41   0.112|}$\pm${\verb| 0.26  -3.181|}$\pm${\verb| 0.12|}\\
{\verb|0.3 1.37 0.751  82.5   3.711|}$\pm${\verb| 0.27  -0.197|}$\pm${\verb| 0.20  -2.611|}$\pm${\verb| 0.08|}\\
{\verb|0.3 1.37 0.751  97.5   2.491|}$\pm${\verb| 0.16  -0.415|}$\pm${\verb| 0.05  -1.817|}$\pm${\verb| 0.01|}\\
{\verb|0.3 1.37 0.751 112.5   1.770|}$\pm${\verb| 0.12  -0.304|}$\pm${\verb| 0.08  -0.799|}$\pm${\verb| 0.02|}\\
{\verb|0.3 1.37 0.751 127.5   1.128|}$\pm${\verb| 0.11  -0.250|}$\pm${\verb| 0.06  -0.469|}$\pm${\verb| 0.01|}\\
{\verb|0.3 1.37 0.751 142.5   1.216|}$\pm${\verb| 0.52  -0.617|}$\pm${\verb| 0.50   0.209|}$\pm${\verb| 0.34|}\\
{\verb|0.3 1.39 0.734   7.5  13.981|}$\pm${\verb| 0.91   0.898|}$\pm${\verb| 0.50  -0.002|}$\pm${\verb| 0.30|}\\
{\verb|0.3 1.39 0.734  37.5   8.470|}$\pm${\verb| 0.70   1.782|}$\pm${\verb| 0.35  -1.980|}$\pm${\verb| 0.18|}\\
{\verb|0.3 1.39 0.734  52.5   6.659|}$\pm${\verb| 0.50   0.646|}$\pm${\verb| 0.33  -3.588|}$\pm${\verb| 0.18|}\\
{\verb|0.3 1.39 0.734  67.5   4.851|}$\pm${\verb| 0.48  -0.032|}$\pm${\verb| 0.28  -3.155|}$\pm${\verb| 0.14|}\\
{\verb|0.3 1.39 0.734  82.5   3.296|}$\pm${\verb| 0.23  -0.436|}$\pm${\verb| 0.10  -2.655|}$\pm${\verb| 0.03|}\\
{\verb|0.3 1.39 0.734  97.5   2.383|}$\pm${\verb| 0.15  -0.210|}$\pm${\verb| 0.06  -1.684|}$\pm${\verb| 0.01|}\\
{\verb|0.3 1.39 0.734 112.5   1.515|}$\pm${\verb| 0.11  -0.283|}$\pm${\verb| 0.06  -0.810|}$\pm${\verb| 0.01|}\\
{\verb|0.3 1.39 0.734 127.5   0.922|}$\pm${\verb| 0.08  -0.167|}$\pm${\verb| 0.06  -0.559|}$\pm${\verb| 0.01|}\\
{\verb|0.3 1.39 0.734 142.5   1.454|}$\pm${\verb| 1.47  -0.922|}$\pm${\verb| 1.37   0.556|}$\pm${\verb| 1.54|}\\
{\verb|0.3 1.41 0.715   7.5  12.408|}$\pm${\verb| 0.84  -0.085|}$\pm${\verb| 0.32  -1.107|}$\pm${\verb| 0.18|}\\
{\verb|0.3 1.41 0.715  22.5   9.119|}$\pm${\verb| 1.09   0.371|}$\pm${\verb| 0.69  -1.825|}$\pm${\verb| 0.56|}\\
{\verb|0.3 1.41 0.715  37.5   7.486|}$\pm${\verb| 0.64   1.512|}$\pm${\verb| 0.29  -2.465|}$\pm${\verb| 0.14|}\\
{\verb|0.3 1.41 0.715  52.5   6.129|}$\pm${\verb| 0.44   0.571|}$\pm${\verb| 0.33  -3.617|}$\pm${\verb| 0.16|}\\
{\verb|0.3 1.41 0.715  67.5   4.480|}$\pm${\verb| 0.33  -0.168|}$\pm${\verb| 0.20  -3.726|}$\pm${\verb| 0.08|}\\
{\verb|0.3 1.41 0.715  82.5   3.244|}$\pm${\verb| 0.23  -0.240|}$\pm${\verb| 0.11  -2.530|}$\pm${\verb| 0.03|}\\
{\verb|0.3 1.41 0.715  97.5   2.297|}$\pm${\verb| 0.15  -0.158|}$\pm${\verb| 0.05  -1.731|}$\pm${\verb| 0.01|}\\
{\verb|0.3 1.41 0.715 112.5   1.425|}$\pm${\verb| 0.09  -0.183|}$\pm${\verb| 0.04  -1.010|}$\pm${\verb| 0.01|}\\
{\verb|0.3 1.41 0.715 127.5   1.015|}$\pm${\verb| 0.12  -0.250|}$\pm${\verb| 0.09  -0.453|}$\pm${\verb| 0.02|}\\
{\verb|0.3 1.41 0.715 142.5   0.785|}$\pm${\verb| 0.65  -0.233|}$\pm${\verb| 0.46  -0.090|}$\pm${\verb| 0.31|}\\
{\verb|0.3 1.43 0.695   7.5  11.388|}$\pm${\verb| 0.89   1.796|}$\pm${\verb| 0.30  -0.813|}$\pm${\verb| 0.15|}\\
{\verb|0.3 1.43 0.695  22.5   7.441|}$\pm${\verb| 0.80   0.228|}$\pm${\verb| 0.66  -2.310|}$\pm${\verb| 0.51|}\\
{\verb|0.3 1.43 0.695  52.5   5.876|}$\pm${\verb| 0.59   0.177|}$\pm${\verb| 0.35  -4.535|}$\pm${\verb| 0.18|}\\
{\verb|0.3 1.43 0.695  67.5   4.644|}$\pm${\verb| 0.31  -0.058|}$\pm${\verb| 0.11  -3.828|}$\pm${\verb| 0.03|}\\
{\verb|0.3 1.43 0.695  82.5   3.518|}$\pm${\verb| 0.26  -0.149|}$\pm${\verb| 0.19  -2.869|}$\pm${\verb| 0.07|}\\
{\verb|0.3 1.43 0.695  97.5   2.345|}$\pm${\verb| 0.16  -0.235|}$\pm${\verb| 0.05  -1.914|}$\pm${\verb| 0.01|}\\
{\verb|0.3 1.43 0.695 112.5   1.574|}$\pm${\verb| 0.11  -0.138|}$\pm${\verb| 0.04  -1.259|}$\pm${\verb| 0.01|}\\
{\verb|0.3 1.43 0.695 127.5   0.980|}$\pm${\verb| 0.07  -0.132|}$\pm${\verb| 0.05  -0.671|}$\pm${\verb| 0.01|}\\
{\verb|0.3 1.43 0.695 142.5   0.856|}$\pm${\verb| 0.58  -0.339|}$\pm${\verb| 0.58  -0.134|}$\pm${\verb| 0.39|}\\
{\verb|0.3 1.45 0.674   7.5   9.761|}$\pm${\verb| 0.72   1.533|}$\pm${\verb| 0.26  -1.459|}$\pm${\verb| 0.12|}\\
{\verb|0.3 1.45 0.674  22.5   6.695|}$\pm${\verb| 1.33   0.006|}$\pm${\verb| 1.30  -2.646|}$\pm${\verb| 1.45|}\\
{\verb|0.3 1.45 0.674  37.5   8.181|}$\pm${\verb| 1.35   0.364|}$\pm${\verb| 1.01  -2.126|}$\pm${\verb| 1.00|}\\
{\verb|0.3 1.45 0.674  52.5   6.718|}$\pm${\verb| 0.60   1.298|}$\pm${\verb| 0.49  -3.392|}$\pm${\verb| 0.31|}\\
{\verb|0.3 1.45 0.674  67.5   4.985|}$\pm${\verb| 0.37  -0.037|}$\pm${\verb| 0.09  -4.081|}$\pm${\verb| 0.02|}\\
{\verb|0.3 1.45 0.674  82.5   3.627|}$\pm${\verb| 0.26  -0.158|}$\pm${\verb| 0.10  -2.970|}$\pm${\verb| 0.03|}\\
{\verb|0.3 1.45 0.674  97.5   2.730|}$\pm${\verb| 0.19  -0.212|}$\pm${\verb| 0.05  -2.476|}$\pm${\verb| 0.01|}\\
{\verb|0.3 1.45 0.674 112.5   1.883|}$\pm${\verb| 0.13  -0.207|}$\pm${\verb| 0.04  -1.507|}$\pm${\verb| 0.01|}\\
{\verb|0.3 1.45 0.674 127.5   1.269|}$\pm${\verb| 0.09  -0.209|}$\pm${\verb| 0.05  -0.806|}$\pm${\verb| 0.01|}\\
{\verb|0.3 1.45 0.674 142.5   1.520|}$\pm${\verb| 1.20  -0.751|}$\pm${\verb| 1.20   0.099|}$\pm${\verb| 1.11|}\\
{\verb|0.3 1.47 0.652   7.5  10.384|}$\pm${\verb| 0.71   1.601|}$\pm${\verb| 0.30  -0.872|}$\pm${\verb| 0.15|}\\
{\verb|0.3 1.47 0.652  22.5   8.966|}$\pm${\verb| 1.90   2.585|}$\pm${\verb| 1.97  -0.794|}$\pm${\verb| 2.77|}\\
{\verb|0.3 1.47 0.652  37.5   7.325|}$\pm${\verb| 0.55   1.430|}$\pm${\verb| 0.40  -3.595|}$\pm${\verb| 0.23|}\\
{\verb|0.3 1.47 0.652  52.5   6.294|}$\pm${\verb| 0.47   0.016|}$\pm${\verb| 0.34  -5.393|}$\pm${\verb| 0.17|}\\
{\verb|0.3 1.47 0.652  67.5   5.710|}$\pm${\verb| 0.39   0.058|}$\pm${\verb| 0.11  -4.814|}$\pm${\verb| 0.03|}\\
{\verb|0.3 1.47 0.652  82.5   4.442|}$\pm${\verb| 0.34   0.016|}$\pm${\verb| 0.23  -3.808|}$\pm${\verb| 0.10|}\\
{\verb|0.3 1.47 0.652  97.5   3.550|}$\pm${\verb| 0.23  -0.262|}$\pm${\verb| 0.07  -2.739|}$\pm${\verb| 0.02|}\\
{\verb|0.3 1.47 0.652 112.5   2.577|}$\pm${\verb| 0.17  -0.307|}$\pm${\verb| 0.07  -1.997|}$\pm${\verb| 0.01|}\\
{\verb|0.3 1.47 0.652 127.5   1.916|}$\pm${\verb| 0.14  -0.215|}$\pm${\verb| 0.06  -1.358|}$\pm${\verb| 0.01|}\\
{\verb|0.3 1.47 0.652 142.5   1.586|}$\pm${\verb| 0.67  -0.471|}$\pm${\verb| 0.80  -0.438|}$\pm${\verb| 0.61|}\\
{\verb|0.3 1.49 0.628   7.5  10.202|}$\pm${\verb| 0.70   1.896|}$\pm${\verb| 0.37   0.293|}$\pm${\verb| 0.19|}\\
{\verb|0.3 1.49 0.628  22.5   6.937|}$\pm${\verb| 1.34   0.655|}$\pm${\verb| 1.26  -2.310|}$\pm${\verb| 1.35|}\\
{\verb|0.3 1.49 0.628  52.5   6.063|}$\pm${\verb| 0.44  -0.040|}$\pm${\verb| 0.35  -4.892|}$\pm${\verb| 0.19|}\\
{\verb|0.3 1.49 0.628  67.5   5.706|}$\pm${\verb| 0.42   0.085|}$\pm${\verb| 0.14  -4.624|}$\pm${\verb| 0.04|}\\
{\verb|0.3 1.49 0.628  82.5   4.947|}$\pm${\verb| 0.36  -0.101|}$\pm${\verb| 0.20  -4.544|}$\pm${\verb| 0.09|}\\
{\verb|0.3 1.49 0.628  97.5   4.199|}$\pm${\verb| 0.29  -0.354|}$\pm${\verb| 0.07  -3.412|}$\pm${\verb| 0.02|}\\
{\verb|0.3 1.49 0.628 112.5   3.542|}$\pm${\verb| 0.25  -0.387|}$\pm${\verb| 0.06  -3.278|}$\pm${\verb| 0.01|}\\
{\verb|0.3 1.49 0.628 127.5   2.678|}$\pm${\verb| 0.20  -0.392|}$\pm${\verb| 0.10  -2.050|}$\pm${\verb| 0.03|}\\
{\verb|0.3 1.49 0.628 142.5   2.122|}$\pm${\verb| 0.35  -0.400|}$\pm${\verb| 0.40  -1.334|}$\pm${\verb| 0.21|}\\
{\verb|0.3 1.51 0.602   7.5  11.363|}$\pm${\verb| 0.81   1.969|}$\pm${\verb| 0.46   0.260|}$\pm${\verb| 0.27|}\\
{\verb|0.3 1.51 0.602  22.5   8.248|}$\pm${\verb| 1.51   1.289|}$\pm${\verb| 1.59  -1.480|}$\pm${\verb| 1.98|}\\
{\verb|0.3 1.51 0.602  37.5   7.380|}$\pm${\verb| 1.11   1.315|}$\pm${\verb| 0.78  -2.254|}$\pm${\verb| 0.67|}\\
{\verb|0.3 1.51 0.602  52.5   5.285|}$\pm${\verb| 0.41   0.255|}$\pm${\verb| 0.39  -3.690|}$\pm${\verb| 0.23|}\\
{\verb|0.3 1.51 0.602  67.5   4.250|}$\pm${\verb| 0.27  -0.111|}$\pm${\verb| 0.08  -3.596|}$\pm${\verb| 0.02|}\\
{\verb|0.3 1.51 0.602  82.5   3.825|}$\pm${\verb| 0.30  -0.230|}$\pm${\verb| 0.19  -3.628|}$\pm${\verb| 0.07|}\\
{\verb|0.3 1.51 0.602  97.5   3.590|}$\pm${\verb| 0.24  -0.353|}$\pm${\verb| 0.06  -3.310|}$\pm${\verb| 0.01|}\\
{\verb|0.3 1.51 0.602 112.5   3.379|}$\pm${\verb| 0.23  -0.294|}$\pm${\verb| 0.07  -2.952|}$\pm${\verb| 0.02|}\\
{\verb|0.3 1.51 0.602 127.5   2.683|}$\pm${\verb| 0.20  -0.263|}$\pm${\verb| 0.08  -1.985|}$\pm${\verb| 0.02|}\\
{\verb|0.3 1.51 0.602 142.5   2.002|}$\pm${\verb| 0.53   0.042|}$\pm${\verb| 0.62  -1.599|}$\pm${\verb| 0.48|}\\
{\verb|0.3 1.53 0.575   7.5  13.190|}$\pm${\verb| 1.08   2.641|}$\pm${\verb| 0.65   1.165|}$\pm${\verb| 0.47|}\\
{\verb|0.3 1.53 0.575  22.5   5.607|}$\pm${\verb| 1.20  -2.410|}$\pm${\verb| 1.61  -5.446|}$\pm${\verb| 1.84|}\\
{\verb|0.3 1.53 0.575  37.5   6.150|}$\pm${\verb| 0.57   0.443|}$\pm${\verb| 0.57  -1.619|}$\pm${\verb| 0.40|}\\
{\verb|0.3 1.53 0.575  67.5   2.767|}$\pm${\verb| 0.18  -0.088|}$\pm${\verb| 0.10  -1.846|}$\pm${\verb| 0.03|}\\
{\verb|0.3 1.53 0.575  82.5   2.494|}$\pm${\verb| 0.19  -0.103|}$\pm${\verb| 0.12  -2.088|}$\pm${\verb| 0.04|}\\
{\verb|0.3 1.53 0.575  97.5   2.621|}$\pm${\verb| 0.16  -0.178|}$\pm${\verb| 0.06  -2.316|}$\pm${\verb| 0.01|}\\
{\verb|0.3 1.53 0.575 112.5   2.732|}$\pm${\verb| 0.18  -0.043|}$\pm${\verb| 0.06  -2.315|}$\pm${\verb| 0.01|}\\
{\verb|0.3 1.53 0.575 127.5   2.528|}$\pm${\verb| 0.17  -0.104|}$\pm${\verb| 0.09  -1.536|}$\pm${\verb| 0.03|}\\
{\verb|0.3 1.53 0.575 142.5   2.217|}$\pm${\verb| 0.36  -0.119|}$\pm${\verb| 0.35  -1.366|}$\pm${\verb| 0.18|}\\
{\verb|0.3 1.55 0.547   7.5  14.342|}$\pm${\verb| 1.48   3.682|}$\pm${\verb| 1.43  -0.848|}$\pm${\verb| 1.47|}\\
{\verb|0.3 1.55 0.547  22.5   8.640|}$\pm${\verb| 1.77   1.923|}$\pm${\verb| 1.77  -0.618|}$\pm${\verb| 2.23|}\\
 {$Q^2$~~$W$~~~$\epsilon$~~~~~$\theta$~~~~~$\sigma_{T}+\epsilon\sigma_{L}$~~~~~~$\sigma_{TL}$~~~~~~~~~$\sigma_{TT}$ } \\ 
 \underline{\hspace{6.15cm}} \\ \vspace{0.1cm}
{\verb|0.3 1.55 0.547  37.5   4.925|}$\pm${\verb| 0.47   0.071|}$\pm${\verb| 0.54  -2.457|}$\pm${\verb| 0.36|}\\
{\verb|0.3 1.55 0.547  52.5   2.717|}$\pm${\verb| 0.62   0.070|}$\pm${\verb| 0.59  -2.645|}$\pm${\verb| 0.45|}\\
{\verb|0.3 1.55 0.547  67.5   1.893|}$\pm${\verb| 0.13  -0.171|}$\pm${\verb| 0.08  -1.520|}$\pm${\verb| 0.02|}\\
{\verb|0.3 1.55 0.547  82.5   1.433|}$\pm${\verb| 0.14  -0.117|}$\pm${\verb| 0.12  -1.254|}$\pm${\verb| 0.04|}\\
{\verb|0.3 1.55 0.547  97.5   1.620|}$\pm${\verb| 0.11   0.009|}$\pm${\verb| 0.05  -1.227|}$\pm${\verb| 0.01|}\\
{\verb|0.3 1.55 0.547 112.5   1.929|}$\pm${\verb| 0.14   0.006|}$\pm${\verb| 0.05  -1.558|}$\pm${\verb| 0.01|}\\
{\verb|0.3 1.55 0.547 127.5   1.872|}$\pm${\verb| 0.14   0.045|}$\pm${\verb| 0.06  -1.166|}$\pm${\verb| 0.01|}\\
{\verb|0.3 1.55 0.547 142.5   1.628|}$\pm${\verb| 0.21   0.095|}$\pm${\verb| 0.13  -0.840|}$\pm${\verb| 0.04|}\\
{\verb|0.3 1.57 0.516   7.5  11.843|}$\pm${\verb| 2.00   1.512|}$\pm${\verb| 2.37   1.356|}$\pm${\verb| 3.65|}\\
{\verb|0.3 1.57 0.516  22.5   7.309|}$\pm${\verb| 1.78   0.685|}$\pm${\verb| 2.31  -2.492|}$\pm${\verb| 3.51|}\\
{\verb|0.3 1.57 0.516  37.5   3.701|}$\pm${\verb| 0.41  -0.790|}$\pm${\verb| 0.53  -3.581|}$\pm${\verb| 0.37|}\\
{\verb|0.3 1.57 0.516  52.5   2.138|}$\pm${\verb| 0.44   0.070|}$\pm${\verb| 0.44  -3.372|}$\pm${\verb| 0.29|}\\
{\verb|0.3 1.57 0.516  67.5   1.460|}$\pm${\verb| 0.12  -0.207|}$\pm${\verb| 0.11  -0.869|}$\pm${\verb| 0.03|}\\
{\verb|0.3 1.57 0.516  82.5   0.992|}$\pm${\verb| 0.11  -0.041|}$\pm${\verb| 0.12  -0.346|}$\pm${\verb| 0.04|}\\
{\verb|0.3 1.57 0.516  97.5   1.036|}$\pm${\verb| 0.08   0.000|}$\pm${\verb| 0.09  -0.575|}$\pm${\verb| 0.02|}\\
{\verb|0.3 1.57 0.516 112.5   1.491|}$\pm${\verb| 0.14   0.111|}$\pm${\verb| 0.06  -0.992|}$\pm${\verb| 0.01|}\\
{\verb|0.3 1.57 0.516 127.5   1.492|}$\pm${\verb| 0.11   0.148|}$\pm${\verb| 0.09  -0.626|}$\pm${\verb| 0.02|}\\
{\verb|0.3 1.57 0.516 142.5   1.293|}$\pm${\verb| 0.16   0.401|}$\pm${\verb| 0.17  -0.876|}$\pm${\verb| 0.07|}\\
{\verb|0.4 1.11 0.850   7.5   3.026|}$\pm${\verb| 0.56  -0.592|}$\pm${\verb| 0.44   1.302|}$\pm${\verb| 0.25|}\\
{\verb|0.4 1.11 0.850  22.5   3.448|}$\pm${\verb| 0.27  -0.265|}$\pm${\verb| 0.14  -0.159|}$\pm${\verb| 0.05|}\\
{\verb|0.4 1.11 0.850  37.5   3.315|}$\pm${\verb| 0.27  -0.698|}$\pm${\verb| 0.14  -0.057|}$\pm${\verb| 0.05|}\\
{\verb|0.4 1.11 0.850  52.5   3.303|}$\pm${\verb| 0.32  -1.003|}$\pm${\verb| 0.23  -0.247|}$\pm${\verb| 0.10|}\\
{\verb|0.4 1.11 0.850  67.5   2.823|}$\pm${\verb| 0.27  -1.005|}$\pm${\verb| 0.24  -1.012|}$\pm${\verb| 0.10|}\\
{\verb|0.4 1.11 0.850  82.5   3.131|}$\pm${\verb| 0.24  -0.668|}$\pm${\verb| 0.16  -0.816|}$\pm${\verb| 0.06|}\\
{\verb|0.4 1.11 0.850  97.5   3.014|}$\pm${\verb| 0.34  -0.692|}$\pm${\verb| 0.27  -0.961|}$\pm${\verb| 0.12|}\\
{\verb|0.4 1.11 0.850 112.5   3.226|}$\pm${\verb| 1.19  -0.409|}$\pm${\verb| 1.17  -0.380|}$\pm${\verb| 1.21|}\\
{\verb|0.4 1.13 0.843   7.5   5.045|}$\pm${\verb| 0.66  -0.204|}$\pm${\verb| 0.35  -0.503|}$\pm${\verb| 0.18|}\\
{\verb|0.4 1.13 0.843  22.5   4.058|}$\pm${\verb| 0.30  -0.729|}$\pm${\verb| 0.14  -0.247|}$\pm${\verb| 0.05|}\\
{\verb|0.4 1.13 0.843  37.5   4.385|}$\pm${\verb| 0.32  -0.992|}$\pm${\verb| 0.16  -0.106|}$\pm${\verb| 0.06|}\\
{\verb|0.4 1.13 0.843  52.5   4.441|}$\pm${\verb| 0.42  -0.994|}$\pm${\verb| 0.21  -1.153|}$\pm${\verb| 0.08|}\\
{\verb|0.4 1.13 0.843  67.5   4.314|}$\pm${\verb| 0.30  -0.990|}$\pm${\verb| 0.15  -1.460|}$\pm${\verb| 0.05|}\\
{\verb|0.4 1.13 0.843  82.5   4.453|}$\pm${\verb| 0.31  -1.035|}$\pm${\verb| 0.15  -1.705|}$\pm${\verb| 0.05|}\\
{\verb|0.4 1.13 0.843  97.5   4.093|}$\pm${\verb| 0.34  -1.080|}$\pm${\verb| 0.26  -2.391|}$\pm${\verb| 0.11|}\\
{\verb|0.4 1.13 0.843 112.5   4.252|}$\pm${\verb| 0.33  -0.961|}$\pm${\verb| 0.19  -1.547|}$\pm${\verb| 0.08|}\\
{\verb|0.4 1.13 0.843 127.5   4.292|}$\pm${\verb| 0.42   0.021|}$\pm${\verb| 0.27  -3.695|}$\pm${\verb| 0.12|}\\
{\verb|0.4 1.15 0.834   7.5   4.608|}$\pm${\verb| 0.44  -0.048|}$\pm${\verb| 0.24   0.481|}$\pm${\verb| 0.11|}\\
{\verb|0.4 1.15 0.834  22.5   4.839|}$\pm${\verb| 0.33  -0.842|}$\pm${\verb| 0.14  -0.648|}$\pm${\verb| 0.05|}\\
{\verb|0.4 1.15 0.834  37.5   5.180|}$\pm${\verb| 0.38  -1.272|}$\pm${\verb| 0.14  -1.102|}$\pm${\verb| 0.05|}\\
{\verb|0.4 1.15 0.834  52.5   5.254|}$\pm${\verb| 0.38  -1.879|}$\pm${\verb| 0.22  -1.928|}$\pm${\verb| 0.09|}\\
{\verb|0.4 1.15 0.834  67.5   6.121|}$\pm${\verb| 0.39  -1.328|}$\pm${\verb| 0.18  -2.266|}$\pm${\verb| 0.06|}\\
{\verb|0.4 1.15 0.834  82.5   6.501|}$\pm${\verb| 0.44  -1.051|}$\pm${\verb| 0.22  -2.676|}$\pm${\verb| 0.09|}\\
{\verb|0.4 1.15 0.834  97.5   6.084|}$\pm${\verb| 0.46  -1.073|}$\pm${\verb| 0.24  -3.514|}$\pm${\verb| 0.11|}\\
{\verb|0.4 1.15 0.834 112.5   6.052|}$\pm${\verb| 0.41  -0.554|}$\pm${\verb| 0.18  -2.880|}$\pm${\verb| 0.06|}\\
{\verb|0.4 1.15 0.834 127.5   6.371|}$\pm${\verb| 0.59  -0.344|}$\pm${\verb| 0.44  -2.613|}$\pm${\verb| 0.25|}\\
{\verb|0.4 1.15 0.834 142.5   7.645|}$\pm${\verb| 2.36  -1.068|}$\pm${\verb| 2.05  -0.795|}$\pm${\verb| 2.94|}\\
{\verb|0.4 1.17 0.826   7.5   5.183|}$\pm${\verb| 0.49  -0.969|}$\pm${\verb| 0.26   0.026|}$\pm${\verb| 0.12|}\\
{\verb|0.4 1.17 0.826  22.5   5.743|}$\pm${\verb| 0.39  -1.429|}$\pm${\verb| 0.16  -0.870|}$\pm${\verb| 0.05|}\\
{\verb|0.4 1.17 0.826  37.5   6.654|}$\pm${\verb| 0.66  -1.252|}$\pm${\verb| 0.45  -0.340|}$\pm${\verb| 0.29|}\\
{\verb|0.4 1.17 0.826  52.5   7.210|}$\pm${\verb| 0.46  -2.289|}$\pm${\verb| 0.23  -3.481|}$\pm${\verb| 0.10|}\\
{\verb|0.4 1.17 0.826  67.5   8.043|}$\pm${\verb| 0.52  -1.466|}$\pm${\verb| 0.23  -3.867|}$\pm${\verb| 0.10|}\\
{\verb|0.4 1.17 0.826  82.5   8.874|}$\pm${\verb| 0.61  -1.391|}$\pm${\verb| 0.27  -4.932|}$\pm${\verb| 0.12|}\\
{\verb|0.4 1.17 0.826  97.5   8.568|}$\pm${\verb| 0.58  -1.323|}$\pm${\verb| 0.25  -5.178|}$\pm${\verb| 0.11|}\\
{\verb|0.4 1.17 0.826 112.5   8.755|}$\pm${\verb| 0.55  -0.646|}$\pm${\verb| 0.15  -4.590|}$\pm${\verb| 0.05|}\\
{\verb|0.4 1.17 0.826 127.5   8.386|}$\pm${\verb| 0.60  -0.424|}$\pm${\verb| 0.40  -2.863|}$\pm${\verb| 0.21|}\\
{\verb|0.4 1.17 0.826 142.5   6.490|}$\pm${\verb| 2.24   3.447|}$\pm${\verb| 2.60  -5.914|}$\pm${\verb| 3.53|}\\
{\verb|0.4 1.19 0.816   7.5   7.686|}$\pm${\verb| 0.64  -0.368|}$\pm${\verb| 0.24  -0.778|}$\pm${\verb| 0.11|}\\
{\verb|0.4 1.19 0.816  22.5   7.180|}$\pm${\verb| 0.58  -2.041|}$\pm${\verb| 0.29  -2.011|}$\pm${\verb| 0.14|}\\
{\verb|0.4 1.19 0.816  37.5   8.637|}$\pm${\verb| 0.61  -1.994|}$\pm${\verb| 0.33  -2.056|}$\pm${\verb| 0.16|}\\
{\verb|0.4 1.19 0.816  52.5   9.435|}$\pm${\verb| 0.69  -2.779|}$\pm${\verb| 0.49  -4.851|}$\pm${\verb| 0.29|}\\
{\verb|0.4 1.19 0.816  67.5  11.654|}$\pm${\verb| 0.74  -2.144|}$\pm${\verb| 0.31  -6.047|}$\pm${\verb| 0.15|}\\
{\verb|0.4 1.19 0.816  82.5  12.359|}$\pm${\verb| 0.79  -1.784|}$\pm${\verb| 0.34  -7.104|}$\pm${\verb| 0.17|}\\
{\verb|0.4 1.19 0.816  97.5  12.523|}$\pm${\verb| 0.76  -0.764|}$\pm${\verb| 0.16  -6.557|}$\pm${\verb| 0.06|}\\
{\verb|0.4 1.19 0.816 112.5  12.034|}$\pm${\verb| 0.74  -0.268|}$\pm${\verb| 0.19  -5.973|}$\pm${\verb| 0.07|}\\
{\verb|0.4 1.19 0.816 127.5  11.437|}$\pm${\verb| 0.74  -0.509|}$\pm${\verb| 0.32  -3.368|}$\pm${\verb| 0.16|}\\
{\verb|0.4 1.19 0.816 142.5   6.464|}$\pm${\verb| 1.96   4.437|}$\pm${\verb| 1.80  -8.293|}$\pm${\verb| 2.42|}\\
{\verb|0.4 1.21 0.806   7.5  10.629|}$\pm${\verb| 0.97  -1.161|}$\pm${\verb| 0.66  -1.591|}$\pm${\verb| 0.52|}\\
{\verb|0.4 1.21 0.806  22.5  10.347|}$\pm${\verb| 0.72  -1.708|}$\pm${\verb| 0.33  -0.492|}$\pm${\verb| 0.16|}\\
{\verb|0.4 1.21 0.806  37.5  11.336|}$\pm${\verb| 0.92  -2.215|}$\pm${\verb| 0.41  -2.583|}$\pm${\verb| 0.22|}\\
{\verb|0.4 1.21 0.806  52.5  13.319|}$\pm${\verb| 1.05  -1.537|}$\pm${\verb| 0.76  -4.211|}$\pm${\verb| 0.61|}\\
{\verb|0.4 1.21 0.806  67.5  14.824|}$\pm${\verb| 0.99  -1.601|}$\pm${\verb| 0.50  -7.055|}$\pm${\verb| 0.33|}\\
{\verb|0.4 1.21 0.806  82.5  14.388|}$\pm${\verb| 0.95  -2.225|}$\pm${\verb| 0.36  -9.401|}$\pm${\verb| 0.19|}\\
{\verb|0.4 1.21 0.806  97.5  15.504|}$\pm${\verb| 0.93  -0.987|}$\pm${\verb| 0.23  -8.303|}$\pm${\verb| 0.10|}\\
{\verb|0.4 1.21 0.806 112.5  13.911|}$\pm${\verb| 0.85  -0.142|}$\pm${\verb| 0.16  -7.085|}$\pm${\verb| 0.07|}\\
{\verb|0.4 1.21 0.806 127.5  12.328|}$\pm${\verb| 0.79   0.116|}$\pm${\verb| 0.27  -5.073|}$\pm${\verb| 0.13|}\\
{\verb|0.4 1.21 0.806 142.5  12.066|}$\pm${\verb| 1.95  -0.743|}$\pm${\verb| 1.63  -1.899|}$\pm${\verb| 1.94|}\\
{\verb|0.4 1.23 0.795   7.5  11.708|}$\pm${\verb| 3.72  -0.856|}$\pm${\verb| 3.40  -1.593|}$\pm${\verb| 6.26|}\\
{\verb|0.4 1.23 0.795  22.5  13.914|}$\pm${\verb| 1.04  -0.465|}$\pm${\verb| 0.32  -1.624|}$\pm${\verb| 0.16|}\\
{\verb|0.4 1.23 0.795  37.5  12.940|}$\pm${\verb| 0.90  -1.533|}$\pm${\verb| 0.39  -3.270|}$\pm${\verb| 0.22|}\\
{\verb|0.4 1.23 0.795  52.5  13.269|}$\pm${\verb| 0.87  -1.966|}$\pm${\verb| 0.47  -4.924|}$\pm${\verb| 0.30|}\\
{\verb|0.4 1.23 0.795  67.5  14.214|}$\pm${\verb| 0.91  -1.158|}$\pm${\verb| 0.37  -5.951|}$\pm${\verb| 0.21|}\\
{\verb|0.4 1.23 0.795  82.5  13.309|}$\pm${\verb| 0.86  -1.329|}$\pm${\verb| 0.43  -7.421|}$\pm${\verb| 0.23|}\\
{\verb|0.4 1.23 0.795  97.5  13.116|}$\pm${\verb| 0.79  -0.789|}$\pm${\verb| 0.21  -7.061|}$\pm${\verb| 0.08|}\\
{\verb|0.4 1.23 0.795 112.5  11.843|}$\pm${\verb| 0.71  -0.130|}$\pm${\verb| 0.15  -5.314|}$\pm${\verb| 0.06|}\\
{\verb|0.4 1.23 0.795 127.5   9.943|}$\pm${\verb| 0.61   0.488|}$\pm${\verb| 0.19  -3.987|}$\pm${\verb| 0.08|}\\
{\verb|0.4 1.23 0.795 142.5   8.864|}$\pm${\verb| 1.56  -0.666|}$\pm${\verb| 1.40  -0.921|}$\pm${\verb| 1.41|}\\
{\verb|0.4 1.25 0.783  22.5  13.387|}$\pm${\verb| 0.91  -0.006|}$\pm${\verb| 0.29  -0.282|}$\pm${\verb| 0.14|}\\
{\verb|0.4 1.25 0.783  37.5  12.410|}$\pm${\verb| 0.95  -0.725|}$\pm${\verb| 0.52  -1.046|}$\pm${\verb| 0.36|}\\
{\verb|0.4 1.25 0.783  52.5  12.555|}$\pm${\verb| 0.86  -0.551|}$\pm${\verb| 0.44  -3.525|}$\pm${\verb| 0.29|}\\
{\verb|0.4 1.25 0.783  67.5  11.558|}$\pm${\verb| 0.79  -0.931|}$\pm${\verb| 0.42  -4.732|}$\pm${\verb| 0.26|}\\
{\verb|0.4 1.25 0.783  82.5  10.427|}$\pm${\verb| 0.69  -1.245|}$\pm${\verb| 0.30  -5.298|}$\pm${\verb| 0.15|}\\
{\verb|0.4 1.25 0.783  97.5   9.799|}$\pm${\verb| 0.60  -0.698|}$\pm${\verb| 0.16  -5.462|}$\pm${\verb| 0.05|}\\
{\verb|0.4 1.25 0.783 112.5   8.530|}$\pm${\verb| 0.53  -0.211|}$\pm${\verb| 0.12  -4.097|}$\pm${\verb| 0.04|}\\
{\verb|0.4 1.25 0.783 127.5   6.740|}$\pm${\verb| 0.46   0.244|}$\pm${\verb| 0.17  -2.504|}$\pm${\verb| 0.06|}\\
{\verb|0.4 1.25 0.783 142.5   4.298|}$\pm${\verb| 1.06   0.813|}$\pm${\verb| 0.95  -2.804|}$\pm${\verb| 0.84|}\\
{\verb|0.4 1.27 0.770  22.5  13.365|}$\pm${\verb| 0.94   0.430|}$\pm${\verb| 0.29   1.435|}$\pm${\verb| 0.15|}\\
{\verb|0.4 1.27 0.770  37.5  12.235|}$\pm${\verb| 1.03  -0.357|}$\pm${\verb| 0.53  -2.254|}$\pm${\verb| 0.34|}\\
{\verb|0.4 1.27 0.770  52.5  11.206|}$\pm${\verb| 0.75  -0.323|}$\pm${\verb| 0.32  -2.769|}$\pm${\verb| 0.17|}\\
{\verb|0.4 1.27 0.770  67.5   9.097|}$\pm${\verb| 0.63  -1.080|}$\pm${\verb| 0.38  -4.248|}$\pm${\verb| 0.22|}\\
{\verb|0.4 1.27 0.770  82.5   8.357|}$\pm${\verb| 0.54  -0.789|}$\pm${\verb| 0.24  -4.072|}$\pm${\verb| 0.11|}\\
{\verb|0.4 1.27 0.770  97.5   7.057|}$\pm${\verb| 0.45  -0.654|}$\pm${\verb| 0.16  -3.561|}$\pm${\verb| 0.05|}\\
{\verb|0.4 1.27 0.770 112.5   5.728|}$\pm${\verb| 0.35  -0.335|}$\pm${\verb| 0.10  -2.374|}$\pm${\verb| 0.03|}\\
{\verb|0.4 1.27 0.770 127.5   4.406|}$\pm${\verb| 0.38  -0.137|}$\pm${\verb| 0.26  -2.038|}$\pm${\verb| 0.12|}\\
{\verb|0.4 1.27 0.770 142.5   3.406|}$\pm${\verb| 1.13  -0.057|}$\pm${\verb| 1.04  -1.007|}$\pm${\verb| 0.90|}\\
{\verb|0.4 1.29 0.757   7.5  20.088|}$\pm${\verb|10.46  -4.928|}$\pm${\verb| 9.92   5.324|}$\pm${\verb|31.23|}\\
{\verb|0.4 1.29 0.757  22.5  13.465|}$\pm${\verb| 1.08   0.290|}$\pm${\verb| 0.29  -2.148|}$\pm${\verb| 0.14|}\\
{\verb|0.4 1.29 0.757  37.5  10.746|}$\pm${\verb| 0.89  -0.167|}$\pm${\verb| 0.55  -1.414|}$\pm${\verb| 0.35|}\\
{\verb|0.4 1.29 0.757  52.5   9.464|}$\pm${\verb| 0.61   0.001|}$\pm${\verb| 0.18  -2.089|}$\pm${\verb| 0.07|}\\
{\verb|0.4 1.29 0.757  67.5   7.110|}$\pm${\verb| 0.55  -1.180|}$\pm${\verb| 0.38  -3.252|}$\pm${\verb| 0.20|}\\
 {$Q^2$~~$W$~~~$\epsilon$~~~~~$\theta$~~~~~$\sigma_{T}+\epsilon\sigma_{L}$~~~~~~$\sigma_{TL}$~~~~~~~~~$\sigma_{TT}$ } \\ 
 \underline{\hspace{6.15cm}} \\ \vspace{0.1cm}
{\verb|0.4 1.29 0.757  82.5   6.630|}$\pm${\verb| 0.41  -0.729|}$\pm${\verb| 0.13  -2.639|}$\pm${\verb| 0.04|}\\
{\verb|0.4 1.29 0.757  97.5   5.024|}$\pm${\verb| 0.33  -0.653|}$\pm${\verb| 0.18  -2.176|}$\pm${\verb| 0.06|}\\
{\verb|0.4 1.29 0.757 112.5   3.862|}$\pm${\verb| 0.24  -0.390|}$\pm${\verb| 0.09  -1.519|}$\pm${\verb| 0.02|}\\
{\verb|0.4 1.29 0.757 127.5   2.770|}$\pm${\verb| 0.20  -0.207|}$\pm${\verb| 0.08  -1.215|}$\pm${\verb| 0.02|}\\
{\verb|0.4 1.29 0.757 142.5   1.998|}$\pm${\verb| 0.81  -0.212|}$\pm${\verb| 0.79  -0.727|}$\pm${\verb| 0.65|}\\
{\verb|0.4 1.31 0.742   7.5  16.645|}$\pm${\verb| 2.87  -2.076|}$\pm${\verb| 2.94   1.145|}$\pm${\verb| 4.34|}\\
{\verb|0.4 1.31 0.742  22.5  13.185|}$\pm${\verb| 1.23  -0.187|}$\pm${\verb| 0.52  -1.300|}$\pm${\verb| 0.36|}\\
{\verb|0.4 1.31 0.742  37.5   9.716|}$\pm${\verb| 1.14  -0.207|}$\pm${\verb| 0.94  -2.259|}$\pm${\verb| 0.79|}\\
{\verb|0.4 1.31 0.742  52.5   8.482|}$\pm${\verb| 0.57  -0.116|}$\pm${\verb| 0.23  -2.050|}$\pm${\verb| 0.10|}\\
{\verb|0.4 1.31 0.742  67.5   5.575|}$\pm${\verb| 0.44  -1.115|}$\pm${\verb| 0.33  -3.102|}$\pm${\verb| 0.18|}\\
{\verb|0.4 1.31 0.742  82.5   4.906|}$\pm${\verb| 0.32  -0.621|}$\pm${\verb| 0.17  -2.543|}$\pm${\verb| 0.06|}\\
{\verb|0.4 1.31 0.742  97.5   4.012|}$\pm${\verb| 0.27  -0.468|}$\pm${\verb| 0.12  -1.918|}$\pm${\verb| 0.03|}\\
{\verb|0.4 1.31 0.742 112.5   2.746|}$\pm${\verb| 0.18  -0.414|}$\pm${\verb| 0.08  -1.062|}$\pm${\verb| 0.02|}\\
{\verb|0.4 1.31 0.742 127.5   1.790|}$\pm${\verb| 0.13  -0.155|}$\pm${\verb| 0.06  -0.999|}$\pm${\verb| 0.01|}\\
{\verb|0.4 1.31 0.742 142.5   1.005|}$\pm${\verb| 0.34   0.292|}$\pm${\verb| 0.35  -0.540|}$\pm${\verb| 0.19|}\\
{\verb|0.4 1.33 0.726   7.5  14.481|}$\pm${\verb| 1.03  -0.836|}$\pm${\verb| 0.62  -0.892|}$\pm${\verb| 0.45|}\\
{\verb|0.4 1.33 0.726  22.5  12.583|}$\pm${\verb| 2.70  -0.580|}$\pm${\verb| 2.78  -1.031|}$\pm${\verb| 4.61|}\\
{\verb|0.4 1.33 0.726  37.5   9.506|}$\pm${\verb| 0.72   0.043|}$\pm${\verb| 0.42  -2.850|}$\pm${\verb| 0.26|}\\
{\verb|0.4 1.33 0.726  52.5   7.160|}$\pm${\verb| 0.48  -0.226|}$\pm${\verb| 0.20  -2.266|}$\pm${\verb| 0.08|}\\
{\verb|0.4 1.33 0.726  67.5   4.769|}$\pm${\verb| 0.37  -1.118|}$\pm${\verb| 0.24  -2.891|}$\pm${\verb| 0.11|}\\
{\verb|0.4 1.33 0.726  82.5   4.537|}$\pm${\verb| 0.32  -0.250|}$\pm${\verb| 0.18  -1.643|}$\pm${\verb| 0.07|}\\
{\verb|0.4 1.33 0.726  97.5   3.176|}$\pm${\verb| 0.20  -0.669|}$\pm${\verb| 0.09  -1.843|}$\pm${\verb| 0.02|}\\
{\verb|0.4 1.33 0.726 112.5   2.301|}$\pm${\verb| 0.15  -0.519|}$\pm${\verb| 0.07  -1.041|}$\pm${\verb| 0.02|}\\
{\verb|0.4 1.33 0.726 127.5   1.401|}$\pm${\verb| 0.15  -0.283|}$\pm${\verb| 0.10  -0.669|}$\pm${\verb| 0.03|}\\
{\verb|0.4 1.33 0.726 142.5   1.480|}$\pm${\verb| 1.07  -0.816|}$\pm${\verb| 1.06   0.269|}$\pm${\verb| 1.03|}\\
{\verb|0.4 1.35 0.709   7.5  12.538|}$\pm${\verb| 0.87   0.038|}$\pm${\verb| 0.38  -0.776|}$\pm${\verb| 0.23|}\\
{\verb|0.4 1.35 0.709  22.5  12.641|}$\pm${\verb| 1.70  -0.736|}$\pm${\verb| 1.81   1.983|}$\pm${\verb| 2.41|}\\
{\verb|0.4 1.35 0.709  37.5   8.794|}$\pm${\verb| 0.63   1.050|}$\pm${\verb| 0.44  -1.406|}$\pm${\verb| 0.26|}\\
{\verb|0.4 1.35 0.709  52.5   6.262|}$\pm${\verb| 0.47  -0.205|}$\pm${\verb| 0.18  -2.630|}$\pm${\verb| 0.07|}\\
{\verb|0.4 1.35 0.709  67.5   5.251|}$\pm${\verb| 0.42   0.126|}$\pm${\verb| 0.28  -2.136|}$\pm${\verb| 0.15|}\\
{\verb|0.4 1.35 0.709  82.5   3.660|}$\pm${\verb| 0.26  -0.361|}$\pm${\verb| 0.17  -1.911|}$\pm${\verb| 0.06|}\\
{\verb|0.4 1.35 0.709  97.5   2.649|}$\pm${\verb| 0.17  -0.504|}$\pm${\verb| 0.07  -1.500|}$\pm${\verb| 0.02|}\\
{\verb|0.4 1.35 0.709 112.5   1.839|}$\pm${\verb| 0.12  -0.371|}$\pm${\verb| 0.05  -0.878|}$\pm${\verb| 0.01|}\\
{\verb|0.4 1.35 0.709 127.5   1.093|}$\pm${\verb| 0.11  -0.294|}$\pm${\verb| 0.10  -0.411|}$\pm${\verb| 0.03|}\\
{\verb|0.4 1.35 0.709 142.5   0.923|}$\pm${\verb| 0.29  -0.366|}$\pm${\verb| 0.28  -0.064|}$\pm${\verb| 0.14|}\\
{\verb|0.4 1.37 0.691   7.5  10.500|}$\pm${\verb| 0.75  -0.023|}$\pm${\verb| 0.30  -0.477|}$\pm${\verb| 0.16|}\\
{\verb|0.4 1.37 0.691  22.5  10.651|}$\pm${\verb| 1.23   0.178|}$\pm${\verb| 1.28   0.962|}$\pm${\verb| 1.40|}\\
{\verb|0.4 1.37 0.691  37.5   7.576|}$\pm${\verb| 0.55   0.992|}$\pm${\verb| 0.38  -1.692|}$\pm${\verb| 0.22|}\\
{\verb|0.4 1.37 0.691  52.5   6.200|}$\pm${\verb| 0.48   0.363|}$\pm${\verb| 0.28  -2.621|}$\pm${\verb| 0.12|}\\
{\verb|0.4 1.37 0.691  67.5   4.303|}$\pm${\verb| 0.38  -0.423|}$\pm${\verb| 0.22  -2.805|}$\pm${\verb| 0.09|}\\
{\verb|0.4 1.37 0.691  82.5   3.357|}$\pm${\verb| 0.22  -0.345|}$\pm${\verb| 0.11  -2.107|}$\pm${\verb| 0.03|}\\
{\verb|0.4 1.37 0.691  97.5   2.200|}$\pm${\verb| 0.15  -0.423|}$\pm${\verb| 0.07  -1.432|}$\pm${\verb| 0.02|}\\
{\verb|0.4 1.37 0.691 112.5   1.392|}$\pm${\verb| 0.09  -0.212|}$\pm${\verb| 0.06  -0.803|}$\pm${\verb| 0.01|}\\
{\verb|0.4 1.37 0.691 127.5   0.919|}$\pm${\verb| 0.08  -0.240|}$\pm${\verb| 0.05  -0.404|}$\pm${\verb| 0.01|}\\
{\verb|0.4 1.37 0.691 142.5   0.607|}$\pm${\verb| 0.21  -0.129|}$\pm${\verb| 0.21  -0.213|}$\pm${\verb| 0.08|}\\
{\verb|0.4 1.39 0.672   7.5  10.145|}$\pm${\verb| 0.72   0.725|}$\pm${\verb| 0.33  -1.417|}$\pm${\verb| 0.17|}\\
{\verb|0.4 1.39 0.672  22.5   8.519|}$\pm${\verb| 0.87   0.344|}$\pm${\verb| 0.56  -1.264|}$\pm${\verb| 0.37|}\\
{\verb|0.4 1.39 0.672  37.5   6.839|}$\pm${\verb| 0.57   1.172|}$\pm${\verb| 0.43  -1.647|}$\pm${\verb| 0.24|}\\
{\verb|0.4 1.39 0.672  52.5   5.273|}$\pm${\verb| 0.40   0.282|}$\pm${\verb| 0.22  -2.464|}$\pm${\verb| 0.10|}\\
{\verb|0.4 1.39 0.672  67.5   4.353|}$\pm${\verb| 0.39  -0.104|}$\pm${\verb| 0.21  -2.589|}$\pm${\verb| 0.09|}\\
{\verb|0.4 1.39 0.672  82.5   2.844|}$\pm${\verb| 0.19  -0.323|}$\pm${\verb| 0.10  -2.192|}$\pm${\verb| 0.03|}\\
{\verb|0.4 1.39 0.672  97.5   2.062|}$\pm${\verb| 0.16  -0.303|}$\pm${\verb| 0.07  -1.166|}$\pm${\verb| 0.02|}\\
{\verb|0.4 1.39 0.672 112.5   1.327|}$\pm${\verb| 0.09  -0.244|}$\pm${\verb| 0.04  -0.817|}$\pm${\verb| 0.01|}\\
{\verb|0.4 1.39 0.672 127.5   0.841|}$\pm${\verb| 0.10  -0.191|}$\pm${\verb| 0.04  -0.397|}$\pm${\verb| 0.01|}\\
{\verb|0.4 1.39 0.672 142.5   0.478|}$\pm${\verb| 0.33  -0.100|}$\pm${\verb| 0.34  -0.119|}$\pm${\verb| 0.18|}\\
{\verb|0.4 1.41 0.652   7.5   9.513|}$\pm${\verb| 0.75   0.998|}$\pm${\verb| 0.30  -0.006|}$\pm${\verb| 0.16|}\\
{\verb|0.4 1.41 0.652  22.5   7.418|}$\pm${\verb| 0.58   0.733|}$\pm${\verb| 0.40  -1.663|}$\pm${\verb| 0.22|}\\
{\verb|0.4 1.41 0.652  37.5   7.764|}$\pm${\verb| 1.57  -0.336|}$\pm${\verb| 1.38   0.322|}$\pm${\verb| 1.61|}\\
{\verb|0.4 1.41 0.652  52.5   5.225|}$\pm${\verb| 0.39   0.312|}$\pm${\verb| 0.25  -3.468|}$\pm${\verb| 0.12|}\\
{\verb|0.4 1.41 0.652  67.5   4.122|}$\pm${\verb| 0.29   0.035|}$\pm${\verb| 0.11  -2.973|}$\pm${\verb| 0.03|}\\
{\verb|0.4 1.41 0.652  82.5   2.937|}$\pm${\verb| 0.20  -0.301|}$\pm${\verb| 0.08  -1.971|}$\pm${\verb| 0.02|}\\
{\verb|0.4 1.41 0.652  97.5   2.051|}$\pm${\verb| 0.14  -0.228|}$\pm${\verb| 0.07  -1.168|}$\pm${\verb| 0.02|}\\
{\verb|0.4 1.41 0.652 112.5   1.245|}$\pm${\verb| 0.12  -0.270|}$\pm${\verb| 0.07  -0.683|}$\pm${\verb| 0.02|}\\
{\verb|0.4 1.41 0.652 127.5   0.776|}$\pm${\verb| 0.06  -0.167|}$\pm${\verb| 0.03  -0.456|}$\pm${\verb| 0.01|}\\
{\verb|0.4 1.41 0.652 142.5   0.699|}$\pm${\verb| 0.21  -0.353|}$\pm${\verb| 0.21   0.051|}$\pm${\verb| 0.09|}\\
{\verb|0.4 1.43 0.630   7.5   8.054|}$\pm${\verb| 0.82   0.720|}$\pm${\verb| 0.30   0.173|}$\pm${\verb| 0.14|}\\
{\verb|0.4 1.43 0.630  22.5   7.594|}$\pm${\verb| 0.71   1.440|}$\pm${\verb| 0.64  -0.764|}$\pm${\verb| 0.46|}\\
{\verb|0.4 1.43 0.630  37.5   5.185|}$\pm${\verb| 1.39   2.236|}$\pm${\verb| 1.29  -3.661|}$\pm${\verb| 1.44|}\\
{\verb|0.4 1.43 0.630  52.5   5.248|}$\pm${\verb| 0.51   0.577|}$\pm${\verb| 0.35  -2.945|}$\pm${\verb| 0.17|}\\
{\verb|0.4 1.43 0.630  67.5   4.362|}$\pm${\verb| 0.29   0.078|}$\pm${\verb| 0.11  -3.157|}$\pm${\verb| 0.03|}\\
{\verb|0.4 1.43 0.630  82.5   2.839|}$\pm${\verb| 0.19  -0.379|}$\pm${\verb| 0.07  -2.452|}$\pm${\verb| 0.02|}\\
{\verb|0.4 1.43 0.630  97.5   2.128|}$\pm${\verb| 0.15  -0.313|}$\pm${\verb| 0.07  -1.625|}$\pm${\verb| 0.02|}\\
{\verb|0.4 1.43 0.630 112.5   1.394|}$\pm${\verb| 0.10  -0.224|}$\pm${\verb| 0.06  -0.882|}$\pm${\verb| 0.01|}\\
{\verb|0.4 1.43 0.630 127.5   0.922|}$\pm${\verb| 0.08  -0.171|}$\pm${\verb| 0.05  -0.366|}$\pm${\verb| 0.01|}\\
{\verb|0.4 1.43 0.630 142.5   0.681|}$\pm${\verb| 0.21  -0.258|}$\pm${\verb| 0.25  -0.101|}$\pm${\verb| 0.11|}\\
{\verb|0.4 1.45 0.607   7.5   7.515|}$\pm${\verb| 0.59   0.441|}$\pm${\verb| 0.29  -0.007|}$\pm${\verb| 0.13|}\\
{\verb|0.4 1.45 0.607  22.5   7.144|}$\pm${\verb| 0.92   1.425|}$\pm${\verb| 0.83  -1.704|}$\pm${\verb| 0.64|}\\
{\verb|0.4 1.45 0.607  37.5   7.303|}$\pm${\verb| 0.72   0.507|}$\pm${\verb| 0.55  -1.910|}$\pm${\verb| 0.36|}\\
{\verb|0.4 1.45 0.607  52.5   4.973|}$\pm${\verb| 0.40  -0.047|}$\pm${\verb| 0.24  -4.138|}$\pm${\verb| 0.11|}\\
{\verb|0.4 1.45 0.607  67.5   4.293|}$\pm${\verb| 0.32   0.097|}$\pm${\verb| 0.11  -3.189|}$\pm${\verb| 0.04|}\\
{\verb|0.4 1.45 0.607  82.5   3.317|}$\pm${\verb| 0.23  -0.301|}$\pm${\verb| 0.09  -2.808|}$\pm${\verb| 0.02|}\\
{\verb|0.4 1.45 0.607  97.5   2.377|}$\pm${\verb| 0.18  -0.281|}$\pm${\verb| 0.10  -1.857|}$\pm${\verb| 0.02|}\\
{\verb|0.4 1.45 0.607 112.5   1.661|}$\pm${\verb| 0.12  -0.265|}$\pm${\verb| 0.06  -1.127|}$\pm${\verb| 0.01|}\\
{\verb|0.4 1.45 0.607 127.5   1.140|}$\pm${\verb| 0.08  -0.213|}$\pm${\verb| 0.05  -0.786|}$\pm${\verb| 0.01|}\\
{\verb|0.4 1.45 0.607 142.5   0.718|}$\pm${\verb| 0.08  -0.097|}$\pm${\verb| 0.08  -0.624|}$\pm${\verb| 0.02|}\\
{\verb|0.4 1.47 0.582   7.5   7.799|}$\pm${\verb| 0.58   0.462|}$\pm${\verb| 0.25  -0.124|}$\pm${\verb| 0.12|}\\
{\verb|0.4 1.47 0.582  22.5   7.598|}$\pm${\verb| 1.02   2.182|}$\pm${\verb| 0.89  -0.943|}$\pm${\verb| 0.73|}\\
{\verb|0.4 1.47 0.582  37.5   6.561|}$\pm${\verb| 0.64   1.589|}$\pm${\verb| 0.43  -2.271|}$\pm${\verb| 0.26|}\\
{\verb|0.4 1.47 0.582  52.5   5.648|}$\pm${\verb| 0.40   0.341|}$\pm${\verb| 0.38  -4.398|}$\pm${\verb| 0.23|}\\
{\verb|0.4 1.47 0.582  67.5   4.635|}$\pm${\verb| 0.33   0.008|}$\pm${\verb| 0.12  -3.282|}$\pm${\verb| 0.04|}\\
{\verb|0.4 1.47 0.582  82.5   3.632|}$\pm${\verb| 0.29  -0.479|}$\pm${\verb| 0.08  -3.444|}$\pm${\verb| 0.02|}\\
{\verb|0.4 1.47 0.582  97.5   3.079|}$\pm${\verb| 0.21  -0.462|}$\pm${\verb| 0.10  -2.236|}$\pm${\verb| 0.03|}\\
{\verb|0.4 1.47 0.582 112.5   2.338|}$\pm${\verb| 0.17  -0.377|}$\pm${\verb| 0.05  -1.632|}$\pm${\verb| 0.01|}\\
{\verb|0.4 1.47 0.582 127.5   1.668|}$\pm${\verb| 0.12  -0.373|}$\pm${\verb| 0.06  -1.076|}$\pm${\verb| 0.01|}\\
{\verb|0.4 1.47 0.582 142.5   1.039|}$\pm${\verb| 0.33  -0.154|}$\pm${\verb| 0.33  -0.632|}$\pm${\verb| 0.18|}\\
{\verb|0.4 1.49 0.557   7.5   8.104|}$\pm${\verb| 0.88   0.964|}$\pm${\verb| 0.29   0.314|}$\pm${\verb| 0.15|}\\
{\verb|0.4 1.49 0.557  22.5   5.764|}$\pm${\verb| 1.00   0.100|}$\pm${\verb| 1.08  -3.494|}$\pm${\verb| 1.09|}\\
{\verb|0.4 1.49 0.557  37.5   6.555|}$\pm${\verb| 1.04   1.148|}$\pm${\verb| 1.01  -2.172|}$\pm${\verb| 0.88|}\\
{\verb|0.4 1.49 0.557  52.5   5.826|}$\pm${\verb| 0.40   0.497|}$\pm${\verb| 0.37  -4.188|}$\pm${\verb| 0.21|}\\
{\verb|0.4 1.49 0.557  67.5   5.013|}$\pm${\verb| 0.37   0.170|}$\pm${\verb| 0.14  -4.030|}$\pm${\verb| 0.04|}\\
{\verb|0.4 1.49 0.557  82.5   4.242|}$\pm${\verb| 0.30  -0.380|}$\pm${\verb| 0.12  -3.697|}$\pm${\verb| 0.04|}\\
{\verb|0.4 1.49 0.557  97.5   3.545|}$\pm${\verb| 0.25  -0.422|}$\pm${\verb| 0.09  -2.900|}$\pm${\verb| 0.02|}\\
{\verb|0.4 1.49 0.557 112.5   3.114|}$\pm${\verb| 0.24  -0.541|}$\pm${\verb| 0.08  -2.456|}$\pm${\verb| 0.02|}\\
{\verb|0.4 1.49 0.557 127.5   2.446|}$\pm${\verb| 0.17  -0.339|}$\pm${\verb| 0.13  -1.694|}$\pm${\verb| 0.04|}\\
{\verb|0.4 1.49 0.557 142.5   1.685|}$\pm${\verb| 0.19  -0.357|}$\pm${\verb| 0.19  -0.875|}$\pm${\verb| 0.07|}\\
{\verb|0.4 1.51 0.529   7.5   9.273|}$\pm${\verb| 0.67   1.069|}$\pm${\verb| 0.43   0.303|}$\pm${\verb| 0.24|}\\
{\verb|0.4 1.51 0.529  22.5   7.981|}$\pm${\verb| 1.79   2.924|}$\pm${\verb| 1.98  -0.555|}$\pm${\verb| 2.77|}\\
{\verb|0.4 1.51 0.529  37.5   5.743|}$\pm${\verb| 0.57   0.985|}$\pm${\verb| 0.55  -1.511|}$\pm${\verb| 0.39|}\\
 {$Q^2$~~$W$~~~$\epsilon$~~~~~$\theta$~~~~~$\sigma_{T}+\epsilon\sigma_{L}$~~~~~~$\sigma_{TL}$~~~~~~~~~$\sigma_{TT}$ } \\ 
 \underline{\hspace{6.15cm}} \\ \vspace{0.1cm}
{\verb|0.4 1.51 0.529  67.5   3.641|}$\pm${\verb| 0.25   0.185|}$\pm${\verb| 0.10  -2.581|}$\pm${\verb| 0.03|}\\
{\verb|0.4 1.51 0.529  82.5   3.220|}$\pm${\verb| 0.21  -0.331|}$\pm${\verb| 0.09  -3.287|}$\pm${\verb| 0.03|}\\
{\verb|0.4 1.51 0.529  97.5   3.228|}$\pm${\verb| 0.22  -0.193|}$\pm${\verb| 0.13  -2.325|}$\pm${\verb| 0.04|}\\
{\verb|0.4 1.51 0.529 112.5   2.926|}$\pm${\verb| 0.22  -0.331|}$\pm${\verb| 0.06  -2.279|}$\pm${\verb| 0.01|}\\
{\verb|0.4 1.51 0.529 127.5   2.497|}$\pm${\verb| 0.17  -0.360|}$\pm${\verb| 0.08  -1.452|}$\pm${\verb| 0.02|}\\
{\verb|0.4 1.51 0.529 142.5   1.833|}$\pm${\verb| 0.28  -0.408|}$\pm${\verb| 0.20  -1.176|}$\pm${\verb| 0.07|}\\
{\verb|0.4 1.53 0.501   7.5  11.265|}$\pm${\verb| 0.91   1.667|}$\pm${\verb| 0.75   1.800|}$\pm${\verb| 0.59|}\\
{\verb|0.4 1.53 0.501  22.5   7.033|}$\pm${\verb| 1.59   0.924|}$\pm${\verb| 1.80  -1.253|}$\pm${\verb| 2.42|}\\
{\verb|0.4 1.53 0.501  37.5   4.894|}$\pm${\verb| 0.50   0.400|}$\pm${\verb| 0.60  -1.814|}$\pm${\verb| 0.42|}\\
{\verb|0.4 1.53 0.501  52.5   4.648|}$\pm${\verb| 0.73  -1.267|}$\pm${\verb| 0.71   0.770|}$\pm${\verb| 0.59|}\\
{\verb|0.4 1.53 0.501  67.5   2.452|}$\pm${\verb| 0.19   0.118|}$\pm${\verb| 0.08  -1.947|}$\pm${\verb| 0.02|}\\
{\verb|0.4 1.53 0.501  82.5   2.171|}$\pm${\verb| 0.15  -0.078|}$\pm${\verb| 0.09  -1.888|}$\pm${\verb| 0.03|}\\
{\verb|0.4 1.53 0.501  97.5   2.356|}$\pm${\verb| 0.20  -0.318|}$\pm${\verb| 0.11  -2.149|}$\pm${\verb| 0.03|}\\
{\verb|0.4 1.53 0.501 112.5   2.486|}$\pm${\verb| 0.17  -0.289|}$\pm${\verb| 0.11  -2.218|}$\pm${\verb| 0.03|}\\
{\verb|0.4 1.53 0.501 127.5   2.140|}$\pm${\verb| 0.16  -0.167|}$\pm${\verb| 0.08  -1.534|}$\pm${\verb| 0.02|}\\
{\verb|0.4 1.53 0.501 142.5   1.759|}$\pm${\verb| 0.20  -0.260|}$\pm${\verb| 0.17  -0.697|}$\pm${\verb| 0.06|}\\
{\verb|0.4 1.55 0.470  97.5   1.236|}$\pm${\verb| 0.23   0.242|}$\pm${\verb| 0.26  -1.528|}$\pm${\verb| 0.13|}\\
{\verb|0.5 1.11 0.808  22.5   2.702|}$\pm${\verb| 0.33  -0.322|}$\pm${\verb| 0.20   0.624|}$\pm${\verb| 0.08|}\\
{\verb|0.5 1.11 0.808  37.5   2.709|}$\pm${\verb| 0.25  -0.169|}$\pm${\verb| 0.13  -0.182|}$\pm${\verb| 0.04|}\\
{\verb|0.5 1.11 0.808  52.5   2.718|}$\pm${\verb| 0.30  -0.316|}$\pm${\verb| 0.21   0.161|}$\pm${\verb| 0.09|}\\
{\verb|0.5 1.11 0.808  67.5   2.726|}$\pm${\verb| 0.26  -0.892|}$\pm${\verb| 0.23  -0.922|}$\pm${\verb| 0.10|}\\
{\verb|0.5 1.11 0.808  82.5   2.462|}$\pm${\verb| 0.21  -0.551|}$\pm${\verb| 0.19  -0.173|}$\pm${\verb| 0.07|}\\
{\verb|0.5 1.11 0.808  97.5   2.104|}$\pm${\verb| 0.24  -0.360|}$\pm${\verb| 0.21  -0.695|}$\pm${\verb| 0.09|}\\
{\verb|0.5 1.11 0.808 112.5   3.745|}$\pm${\verb| 0.64   0.926|}$\pm${\verb| 0.58   1.033|}$\pm${\verb| 0.42|}\\
{\verb|0.5 1.11 0.808 127.5   2.589|}$\pm${\verb| 0.58  -0.631|}$\pm${\verb| 0.51  -1.897|}$\pm${\verb| 0.32|}\\
{\verb|0.5 1.13 0.799   7.5   3.062|}$\pm${\verb| 0.44  -0.520|}$\pm${\verb| 0.34   0.161|}$\pm${\verb| 0.18|}\\
{\verb|0.5 1.13 0.799  22.5   3.523|}$\pm${\verb| 0.33  -0.028|}$\pm${\verb| 0.23  -0.095|}$\pm${\verb| 0.09|}\\
{\verb|0.5 1.13 0.799  37.5   3.282|}$\pm${\verb| 0.29  -0.697|}$\pm${\verb| 0.14  -0.308|}$\pm${\verb| 0.05|}\\
{\verb|0.5 1.13 0.799  52.5   3.233|}$\pm${\verb| 0.28  -0.972|}$\pm${\verb| 0.13  -0.608|}$\pm${\verb| 0.04|}\\
{\verb|0.5 1.13 0.799  67.5   3.368|}$\pm${\verb| 0.25  -1.096|}$\pm${\verb| 0.18  -1.802|}$\pm${\verb| 0.06|}\\
{\verb|0.5 1.13 0.799  82.5   3.362|}$\pm${\verb| 0.24  -0.960|}$\pm${\verb| 0.17  -1.786|}$\pm${\verb| 0.06|}\\
{\verb|0.5 1.13 0.799  97.5   3.301|}$\pm${\verb| 0.31  -0.615|}$\pm${\verb| 0.17  -1.752|}$\pm${\verb| 0.06|}\\
{\verb|0.5 1.13 0.799 112.5   3.446|}$\pm${\verb| 0.36  -0.290|}$\pm${\verb| 0.30  -1.080|}$\pm${\verb| 0.15|}\\
{\verb|0.5 1.13 0.799 127.5   3.622|}$\pm${\verb| 0.52  -0.989|}$\pm${\verb| 0.47  -2.179|}$\pm${\verb| 0.28|}\\
{\verb|0.5 1.15 0.789   7.5   4.027|}$\pm${\verb| 0.98  -1.328|}$\pm${\verb| 0.66  -0.354|}$\pm${\verb| 0.48|}\\
{\verb|0.5 1.15 0.789  22.5   3.890|}$\pm${\verb| 0.31  -0.573|}$\pm${\verb| 0.13  -0.941|}$\pm${\verb| 0.04|}\\
{\verb|0.5 1.15 0.789  37.5   3.803|}$\pm${\verb| 0.30  -0.864|}$\pm${\verb| 0.21   0.106|}$\pm${\verb| 0.08|}\\
{\verb|0.5 1.15 0.789  52.5   4.452|}$\pm${\verb| 0.36  -0.803|}$\pm${\verb| 0.30  -0.934|}$\pm${\verb| 0.14|}\\
{\verb|0.5 1.15 0.789  67.5   4.902|}$\pm${\verb| 0.33  -0.959|}$\pm${\verb| 0.19  -2.627|}$\pm${\verb| 0.07|}\\
{\verb|0.5 1.15 0.789  82.5   5.105|}$\pm${\verb| 0.34  -0.874|}$\pm${\verb| 0.19  -2.557|}$\pm${\verb| 0.07|}\\
{\verb|0.5 1.15 0.789  97.5   4.938|}$\pm${\verb| 0.35  -0.847|}$\pm${\verb| 0.22  -2.934|}$\pm${\verb| 0.09|}\\
{\verb|0.5 1.15 0.789 112.5   4.938|}$\pm${\verb| 0.34  -0.550|}$\pm${\verb| 0.19  -2.478|}$\pm${\verb| 0.07|}\\
{\verb|0.5 1.15 0.789 127.5   4.939|}$\pm${\verb| 0.83  -1.000|}$\pm${\verb| 0.46  -1.852|}$\pm${\verb| 0.27|}\\
{\verb|0.5 1.15 0.789 142.5   8.678|}$\pm${\verb| 2.20  -3.743|}$\pm${\verb| 2.04   0.562|}$\pm${\verb| 2.90|}\\
{\verb|0.5 1.17 0.779   7.5   3.713|}$\pm${\verb| 0.45  -0.331|}$\pm${\verb| 0.24  -1.059|}$\pm${\verb| 0.11|}\\
{\verb|0.5 1.17 0.779  22.5   4.173|}$\pm${\verb| 0.34  -0.927|}$\pm${\verb| 0.19  -0.374|}$\pm${\verb| 0.07|}\\
{\verb|0.5 1.17 0.779  37.5   4.817|}$\pm${\verb| 0.48  -1.284|}$\pm${\verb| 0.31  -0.806|}$\pm${\verb| 0.14|}\\
{\verb|0.5 1.17 0.779  52.5   5.701|}$\pm${\verb| 0.54  -1.292|}$\pm${\verb| 0.41  -1.645|}$\pm${\verb| 0.23|}\\
{\verb|0.5 1.17 0.779  67.5   7.084|}$\pm${\verb| 0.51  -1.071|}$\pm${\verb| 0.20  -3.452|}$\pm${\verb| 0.08|}\\
{\verb|0.5 1.17 0.779  82.5   8.029|}$\pm${\verb| 0.57  -0.590|}$\pm${\verb| 0.48  -3.990|}$\pm${\verb| 0.28|}\\
{\verb|0.5 1.17 0.779  97.5   7.002|}$\pm${\verb| 0.47  -0.799|}$\pm${\verb| 0.25  -4.587|}$\pm${\verb| 0.11|}\\
{\verb|0.5 1.17 0.779 112.5   7.352|}$\pm${\verb| 0.49  -0.399|}$\pm${\verb| 0.25  -3.888|}$\pm${\verb| 0.10|}\\
{\verb|0.5 1.17 0.779 127.5   6.752|}$\pm${\verb| 0.46  -0.143|}$\pm${\verb| 0.20  -3.269|}$\pm${\verb| 0.09|}\\
{\verb|0.5 1.17 0.779 142.5   6.396|}$\pm${\verb| 1.66   0.979|}$\pm${\verb| 1.46  -2.279|}$\pm${\verb| 1.50|}\\
{\verb|0.5 1.19 0.768   7.5   4.683|}$\pm${\verb| 0.65  -1.122|}$\pm${\verb| 0.48   0.560|}$\pm${\verb| 0.33|}\\
{\verb|0.5 1.19 0.768  22.5   5.578|}$\pm${\verb| 0.84  -0.845|}$\pm${\verb| 0.75   0.318|}$\pm${\verb| 0.57|}\\
{\verb|0.5 1.19 0.768  37.5   7.227|}$\pm${\verb| 0.52  -1.475|}$\pm${\verb| 0.26  -1.395|}$\pm${\verb| 0.11|}\\
{\verb|0.5 1.19 0.768  52.5   8.185|}$\pm${\verb| 0.80  -1.536|}$\pm${\verb| 0.63  -3.264|}$\pm${\verb| 0.42|}\\
{\verb|0.5 1.19 0.768  67.5   9.722|}$\pm${\verb| 0.67  -1.555|}$\pm${\verb| 0.45  -5.435|}$\pm${\verb| 0.27|}\\
{\verb|0.5 1.19 0.768  82.5  10.126|}$\pm${\verb| 0.69  -1.280|}$\pm${\verb| 0.28  -6.003|}$\pm${\verb| 0.13|}\\
{\verb|0.5 1.19 0.768  97.5  10.496|}$\pm${\verb| 0.67  -0.687|}$\pm${\verb| 0.21  -6.100|}$\pm${\verb| 0.08|}\\
{\verb|0.5 1.19 0.768 112.5   9.976|}$\pm${\verb| 0.62  -0.411|}$\pm${\verb| 0.18  -5.675|}$\pm${\verb| 0.07|}\\
{\verb|0.5 1.19 0.768 127.5   8.936|}$\pm${\verb| 0.57   0.187|}$\pm${\verb| 0.23  -3.716|}$\pm${\verb| 0.10|}\\
{\verb|0.5 1.19 0.768 142.5   7.607|}$\pm${\verb| 1.74   1.346|}$\pm${\verb| 1.46  -3.774|}$\pm${\verb| 1.48|}\\
{\verb|0.5 1.21 0.757   7.5   6.565|}$\pm${\verb| 2.60  -1.176|}$\pm${\verb| 2.52  -0.567|}$\pm${\verb| 3.85|}\\
{\verb|0.5 1.21 0.757  22.5   9.243|}$\pm${\verb| 0.75  -0.755|}$\pm${\verb| 0.35  -0.731|}$\pm${\verb| 0.18|}\\
{\verb|0.5 1.21 0.757  37.5   8.825|}$\pm${\verb| 0.65  -1.369|}$\pm${\verb| 0.35  -1.456|}$\pm${\verb| 0.19|}\\
{\verb|0.5 1.21 0.757  52.5  10.245|}$\pm${\verb| 0.83  -1.720|}$\pm${\verb| 0.60  -4.074|}$\pm${\verb| 0.43|}\\
{\verb|0.5 1.21 0.757  67.5  12.805|}$\pm${\verb| 0.98  -0.168|}$\pm${\verb| 0.64  -3.635|}$\pm${\verb| 0.43|}\\
{\verb|0.5 1.21 0.757  82.5  11.884|}$\pm${\verb| 0.80  -1.493|}$\pm${\verb| 0.41  -6.450|}$\pm${\verb| 0.24|}\\
{\verb|0.5 1.21 0.757  97.5  12.356|}$\pm${\verb| 0.76  -0.891|}$\pm${\verb| 0.22  -7.339|}$\pm${\verb| 0.08|}\\
{\verb|0.5 1.21 0.757 112.5  11.098|}$\pm${\verb| 0.68  -0.188|}$\pm${\verb| 0.16  -6.412|}$\pm${\verb| 0.06|}\\
{\verb|0.5 1.21 0.757 127.5   9.682|}$\pm${\verb| 0.63   0.402|}$\pm${\verb| 0.20  -4.015|}$\pm${\verb| 0.08|}\\
{\verb|0.5 1.21 0.757 142.5   8.937|}$\pm${\verb| 1.52  -0.244|}$\pm${\verb| 1.27  -2.735|}$\pm${\verb| 1.25|}\\
{\verb|0.5 1.23 0.744  22.5  10.699|}$\pm${\verb| 0.73  -0.976|}$\pm${\verb| 0.23  -0.153|}$\pm${\verb| 0.10|}\\
{\verb|0.5 1.23 0.744  37.5  11.451|}$\pm${\verb| 0.81  -0.448|}$\pm${\verb| 0.36  -1.687|}$\pm${\verb| 0.20|}\\
{\verb|0.5 1.23 0.744  52.5  11.645|}$\pm${\verb| 0.84  -0.569|}$\pm${\verb| 0.47  -2.976|}$\pm${\verb| 0.32|}\\
{\verb|0.5 1.23 0.744  67.5  11.503|}$\pm${\verb| 0.79  -0.787|}$\pm${\verb| 0.48  -4.679|}$\pm${\verb| 0.30|}\\
{\verb|0.5 1.23 0.744  82.5  10.452|}$\pm${\verb| 0.70  -1.752|}$\pm${\verb| 0.39  -6.714|}$\pm${\verb| 0.23|}\\
{\verb|0.5 1.23 0.744  97.5  10.658|}$\pm${\verb| 0.65  -0.643|}$\pm${\verb| 0.19  -6.390|}$\pm${\verb| 0.07|}\\
{\verb|0.5 1.23 0.744 112.5  10.021|}$\pm${\verb| 0.61   0.194|}$\pm${\verb| 0.19  -4.633|}$\pm${\verb| 0.07|}\\
{\verb|0.5 1.23 0.744 127.5   8.074|}$\pm${\verb| 0.51   0.380|}$\pm${\verb| 0.15  -3.657|}$\pm${\verb| 0.06|}\\
{\verb|0.5 1.23 0.744 142.5   6.171|}$\pm${\verb| 0.98   1.215|}$\pm${\verb| 0.88  -4.059|}$\pm${\verb| 0.78|}\\
{\verb|0.5 1.25 0.731  22.5  11.692|}$\pm${\verb| 0.89  -0.082|}$\pm${\verb| 0.27  -0.384|}$\pm${\verb| 0.13|}\\
{\verb|0.5 1.25 0.731  37.5  10.812|}$\pm${\verb| 0.78  -0.804|}$\pm${\verb| 0.41  -1.046|}$\pm${\verb| 0.23|}\\
{\verb|0.5 1.25 0.731  52.5  11.382|}$\pm${\verb| 0.75  -0.432|}$\pm${\verb| 0.35  -2.462|}$\pm${\verb| 0.19|}\\
{\verb|0.5 1.25 0.731  67.5   9.824|}$\pm${\verb| 0.70  -0.931|}$\pm${\verb| 0.45  -3.891|}$\pm${\verb| 0.29|}\\
{\verb|0.5 1.25 0.731  82.5   8.885|}$\pm${\verb| 0.63  -1.124|}$\pm${\verb| 0.42  -4.332|}$\pm${\verb| 0.23|}\\
{\verb|0.5 1.25 0.731  97.5   8.154|}$\pm${\verb| 0.52  -0.590|}$\pm${\verb| 0.22  -4.037|}$\pm${\verb| 0.09|}\\
{\verb|0.5 1.25 0.731 112.5   7.026|}$\pm${\verb| 0.45  -0.234|}$\pm${\verb| 0.13  -2.907|}$\pm${\verb| 0.04|}\\
{\verb|0.5 1.25 0.731 127.5   5.463|}$\pm${\verb| 0.35   0.137|}$\pm${\verb| 0.16  -1.926|}$\pm${\verb| 0.05|}\\
{\verb|0.5 1.25 0.731 142.5   4.194|}$\pm${\verb| 0.66   0.562|}$\pm${\verb| 0.61  -1.339|}$\pm${\verb| 0.46|}\\
{\verb|0.5 1.27 0.716   7.5  58.334|}$\pm${\verb|16.41 -42.845|}$\pm${\verb|15.35  36.013|}$\pm${\verb|60.16|}\\
{\verb|0.5 1.27 0.716  22.5  10.910|}$\pm${\verb| 0.88  -0.023|}$\pm${\verb| 0.29   0.180|}$\pm${\verb| 0.14|}\\
{\verb|0.5 1.27 0.716  37.5  11.031|}$\pm${\verb| 0.88  -0.219|}$\pm${\verb| 0.61  -2.213|}$\pm${\verb| 0.41|}\\
{\verb|0.5 1.27 0.716  52.5   9.362|}$\pm${\verb| 0.60  -0.173|}$\pm${\verb| 0.24  -2.219|}$\pm${\verb| 0.11|}\\
{\verb|0.5 1.27 0.716  67.5   7.570|}$\pm${\verb| 0.62  -1.195|}$\pm${\verb| 0.47  -3.478|}$\pm${\verb| 0.27|}\\
{\verb|0.5 1.27 0.716  82.5   7.194|}$\pm${\verb| 0.52  -0.679|}$\pm${\verb| 0.35  -2.736|}$\pm${\verb| 0.18|}\\
{\verb|0.5 1.27 0.716  97.5   5.943|}$\pm${\verb| 0.37  -0.512|}$\pm${\verb| 0.14  -2.221|}$\pm${\verb| 0.05|}\\
{\verb|0.5 1.27 0.716 112.5   4.658|}$\pm${\verb| 0.29  -0.038|}$\pm${\verb| 0.10  -1.834|}$\pm${\verb| 0.03|}\\
{\verb|0.5 1.27 0.716 127.5   3.607|}$\pm${\verb| 0.25   0.091|}$\pm${\verb| 0.11  -1.316|}$\pm${\verb| 0.03|}\\
{\verb|0.5 1.27 0.716 142.5   2.638|}$\pm${\verb| 0.72   0.134|}$\pm${\verb| 0.63  -1.225|}$\pm${\verb| 0.45|}\\
{\verb|0.5 1.29 0.701   7.5  19.215|}$\pm${\verb| 2.65  -5.363|}$\pm${\verb| 2.79   7.123|}$\pm${\verb| 4.57|}\\
{\verb|0.5 1.29 0.701  22.5  10.859|}$\pm${\verb| 1.00   0.512|}$\pm${\verb| 0.39  -0.954|}$\pm${\verb| 0.22|}\\
{\verb|0.5 1.29 0.701  37.5   9.499|}$\pm${\verb| 0.88   0.410|}$\pm${\verb| 0.61   2.486|}$\pm${\verb| 0.44|}\\
{\verb|0.5 1.29 0.701  52.5   7.943|}$\pm${\verb| 0.53  -0.499|}$\pm${\verb| 0.22  -1.485|}$\pm${\verb| 0.09|}\\
{\verb|0.5 1.29 0.701  67.5   6.771|}$\pm${\verb| 0.61  -0.630|}$\pm${\verb| 0.43  -1.759|}$\pm${\verb| 0.24|}\\
{\verb|0.5 1.29 0.701  82.5   5.331|}$\pm${\verb| 0.36  -0.922|}$\pm${\verb| 0.15  -2.281|}$\pm${\verb| 0.05|}\\
 {$Q^2$~~$W$~~~$\epsilon$~~~~~$\theta$~~~~~$\sigma_{T}+\epsilon\sigma_{L}$~~~~~~$\sigma_{TL}$~~~~~~~~~$\sigma_{TT}$ } \\ 
 \underline{\hspace{6.15cm}} \\ \vspace{0.1cm}
{\verb|0.5 1.29 0.701  97.5   4.211|}$\pm${\verb| 0.27  -0.603|}$\pm${\verb| 0.10  -1.957|}$\pm${\verb| 0.03|}\\
{\verb|0.5 1.29 0.701 112.5   3.357|}$\pm${\verb| 0.21  -0.292|}$\pm${\verb| 0.11  -1.160|}$\pm${\verb| 0.03|}\\
{\verb|0.5 1.29 0.701 127.5   2.430|}$\pm${\verb| 0.18  -0.092|}$\pm${\verb| 0.10  -0.738|}$\pm${\verb| 0.03|}\\
{\verb|0.5 1.29 0.701 142.5   1.418|}$\pm${\verb| 0.19   0.073|}$\pm${\verb| 0.15  -0.573|}$\pm${\verb| 0.05|}\\
{\verb|0.5 1.31 0.684   7.5  13.848|}$\pm${\verb| 1.19  -0.818|}$\pm${\verb| 0.79  -1.479|}$\pm${\verb| 0.70|}\\
{\verb|0.5 1.31 0.684  22.5  20.488|}$\pm${\verb|10.01  -7.712|}$\pm${\verb|10.22   5.929|}$\pm${\verb|32.65|}\\
{\verb|0.5 1.31 0.684  37.5   8.609|}$\pm${\verb| 0.69  -0.560|}$\pm${\verb| 0.58  -2.617|}$\pm${\verb| 0.37|}\\
{\verb|0.5 1.31 0.684  52.5   7.231|}$\pm${\verb| 0.47  -0.058|}$\pm${\verb| 0.20  -1.348|}$\pm${\verb| 0.09|}\\
{\verb|0.5 1.31 0.684  67.5   5.527|}$\pm${\verb| 0.67  -0.848|}$\pm${\verb| 0.62  -2.100|}$\pm${\verb| 0.44|}\\
{\verb|0.5 1.31 0.684  82.5   3.971|}$\pm${\verb| 0.27  -0.929|}$\pm${\verb| 0.15  -1.747|}$\pm${\verb| 0.05|}\\
{\verb|0.5 1.31 0.684  97.5   3.398|}$\pm${\verb| 0.22  -0.446|}$\pm${\verb| 0.11  -1.121|}$\pm${\verb| 0.03|}\\
{\verb|0.5 1.31 0.684 112.5   2.337|}$\pm${\verb| 0.15  -0.439|}$\pm${\verb| 0.10  -0.657|}$\pm${\verb| 0.03|}\\
{\verb|0.5 1.31 0.684 127.5   1.724|}$\pm${\verb| 0.16  -0.267|}$\pm${\verb| 0.10  -0.525|}$\pm${\verb| 0.03|}\\
{\verb|0.5 1.31 0.684 142.5   0.917|}$\pm${\verb| 0.15   0.079|}$\pm${\verb| 0.19  -0.496|}$\pm${\verb| 0.07|}\\
{\verb|0.5 1.33 0.667   7.5  11.443|}$\pm${\verb| 1.16   1.066|}$\pm${\verb| 0.56   0.706|}$\pm${\verb| 0.40|}\\
{\verb|0.5 1.33 0.667  22.5  12.483|}$\pm${\verb| 2.01  -0.364|}$\pm${\verb| 2.26  -0.195|}$\pm${\verb| 3.38|}\\
{\verb|0.5 1.33 0.667  37.5   8.152|}$\pm${\verb| 0.69   0.523|}$\pm${\verb| 0.72  -1.642|}$\pm${\verb| 0.52|}\\
{\verb|0.5 1.33 0.667  52.5   6.382|}$\pm${\verb| 0.44   0.248|}$\pm${\verb| 0.22  -1.649|}$\pm${\verb| 0.09|}\\
{\verb|0.5 1.33 0.667  67.5   4.837|}$\pm${\verb| 0.49  -0.458|}$\pm${\verb| 0.48  -1.746|}$\pm${\verb| 0.28|}\\
{\verb|0.5 1.33 0.667  82.5   3.637|}$\pm${\verb| 0.24  -0.572|}$\pm${\verb| 0.12  -1.864|}$\pm${\verb| 0.04|}\\
{\verb|0.5 1.33 0.667  97.5   2.691|}$\pm${\verb| 0.17  -0.509|}$\pm${\verb| 0.09  -1.353|}$\pm${\verb| 0.02|}\\
{\verb|0.5 1.33 0.667 112.5   1.865|}$\pm${\verb| 0.13  -0.313|}$\pm${\verb| 0.10  -0.793|}$\pm${\verb| 0.03|}\\
{\verb|0.5 1.33 0.667 127.5   1.343|}$\pm${\verb| 0.11  -0.286|}$\pm${\verb| 0.14  -0.588|}$\pm${\verb| 0.05|}\\
{\verb|0.5 1.33 0.667 142.5   0.539|}$\pm${\verb| 0.33   0.037|}$\pm${\verb| 0.31  -0.363|}$\pm${\verb| 0.14|}\\
{\verb|0.5 1.35 0.648   7.5  11.045|}$\pm${\verb| 0.86   1.222|}$\pm${\verb| 0.59  -0.250|}$\pm${\verb| 0.39|}\\
{\verb|0.5 1.35 0.648  22.5   6.222|}$\pm${\verb| 1.27   4.666|}$\pm${\verb| 1.67  -6.508|}$\pm${\verb| 2.12|}\\
{\verb|0.5 1.35 0.648  37.5   7.830|}$\pm${\verb| 0.74   2.475|}$\pm${\verb| 1.08   1.467|}$\pm${\verb| 1.00|}\\
{\verb|0.5 1.35 0.648  52.5   5.956|}$\pm${\verb| 0.42   0.324|}$\pm${\verb| 0.26  -0.830|}$\pm${\verb| 0.12|}\\
{\verb|0.5 1.35 0.648  67.5   4.123|}$\pm${\verb| 0.31  -0.778|}$\pm${\verb| 0.26  -2.022|}$\pm${\verb| 0.12|}\\
{\verb|0.5 1.35 0.648  82.5   3.151|}$\pm${\verb| 0.21  -0.670|}$\pm${\verb| 0.10  -1.665|}$\pm${\verb| 0.03|}\\
{\verb|0.5 1.35 0.648  97.5   2.312|}$\pm${\verb| 0.16  -0.542|}$\pm${\verb| 0.13  -1.326|}$\pm${\verb| 0.04|}\\
{\verb|0.5 1.35 0.648 112.5   1.427|}$\pm${\verb| 0.10  -0.360|}$\pm${\verb| 0.07  -0.323|}$\pm${\verb| 0.02|}\\
{\verb|0.5 1.35 0.648 127.5   0.942|}$\pm${\verb| 0.08  -0.298|}$\pm${\verb| 0.06  -0.308|}$\pm${\verb| 0.01|}\\
{\verb|0.5 1.35 0.648 142.5   0.908|}$\pm${\verb| 0.28  -0.492|}$\pm${\verb| 0.31   0.153|}$\pm${\verb| 0.15|}\\
{\verb|0.5 1.37 0.629   7.5   9.668|}$\pm${\verb| 0.71   1.076|}$\pm${\verb| 0.52   1.288|}$\pm${\verb| 0.32|}\\
{\verb|0.5 1.37 0.629  22.5   8.243|}$\pm${\verb| 0.75   0.313|}$\pm${\verb| 0.92  -0.046|}$\pm${\verb| 0.75|}\\
{\verb|0.5 1.37 0.629  37.5   6.765|}$\pm${\verb| 0.61   0.294|}$\pm${\verb| 0.82  -2.590|}$\pm${\verb| 0.65|}\\
{\verb|0.5 1.37 0.629  52.5   5.883|}$\pm${\verb| 0.47   0.465|}$\pm${\verb| 0.33  -2.107|}$\pm${\verb| 0.18|}\\
{\verb|0.5 1.37 0.629  67.5   3.942|}$\pm${\verb| 0.34  -0.234|}$\pm${\verb| 0.23  -2.457|}$\pm${\verb| 0.10|}\\
{\verb|0.5 1.37 0.629  82.5   2.704|}$\pm${\verb| 0.18  -0.567|}$\pm${\verb| 0.08  -1.429|}$\pm${\verb| 0.02|}\\
{\verb|0.5 1.37 0.629  97.5   2.066|}$\pm${\verb| 0.14  -0.276|}$\pm${\verb| 0.12  -0.577|}$\pm${\verb| 0.04|}\\
{\verb|0.5 1.37 0.629 112.5   1.258|}$\pm${\verb| 0.11  -0.246|}$\pm${\verb| 0.09  -0.580|}$\pm${\verb| 0.02|}\\
{\verb|0.5 1.37 0.629 127.5   0.749|}$\pm${\verb| 0.07  -0.240|}$\pm${\verb| 0.05  -0.213|}$\pm${\verb| 0.01|}\\
{\verb|0.5 1.37 0.629 142.5   0.468|}$\pm${\verb| 0.07  -0.154|}$\pm${\verb| 0.07   0.068|}$\pm${\verb| 0.02|}\\
{\verb|0.5 1.39 0.608   7.5   8.515|}$\pm${\verb| 0.65   0.529|}$\pm${\verb| 0.61   0.570|}$\pm${\verb| 0.40|}\\
{\verb|0.5 1.39 0.608  22.5   7.677|}$\pm${\verb| 0.62   0.498|}$\pm${\verb| 0.43  -0.540|}$\pm${\verb| 0.27|}\\
{\verb|0.5 1.39 0.608  37.5   7.727|}$\pm${\verb| 2.34  -0.621|}$\pm${\verb| 2.31  -0.071|}$\pm${\verb| 3.49|}\\
{\verb|0.5 1.39 0.608  52.5   5.364|}$\pm${\verb| 0.54   0.939|}$\pm${\verb| 0.43  -0.875|}$\pm${\verb| 0.26|}\\
{\verb|0.5 1.39 0.608  67.5   3.877|}$\pm${\verb| 0.31  -0.056|}$\pm${\verb| 0.15  -2.085|}$\pm${\verb| 0.05|}\\
{\verb|0.5 1.39 0.608  82.5   2.564|}$\pm${\verb| 0.19  -0.439|}$\pm${\verb| 0.08  -1.593|}$\pm${\verb| 0.02|}\\
{\verb|0.5 1.39 0.608  97.5   1.896|}$\pm${\verb| 0.13  -0.477|}$\pm${\verb| 0.09  -1.387|}$\pm${\verb| 0.02|}\\
{\verb|0.5 1.39 0.608 112.5   1.198|}$\pm${\verb| 0.12  -0.256|}$\pm${\verb| 0.09  -0.599|}$\pm${\verb| 0.02|}\\
{\verb|0.5 1.39 0.608 127.5   0.759|}$\pm${\verb| 0.12  -0.202|}$\pm${\verb| 0.17  -0.298|}$\pm${\verb| 0.06|}\\
{\verb|0.5 1.39 0.608 142.5   0.412|}$\pm${\verb| 1.20  -0.118|}$\pm${\verb| 1.33  -0.071|}$\pm${\verb| 1.53|}\\
{\verb|0.5 1.41 0.586   7.5   7.814|}$\pm${\verb| 0.66   0.175|}$\pm${\verb| 0.42   0.791|}$\pm${\verb| 0.23|}\\
{\verb|0.5 1.41 0.586  22.5   7.093|}$\pm${\verb| 0.74   1.108|}$\pm${\verb| 0.55  -0.355|}$\pm${\verb| 0.37|}\\
{\verb|0.5 1.41 0.586  37.5   7.844|}$\pm${\verb| 1.54  -1.257|}$\pm${\verb| 1.44   2.432|}$\pm${\verb| 1.69|}\\
{\verb|0.5 1.41 0.586  52.5   4.543|}$\pm${\verb| 0.33   0.358|}$\pm${\verb| 0.31  -2.084|}$\pm${\verb| 0.17|}\\
{\verb|0.5 1.41 0.586  67.5   3.854|}$\pm${\verb| 0.32   0.378|}$\pm${\verb| 0.19  -1.513|}$\pm${\verb| 0.08|}\\
{\verb|0.5 1.41 0.586  82.5   2.551|}$\pm${\verb| 0.18  -0.254|}$\pm${\verb| 0.09  -1.825|}$\pm${\verb| 0.02|}\\
{\verb|0.5 1.41 0.586  97.5   1.964|}$\pm${\verb| 0.15  -0.395|}$\pm${\verb| 0.10  -0.891|}$\pm${\verb| 0.03|}\\
{\verb|0.5 1.41 0.586 112.5   1.121|}$\pm${\verb| 0.08  -0.168|}$\pm${\verb| 0.06  -0.474|}$\pm${\verb| 0.01|}\\
{\verb|0.5 1.41 0.586 127.5   0.659|}$\pm${\verb| 0.06  -0.099|}$\pm${\verb| 0.08  -0.327|}$\pm${\verb| 0.02|}\\
{\verb|0.5 1.41 0.586 142.5   0.406|}$\pm${\verb| 0.11  -0.078|}$\pm${\verb| 0.11  -0.201|}$\pm${\verb| 0.03|}\\
{\verb|0.5 1.43 0.562   7.5   7.009|}$\pm${\verb| 0.65  -0.609|}$\pm${\verb| 0.38   0.415|}$\pm${\verb| 0.22|}\\
{\verb|0.5 1.43 0.562  22.5   6.304|}$\pm${\verb| 0.81   1.279|}$\pm${\verb| 0.91  -0.934|}$\pm${\verb| 0.85|}\\
{\verb|0.5 1.43 0.562  37.5   4.752|}$\pm${\verb| 1.09   1.322|}$\pm${\verb| 0.96  -3.342|}$\pm${\verb| 0.91|}\\
{\verb|0.5 1.43 0.562  52.5   4.515|}$\pm${\verb| 0.35   0.883|}$\pm${\verb| 0.31  -1.512|}$\pm${\verb| 0.15|}\\
{\verb|0.5 1.43 0.562  67.5   3.783|}$\pm${\verb| 0.30   0.150|}$\pm${\verb| 0.12  -2.648|}$\pm${\verb| 0.04|}\\
{\verb|0.5 1.43 0.562  82.5   2.678|}$\pm${\verb| 0.19  -0.229|}$\pm${\verb| 0.11  -1.612|}$\pm${\verb| 0.03|}\\
{\verb|0.5 1.43 0.562  97.5   1.747|}$\pm${\verb| 0.12  -0.462|}$\pm${\verb| 0.10  -1.493|}$\pm${\verb| 0.03|}\\
{\verb|0.5 1.43 0.562 112.5   1.294|}$\pm${\verb| 0.09  -0.218|}$\pm${\verb| 0.06  -0.758|}$\pm${\verb| 0.01|}\\
{\verb|0.5 1.43 0.562 127.5   0.833|}$\pm${\verb| 0.09  -0.183|}$\pm${\verb| 0.06  -0.421|}$\pm${\verb| 0.01|}\\
{\verb|0.5 1.43 0.562 142.5   0.576|}$\pm${\verb| 0.15  -0.247|}$\pm${\verb| 0.18  -0.086|}$\pm${\verb| 0.07|}\\
{\verb|0.5 1.45 0.537   7.5   6.186|}$\pm${\verb| 0.82   0.257|}$\pm${\verb| 0.34   0.880|}$\pm${\verb| 0.19|}\\
{\verb|0.5 1.45 0.537  22.5   5.961|}$\pm${\verb| 0.71   1.035|}$\pm${\verb| 0.75  -1.717|}$\pm${\verb| 0.58|}\\
{\verb|0.5 1.45 0.537  37.5   5.339|}$\pm${\verb| 0.63   0.664|}$\pm${\verb| 0.52  -2.502|}$\pm${\verb| 0.35|}\\
{\verb|0.5 1.45 0.537  52.5   4.806|}$\pm${\verb| 0.40   0.664|}$\pm${\verb| 0.36  -3.082|}$\pm${\verb| 0.20|}\\
{\verb|0.5 1.45 0.537  67.5   3.767|}$\pm${\verb| 0.25   0.052|}$\pm${\verb| 0.10  -2.405|}$\pm${\verb| 0.03|}\\
{\verb|0.5 1.45 0.537  82.5   3.055|}$\pm${\verb| 0.20  -0.228|}$\pm${\verb| 0.15  -2.274|}$\pm${\verb| 0.05|}\\
{\verb|0.5 1.45 0.537  97.5   2.198|}$\pm${\verb| 0.16  -0.382|}$\pm${\verb| 0.09  -1.636|}$\pm${\verb| 0.03|}\\
{\verb|0.5 1.45 0.537 112.5   1.501|}$\pm${\verb| 0.11  -0.230|}$\pm${\verb| 0.08  -0.766|}$\pm${\verb| 0.02|}\\
{\verb|0.5 1.45 0.537 127.5   1.041|}$\pm${\verb| 0.09  -0.290|}$\pm${\verb| 0.05  -0.415|}$\pm${\verb| 0.01|}\\
{\verb|0.5 1.45 0.537 142.5   0.525|}$\pm${\verb| 0.10  -0.074|}$\pm${\verb| 0.13  -0.438|}$\pm${\verb| 0.04|}\\
{\verb|0.5 1.47 0.511   7.5   6.493|}$\pm${\verb| 0.51   1.240|}$\pm${\verb| 0.44   0.757|}$\pm${\verb| 0.27|}\\
{\verb|0.5 1.47 0.511  22.5   5.419|}$\pm${\verb| 0.64   0.281|}$\pm${\verb| 0.59  -2.034|}$\pm${\verb| 0.41|}\\
{\verb|0.5 1.47 0.511  37.5   5.346|}$\pm${\verb| 0.52   1.388|}$\pm${\verb| 0.41  -2.538|}$\pm${\verb| 0.26|}\\
{\verb|0.5 1.47 0.511  52.5   4.663|}$\pm${\verb| 0.46   0.623|}$\pm${\verb| 0.45  -3.479|}$\pm${\verb| 0.26|}\\
{\verb|0.5 1.47 0.511  67.5   4.404|}$\pm${\verb| 0.39   0.171|}$\pm${\verb| 0.09  -2.818|}$\pm${\verb| 0.02|}\\
{\verb|0.5 1.47 0.511  82.5   3.475|}$\pm${\verb| 0.24  -0.421|}$\pm${\verb| 0.13  -2.452|}$\pm${\verb| 0.04|}\\
{\verb|0.5 1.47 0.511  97.5   2.723|}$\pm${\verb| 0.19  -0.429|}$\pm${\verb| 0.14  -2.035|}$\pm${\verb| 0.04|}\\
{\verb|0.5 1.47 0.511 112.5   2.011|}$\pm${\verb| 0.16  -0.440|}$\pm${\verb| 0.07  -1.093|}$\pm${\verb| 0.02|}\\
{\verb|0.5 1.47 0.511 127.5   1.403|}$\pm${\verb| 0.11  -0.249|}$\pm${\verb| 0.08  -0.911|}$\pm${\verb| 0.02|}\\
{\verb|0.5 1.47 0.511 142.5   1.097|}$\pm${\verb| 0.24  -0.283|}$\pm${\verb| 0.29  -0.724|}$\pm${\verb| 0.13|}\\
{\verb|0.5 1.49 0.484   7.5   6.737|}$\pm${\verb| 0.79   1.502|}$\pm${\verb| 0.61  -1.058|}$\pm${\verb| 0.40|}\\
{\verb|0.5 1.49 0.484  22.5   5.545|}$\pm${\verb| 1.30   1.010|}$\pm${\verb| 1.66  -2.434|}$\pm${\verb| 2.12|}\\
{\verb|0.5 1.49 0.484  37.5   5.534|}$\pm${\verb| 0.55   1.060|}$\pm${\verb| 0.57  -0.767|}$\pm${\verb| 0.40|}\\
{\verb|0.5 1.49 0.484  52.5   4.024|}$\pm${\verb| 1.12   1.384|}$\pm${\verb| 1.05  -5.446|}$\pm${\verb| 1.05|}\\
{\verb|0.5 1.49 0.484  67.5   4.338|}$\pm${\verb| 0.33   0.059|}$\pm${\verb| 0.18  -2.964|}$\pm${\verb| 0.06|}\\
{\verb|0.5 1.49 0.484  82.5   3.849|}$\pm${\verb| 0.28  -0.205|}$\pm${\verb| 0.15  -2.743|}$\pm${\verb| 0.06|}\\
{\verb|0.5 1.49 0.484  97.5   3.217|}$\pm${\verb| 0.24  -0.474|}$\pm${\verb| 0.19  -2.787|}$\pm${\verb| 0.07|}\\
{\verb|0.5 1.49 0.484 112.5   2.829|}$\pm${\verb| 0.22  -0.466|}$\pm${\verb| 0.18  -2.051|}$\pm${\verb| 0.06|}\\
{\verb|0.5 1.49 0.484 127.5   2.097|}$\pm${\verb| 0.18  -0.582|}$\pm${\verb| 0.15  -1.631|}$\pm${\verb| 0.05|}\\
{\verb|0.5 1.49 0.484 142.5   1.543|}$\pm${\verb| 0.30  -0.569|}$\pm${\verb| 0.31  -0.823|}$\pm${\verb| 0.14|}\\
{\verb|0.5 1.51 0.455  37.5   3.093|}$\pm${\verb| 1.47  -1.948|}$\pm${\verb| 2.77  -7.737|}$\pm${\verb| 4.61|}\\
{\verb|0.5 1.51 0.455  52.5   1.924|}$\pm${\verb| 1.92   2.315|}$\pm${\verb| 2.22  -8.786|}$\pm${\verb| 3.16|}\\
{\verb|0.5 1.51 0.455  67.5   2.791|}$\pm${\verb| 0.32   0.089|}$\pm${\verb| 0.15  -1.844|}$\pm${\verb| 0.05|}\\
{\verb|0.5 1.51 0.455  82.5   2.629|}$\pm${\verb| 0.57  -0.143|}$\pm${\verb| 0.69  -2.840|}$\pm${\verb| 0.50|}\\
 {$Q^2$~~$W$~~~$\epsilon$~~~~~$\theta$~~~~~$\sigma_{T}+\epsilon\sigma_{L}$~~~~~~$\sigma_{TL}$~~~~~~~~~$\sigma_{TT}$ } \\ 
 \underline{\hspace{6.15cm}} \\ \vspace{0.1cm}
{\verb|0.5 1.51 0.455  97.5   2.471|}$\pm${\verb| 0.37  -0.746|}$\pm${\verb| 0.25  -3.246|}$\pm${\verb| 0.10|}\\
{\verb|0.5 1.51 0.455 112.5   2.252|}$\pm${\verb| 0.25  -0.437|}$\pm${\verb| 0.27  -2.160|}$\pm${\verb| 0.12|}\\
{\verb|0.5 1.51 0.455 127.5   2.035|}$\pm${\verb| 1.03  -0.420|}$\pm${\verb| 1.35  -1.524|}$\pm${\verb| 1.54|}\\
{\verb|0.6 1.11 0.763  22.5   1.694|}$\pm${\verb| 0.28  -0.258|}$\pm${\verb| 0.18   0.546|}$\pm${\verb| 0.07|}\\
{\verb|0.6 1.11 0.763  37.5   2.421|}$\pm${\verb| 0.26  -0.516|}$\pm${\verb| 0.14  -0.484|}$\pm${\verb| 0.05|}\\
{\verb|0.6 1.11 0.763  52.5   2.101|}$\pm${\verb| 0.50  -0.513|}$\pm${\verb| 0.41  -0.604|}$\pm${\verb| 0.24|}\\
{\verb|0.6 1.11 0.763  67.5   2.196|}$\pm${\verb| 0.26  -0.558|}$\pm${\verb| 0.28  -0.288|}$\pm${\verb| 0.13|}\\
{\verb|0.6 1.11 0.763  82.5   1.964|}$\pm${\verb| 0.20  -0.301|}$\pm${\verb| 0.20  -1.035|}$\pm${\verb| 0.08|}\\
{\verb|0.6 1.11 0.763  97.5   1.823|}$\pm${\verb| 0.40  -0.664|}$\pm${\verb| 0.28  -1.277|}$\pm${\verb| 0.13|}\\
{\verb|0.6 1.11 0.763 112.5   1.580|}$\pm${\verb| 0.72  -0.614|}$\pm${\verb| 0.70  -1.297|}$\pm${\verb| 0.55|}\\
{\verb|0.6 1.13 0.752  22.5   2.781|}$\pm${\verb| 0.28  -0.057|}$\pm${\verb| 0.21   0.329|}$\pm${\verb| 0.09|}\\
{\verb|0.6 1.13 0.752  37.5   2.911|}$\pm${\verb| 0.30  -0.573|}$\pm${\verb| 0.16  -0.596|}$\pm${\verb| 0.06|}\\
{\verb|0.6 1.13 0.752  52.5   2.941|}$\pm${\verb| 0.35  -0.387|}$\pm${\verb| 0.31  -0.073|}$\pm${\verb| 0.17|}\\
{\verb|0.6 1.13 0.752  67.5   2.824|}$\pm${\verb| 0.26  -0.673|}$\pm${\verb| 0.22  -1.128|}$\pm${\verb| 0.09|}\\
{\verb|0.6 1.13 0.752  82.5   2.995|}$\pm${\verb| 0.23  -0.586|}$\pm${\verb| 0.12  -1.557|}$\pm${\verb| 0.04|}\\
{\verb|0.6 1.13 0.752  97.5   2.847|}$\pm${\verb| 0.25  -0.502|}$\pm${\verb| 0.20  -1.417|}$\pm${\verb| 0.08|}\\
{\verb|0.6 1.13 0.752 112.5   2.649|}$\pm${\verb| 0.33  -0.377|}$\pm${\verb| 0.31  -1.562|}$\pm${\verb| 0.16|}\\
{\verb|0.6 1.13 0.752 127.5   2.866|}$\pm${\verb| 0.44  -0.650|}$\pm${\verb| 0.41  -1.025|}$\pm${\verb| 0.22|}\\
{\verb|0.6 1.15 0.742  22.5   3.629|}$\pm${\verb| 0.36  -0.644|}$\pm${\verb| 0.21  -1.393|}$\pm${\verb| 0.08|}\\
{\verb|0.6 1.15 0.742  37.5   3.167|}$\pm${\verb| 0.42  -0.912|}$\pm${\verb| 0.43  -1.821|}$\pm${\verb| 0.24|}\\
{\verb|0.6 1.15 0.742  52.5   3.272|}$\pm${\verb| 0.70  -1.202|}$\pm${\verb| 0.75  -2.286|}$\pm${\verb| 0.61|}\\
{\verb|0.6 1.15 0.742  67.5   4.034|}$\pm${\verb| 0.29  -1.075|}$\pm${\verb| 0.18  -1.943|}$\pm${\verb| 0.07|}\\
{\verb|0.6 1.15 0.742  82.5   4.300|}$\pm${\verb| 0.35  -0.364|}$\pm${\verb| 0.19  -1.633|}$\pm${\verb| 0.08|}\\
{\verb|0.6 1.15 0.742  97.5   4.509|}$\pm${\verb| 0.35  -0.374|}$\pm${\verb| 0.26  -2.661|}$\pm${\verb| 0.13|}\\
{\verb|0.6 1.15 0.742 112.5   3.906|}$\pm${\verb| 0.30  -0.564|}$\pm${\verb| 0.17  -2.547|}$\pm${\verb| 0.06|}\\
{\verb|0.6 1.15 0.742 127.5   3.787|}$\pm${\verb| 0.41  -0.392|}$\pm${\verb| 0.22  -1.837|}$\pm${\verb| 0.10|}\\
{\verb|0.6 1.15 0.742 142.5   6.362|}$\pm${\verb| 1.91  -2.414|}$\pm${\verb| 1.88   0.906|}$\pm${\verb| 2.56|}\\
{\verb|0.6 1.17 0.730  22.5   3.899|}$\pm${\verb| 0.79  -0.238|}$\pm${\verb| 0.80   0.531|}$\pm${\verb| 0.68|}\\
{\verb|0.6 1.17 0.730  37.5   3.840|}$\pm${\verb| 0.39  -1.192|}$\pm${\verb| 0.29  -1.054|}$\pm${\verb| 0.14|}\\
{\verb|0.6 1.17 0.730  52.5   5.724|}$\pm${\verb| 1.05  -0.484|}$\pm${\verb| 1.20  -1.972|}$\pm${\verb| 1.11|}\\
{\verb|0.6 1.17 0.730  67.5   5.386|}$\pm${\verb| 0.41  -1.455|}$\pm${\verb| 0.26  -3.852|}$\pm${\verb| 0.11|}\\
{\verb|0.6 1.17 0.730  82.5   6.273|}$\pm${\verb| 0.45  -0.903|}$\pm${\verb| 0.29  -4.025|}$\pm${\verb| 0.13|}\\
{\verb|0.6 1.17 0.730  97.5   5.997|}$\pm${\verb| 0.43  -0.646|}$\pm${\verb| 0.22  -3.741|}$\pm${\verb| 0.10|}\\
{\verb|0.6 1.17 0.730 112.5   5.815|}$\pm${\verb| 0.38   0.137|}$\pm${\verb| 0.20  -3.315|}$\pm${\verb| 0.07|}\\
{\verb|0.6 1.17 0.730 127.5   5.594|}$\pm${\verb| 0.41  -0.267|}$\pm${\verb| 0.32  -2.349|}$\pm${\verb| 0.15|}\\
{\verb|0.6 1.17 0.730 142.5   6.416|}$\pm${\verb| 1.76   0.529|}$\pm${\verb| 1.77  -2.563|}$\pm${\verb| 2.35|}\\
{\verb|0.6 1.19 0.718  22.5   4.580|}$\pm${\verb| 0.57  -0.849|}$\pm${\verb| 0.34  -1.383|}$\pm${\verb| 0.17|}\\
{\verb|0.6 1.19 0.718  37.5   5.755|}$\pm${\verb| 0.58  -0.906|}$\pm${\verb| 0.45  -0.376|}$\pm${\verb| 0.27|}\\
{\verb|0.6 1.19 0.718  52.5   6.580|}$\pm${\verb| 0.69  -1.566|}$\pm${\verb| 0.63  -2.904|}$\pm${\verb| 0.48|}\\
{\verb|0.6 1.19 0.718  67.5   8.403|}$\pm${\verb| 0.68  -0.575|}$\pm${\verb| 0.56  -3.656|}$\pm${\verb| 0.41|}\\
{\verb|0.6 1.19 0.718  82.5   8.310|}$\pm${\verb| 0.56  -1.020|}$\pm${\verb| 0.30  -5.952|}$\pm${\verb| 0.16|}\\
{\verb|0.6 1.19 0.718  97.5   8.933|}$\pm${\verb| 0.61  -0.469|}$\pm${\verb| 0.26  -5.504|}$\pm${\verb| 0.12|}\\
{\verb|0.6 1.19 0.718 112.5   8.264|}$\pm${\verb| 0.54   0.035|}$\pm${\verb| 0.22  -4.498|}$\pm${\verb| 0.09|}\\
{\verb|0.6 1.19 0.718 127.5   7.675|}$\pm${\verb| 0.56   0.218|}$\pm${\verb| 0.31  -3.107|}$\pm${\verb| 0.15|}\\
{\verb|0.6 1.19 0.718 142.5   6.351|}$\pm${\verb| 0.89   0.863|}$\pm${\verb| 0.85  -2.880|}$\pm${\verb| 0.74|}\\
{\verb|0.6 1.21 0.704  22.5   7.259|}$\pm${\verb| 0.80  -1.361|}$\pm${\verb| 0.37  -0.350|}$\pm${\verb| 0.19|}\\
{\verb|0.6 1.21 0.704  37.5   7.749|}$\pm${\verb| 0.63  -0.722|}$\pm${\verb| 0.42  -0.771|}$\pm${\verb| 0.24|}\\
{\verb|0.6 1.21 0.704  52.5   8.603|}$\pm${\verb| 0.71  -1.635|}$\pm${\verb| 0.48  -4.757|}$\pm${\verb| 0.30|}\\
{\verb|0.6 1.21 0.704  67.5   9.952|}$\pm${\verb| 0.73  -1.279|}$\pm${\verb| 0.52  -4.917|}$\pm${\verb| 0.35|}\\
{\verb|0.6 1.21 0.704  82.5  10.180|}$\pm${\verb| 0.72  -0.913|}$\pm${\verb| 0.47  -5.997|}$\pm${\verb| 0.30|}\\
{\verb|0.6 1.21 0.704  97.5  10.446|}$\pm${\verb| 0.66  -0.542|}$\pm${\verb| 0.24  -5.591|}$\pm${\verb| 0.11|}\\
{\verb|0.6 1.21 0.704 112.5   9.808|}$\pm${\verb| 0.66   0.339|}$\pm${\verb| 0.41  -5.239|}$\pm${\verb| 0.23|}\\
{\verb|0.6 1.21 0.704 127.5   8.760|}$\pm${\verb| 0.59   0.590|}$\pm${\verb| 0.23  -4.167|}$\pm${\verb| 0.10|}\\
{\verb|0.6 1.21 0.704 142.5   6.591|}$\pm${\verb| 1.37   0.957|}$\pm${\verb| 1.34  -3.864|}$\pm${\verb| 1.38|}\\
{\verb|0.6 1.23 0.690  22.5   8.618|}$\pm${\verb| 0.84  -0.649|}$\pm${\verb| 0.38  -0.232|}$\pm${\verb| 0.21|}\\
{\verb|0.6 1.23 0.690  37.5   9.234|}$\pm${\verb| 0.70  -0.931|}$\pm${\verb| 0.38  -0.557|}$\pm${\verb| 0.23|}\\
{\verb|0.6 1.23 0.690  52.5  10.844|}$\pm${\verb| 0.82   0.302|}$\pm${\verb| 0.62  -1.578|}$\pm${\verb| 0.42|}\\
{\verb|0.6 1.23 0.690  67.5   9.592|}$\pm${\verb| 0.79  -2.084|}$\pm${\verb| 0.61  -6.395|}$\pm${\verb| 0.44|}\\
{\verb|0.6 1.23 0.690  82.5   8.972|}$\pm${\verb| 0.71  -1.632|}$\pm${\verb| 0.56  -5.479|}$\pm${\verb| 0.37|}\\
{\verb|0.6 1.23 0.690  97.5   8.957|}$\pm${\verb| 0.55  -0.806|}$\pm${\verb| 0.22  -4.975|}$\pm${\verb| 0.09|}\\
{\verb|0.6 1.23 0.690 112.5   8.148|}$\pm${\verb| 0.53   0.305|}$\pm${\verb| 0.20  -4.716|}$\pm${\verb| 0.09|}\\
{\verb|0.6 1.23 0.690 127.5   7.130|}$\pm${\verb| 0.46   0.606|}$\pm${\verb| 0.20  -3.026|}$\pm${\verb| 0.08|}\\
{\verb|0.6 1.23 0.690 142.5   7.504|}$\pm${\verb| 1.05  -1.681|}$\pm${\verb| 1.04   0.629|}$\pm${\verb| 0.99|}\\
{\verb|0.6 1.25 0.675  22.5   8.725|}$\pm${\verb| 0.98   0.322|}$\pm${\verb| 0.41   3.193|}$\pm${\verb| 0.25|}\\
{\verb|0.6 1.25 0.675  37.5  10.228|}$\pm${\verb| 0.85   0.569|}$\pm${\verb| 0.51  -1.810|}$\pm${\verb| 0.33|}\\
{\verb|0.6 1.25 0.675  52.5   9.522|}$\pm${\verb| 0.68   0.043|}$\pm${\verb| 0.39  -1.583|}$\pm${\verb| 0.23|}\\
{\verb|0.6 1.25 0.675  67.5   8.376|}$\pm${\verb| 0.59  -0.634|}$\pm${\verb| 0.38  -1.673|}$\pm${\verb| 0.22|}\\
{\verb|0.6 1.25 0.675  82.5   7.083|}$\pm${\verb| 0.58  -1.476|}$\pm${\verb| 0.50  -5.152|}$\pm${\verb| 0.32|}\\
{\verb|0.6 1.25 0.675  97.5   6.816|}$\pm${\verb| 0.43  -0.535|}$\pm${\verb| 0.16  -3.295|}$\pm${\verb| 0.06|}\\
{\verb|0.6 1.25 0.675 112.5   5.785|}$\pm${\verb| 0.37  -0.205|}$\pm${\verb| 0.25  -2.848|}$\pm${\verb| 0.11|}\\
{\verb|0.6 1.25 0.675 127.5   4.852|}$\pm${\verb| 0.34  -0.047|}$\pm${\verb| 0.16  -1.667|}$\pm${\verb| 0.06|}\\
{\verb|0.6 1.25 0.675 142.5   3.961|}$\pm${\verb| 0.50   0.091|}$\pm${\verb| 0.50  -0.632|}$\pm${\verb| 0.34|}\\
{\verb|0.6 1.27 0.660   7.5   6.519|}$\pm${\verb| 3.15   6.526|}$\pm${\verb| 3.60  -5.081|}$\pm${\verb| 6.56|}\\
{\verb|0.6 1.27 0.660  22.5  12.116|}$\pm${\verb| 1.68   0.158|}$\pm${\verb| 0.59  -5.376|}$\pm${\verb| 0.40|}\\
{\verb|0.6 1.27 0.660  37.5   9.482|}$\pm${\verb| 0.79  -0.165|}$\pm${\verb| 0.47  -1.871|}$\pm${\verb| 0.32|}\\
{\verb|0.6 1.27 0.660  52.5   8.546|}$\pm${\verb| 0.65   0.310|}$\pm${\verb| 0.37  -0.868|}$\pm${\verb| 0.20|}\\
{\verb|0.6 1.27 0.660  67.5   7.293|}$\pm${\verb| 0.68  -0.216|}$\pm${\verb| 0.61  -1.224|}$\pm${\verb| 0.45|}\\
{\verb|0.6 1.27 0.660  82.5   5.670|}$\pm${\verb| 0.46  -1.023|}$\pm${\verb| 0.34  -2.716|}$\pm${\verb| 0.18|}\\
{\verb|0.6 1.27 0.660  97.5   4.810|}$\pm${\verb| 0.36  -1.039|}$\pm${\verb| 0.34  -2.696|}$\pm${\verb| 0.18|}\\
{\verb|0.6 1.27 0.660 112.5   4.218|}$\pm${\verb| 0.29  -0.279|}$\pm${\verb| 0.20  -2.096|}$\pm${\verb| 0.08|}\\
{\verb|0.6 1.27 0.660 127.5   3.083|}$\pm${\verb| 0.25   0.104|}$\pm${\verb| 0.14  -0.906|}$\pm${\verb| 0.05|}\\
{\verb|0.6 1.27 0.660 142.5   2.144|}$\pm${\verb| 0.30   0.038|}$\pm${\verb| 0.28  -0.899|}$\pm${\verb| 0.13|}\\
{\verb|0.6 1.29 0.643   7.5  10.927|}$\pm${\verb| 1.19  -1.076|}$\pm${\verb| 1.00   2.494|}$\pm${\verb| 0.98|}\\
{\verb|0.6 1.29 0.643  22.5  32.875|}$\pm${\verb|16.21 -22.147|}$\pm${\verb|16.45  14.492|}$\pm${\verb|66.68|}\\
{\verb|0.6 1.29 0.643  37.5   9.613|}$\pm${\verb| 1.21   0.795|}$\pm${\verb| 1.29  -1.005|}$\pm${\verb| 1.36|}\\
{\verb|0.6 1.29 0.643  52.5   7.230|}$\pm${\verb| 0.55  -0.272|}$\pm${\verb| 0.31  -1.807|}$\pm${\verb| 0.16|}\\
{\verb|0.6 1.29 0.643  67.5   5.405|}$\pm${\verb| 0.44  -1.205|}$\pm${\verb| 0.38  -1.620|}$\pm${\verb| 0.22|}\\
{\verb|0.6 1.29 0.643  82.5   4.833|}$\pm${\verb| 0.35  -0.551|}$\pm${\verb| 0.23  -1.671|}$\pm${\verb| 0.10|}\\
{\verb|0.6 1.29 0.643  97.5   3.725|}$\pm${\verb| 0.26  -0.528|}$\pm${\verb| 0.20  -1.717|}$\pm${\verb| 0.07|}\\
{\verb|0.6 1.29 0.643 112.5   2.816|}$\pm${\verb| 0.20  -0.286|}$\pm${\verb| 0.09  -1.443|}$\pm${\verb| 0.03|}\\
{\verb|0.6 1.29 0.643 127.5   1.822|}$\pm${\verb| 0.18  -0.109|}$\pm${\verb| 0.11  -0.541|}$\pm${\verb| 0.03|}\\
{\verb|0.6 1.29 0.643 142.5   1.400|}$\pm${\verb| 0.75   0.187|}$\pm${\verb| 0.76  -1.164|}$\pm${\verb| 0.60|}\\
{\verb|0.6 1.31 0.625   7.5   8.989|}$\pm${\verb| 0.98   0.774|}$\pm${\verb| 0.81   1.616|}$\pm${\verb| 0.71|}\\
{\verb|0.6 1.31 0.625  22.5  11.234|}$\pm${\verb| 3.78   0.305|}$\pm${\verb| 4.86  -1.951|}$\pm${\verb| 9.44|}\\
{\verb|0.6 1.31 0.625  37.5   8.146|}$\pm${\verb| 0.71   0.232|}$\pm${\verb| 0.72  -0.283|}$\pm${\verb| 0.52|}\\
{\verb|0.6 1.31 0.625  52.5   6.344|}$\pm${\verb| 0.48  -0.197|}$\pm${\verb| 0.36  -0.865|}$\pm${\verb| 0.18|}\\
{\verb|0.6 1.31 0.625  67.5   5.185|}$\pm${\verb| 0.42  -0.267|}$\pm${\verb| 0.33  -1.387|}$\pm${\verb| 0.17|}\\
{\verb|0.6 1.31 0.625  82.5   3.999|}$\pm${\verb| 0.31  -0.567|}$\pm${\verb| 0.17  -1.620|}$\pm${\verb| 0.07|}\\
{\verb|0.6 1.31 0.625  97.5   2.869|}$\pm${\verb| 0.23  -0.588|}$\pm${\verb| 0.19  -1.406|}$\pm${\verb| 0.07|}\\
{\verb|0.6 1.31 0.625 112.5   2.156|}$\pm${\verb| 0.16  -0.310|}$\pm${\verb| 0.12  -0.579|}$\pm${\verb| 0.04|}\\
{\verb|0.6 1.31 0.625 127.5   1.378|}$\pm${\verb| 0.11  -0.151|}$\pm${\verb| 0.09  -0.642|}$\pm${\verb| 0.02|}\\
{\verb|0.6 1.31 0.625 142.5   0.796|}$\pm${\verb| 0.23   0.085|}$\pm${\verb| 0.15  -0.422|}$\pm${\verb| 0.05|}\\
{\verb|0.6 1.33 0.606   7.5  10.663|}$\pm${\verb| 1.18   0.960|}$\pm${\verb| 0.81   1.251|}$\pm${\verb| 0.65|}\\
{\verb|0.6 1.33 0.606  22.5   8.178|}$\pm${\verb| 1.86   0.488|}$\pm${\verb| 2.25  -0.263|}$\pm${\verb| 3.35|}\\
{\verb|0.6 1.33 0.606  37.5   7.542|}$\pm${\verb| 0.69  -0.130|}$\pm${\verb| 0.55  -2.202|}$\pm${\verb| 0.39|}\\
{\verb|0.6 1.33 0.606  52.5   5.987|}$\pm${\verb| 0.46   0.161|}$\pm${\verb| 0.35  -0.516|}$\pm${\verb| 0.18|}\\
{\verb|0.6 1.33 0.606  67.5   4.482|}$\pm${\verb| 0.41  -0.435|}$\pm${\verb| 0.35  -1.028|}$\pm${\verb| 0.18|}\\
{\verb|0.6 1.33 0.606  82.5   3.344|}$\pm${\verb| 0.25  -0.489|}$\pm${\verb| 0.15  -0.827|}$\pm${\verb| 0.06|}\\
 {$Q^2$~~$W$~~~$\epsilon$~~~~~$\theta$~~~~~$\sigma_{T}+\epsilon\sigma_{L}$~~~~~~$\sigma_{TL}$~~~~~~~~~$\sigma_{TT}$ } \\ 
 \underline{\hspace{6.15cm}} \\ \vspace{0.1cm}
{\verb|0.6 1.33 0.606  97.5   2.464|}$\pm${\verb| 0.20  -0.617|}$\pm${\verb| 0.16  -0.491|}$\pm${\verb| 0.05|}\\
{\verb|0.6 1.33 0.606 112.5   1.637|}$\pm${\verb| 0.27  -0.348|}$\pm${\verb| 0.16  -0.764|}$\pm${\verb| 0.06|}\\
{\verb|0.6 1.33 0.606 127.5   1.076|}$\pm${\verb| 0.10  -0.049|}$\pm${\verb| 0.09  -0.297|}$\pm${\verb| 0.02|}\\
{\verb|0.6 1.33 0.606 142.5   0.596|}$\pm${\verb| 0.41  -0.190|}$\pm${\verb| 0.55  -0.184|}$\pm${\verb| 0.39|}\\
{\verb|0.6 1.35 0.585   7.5  10.280|}$\pm${\verb| 0.95  -1.230|}$\pm${\verb| 1.02   2.510|}$\pm${\verb| 0.88|}\\
{\verb|0.6 1.35 0.585  22.5   8.556|}$\pm${\verb| 1.06   0.161|}$\pm${\verb| 0.83  -0.323|}$\pm${\verb| 0.66|}\\
{\verb|0.6 1.35 0.585  52.5   5.469|}$\pm${\verb| 0.47   0.143|}$\pm${\verb| 0.41  -1.736|}$\pm${\verb| 0.23|}\\
{\verb|0.6 1.35 0.585  67.5   4.306|}$\pm${\verb| 0.47  -0.125|}$\pm${\verb| 0.54  -1.657|}$\pm${\verb| 0.34|}\\
{\verb|0.6 1.35 0.585  82.5   2.913|}$\pm${\verb| 0.21  -0.686|}$\pm${\verb| 0.14  -1.253|}$\pm${\verb| 0.04|}\\
{\verb|0.6 1.35 0.585  97.5   2.072|}$\pm${\verb| 0.16  -0.632|}$\pm${\verb| 0.14  -1.029|}$\pm${\verb| 0.04|}\\
{\verb|0.6 1.35 0.585 112.5   1.391|}$\pm${\verb| 0.12  -0.351|}$\pm${\verb| 0.10  -0.478|}$\pm${\verb| 0.03|}\\
{\verb|0.6 1.35 0.585 127.5   0.899|}$\pm${\verb| 0.11  -0.148|}$\pm${\verb| 0.16  -0.184|}$\pm${\verb| 0.06|}\\
{\verb|0.6 1.35 0.585 142.5   0.470|}$\pm${\verb| 0.10  -0.178|}$\pm${\verb| 0.12   0.027|}$\pm${\verb| 0.03|}\\
{\verb|0.6 1.37 0.564   7.5   7.837|}$\pm${\verb| 0.88  -0.683|}$\pm${\verb| 0.81   0.858|}$\pm${\verb| 0.64|}\\
{\verb|0.6 1.37 0.564  22.5   7.909|}$\pm${\verb| 0.82   0.474|}$\pm${\verb| 0.63  -2.110|}$\pm${\verb| 0.48|}\\
{\verb|0.6 1.37 0.564  37.5   9.252|}$\pm${\verb| 2.69  -3.438|}$\pm${\verb| 2.74   2.210|}$\pm${\verb| 4.48|}\\
{\verb|0.6 1.37 0.564  52.5   5.254|}$\pm${\verb| 0.42   0.213|}$\pm${\verb| 0.47  -1.140|}$\pm${\verb| 0.28|}\\
{\verb|0.6 1.37 0.564  67.5   3.807|}$\pm${\verb| 0.35   0.000|}$\pm${\verb| 0.29  -1.035|}$\pm${\verb| 0.14|}\\
{\verb|0.6 1.37 0.564  82.5   2.637|}$\pm${\verb| 0.20  -0.346|}$\pm${\verb| 0.18  -1.339|}$\pm${\verb| 0.07|}\\
{\verb|0.6 1.37 0.564  97.5   1.979|}$\pm${\verb| 0.18  -0.177|}$\pm${\verb| 0.16   0.033|}$\pm${\verb| 0.06|}\\
{\verb|0.6 1.37 0.564 112.5   1.076|}$\pm${\verb| 0.10  -0.263|}$\pm${\verb| 0.10  -0.284|}$\pm${\verb| 0.03|}\\
{\verb|0.6 1.37 0.564 127.5   0.667|}$\pm${\verb| 0.10  -0.255|}$\pm${\verb| 0.07  -0.205|}$\pm${\verb| 0.02|}\\
{\verb|0.6 1.37 0.564 142.5   0.454|}$\pm${\verb| 0.14  -0.144|}$\pm${\verb| 0.19  -0.152|}$\pm${\verb| 0.07|}\\
{\verb|0.6 1.39 0.541   7.5   7.566|}$\pm${\verb| 1.79   0.416|}$\pm${\verb| 0.65   0.104|}$\pm${\verb| 0.47|}\\
{\verb|0.6 1.39 0.541  22.5   6.792|}$\pm${\verb| 0.70   0.685|}$\pm${\verb| 0.72  -1.268|}$\pm${\verb| 0.56|}\\
{\verb|0.6 1.39 0.541  37.5   8.205|}$\pm${\verb| 2.06  -0.844|}$\pm${\verb| 2.04   4.278|}$\pm${\verb| 2.71|}\\
{\verb|0.6 1.39 0.541  52.5   4.322|}$\pm${\verb| 0.44  -0.049|}$\pm${\verb| 0.51  -2.415|}$\pm${\verb| 0.34|}\\
{\verb|0.6 1.39 0.541  67.5   3.496|}$\pm${\verb| 0.47   0.028|}$\pm${\verb| 0.47  -1.563|}$\pm${\verb| 0.27|}\\
{\verb|0.6 1.39 0.541  82.5   2.549|}$\pm${\verb| 0.21  -0.058|}$\pm${\verb| 0.17  -0.935|}$\pm${\verb| 0.06|}\\
{\verb|0.6 1.39 0.541  97.5   1.713|}$\pm${\verb| 0.20  -0.382|}$\pm${\verb| 0.15  -0.896|}$\pm${\verb| 0.05|}\\
{\verb|0.6 1.39 0.541 112.5   1.138|}$\pm${\verb| 0.20  -0.284|}$\pm${\verb| 0.18  -0.297|}$\pm${\verb| 0.06|}\\
{\verb|0.6 1.39 0.541 127.5   0.542|}$\pm${\verb| 0.09  -0.130|}$\pm${\verb| 0.11  -0.097|}$\pm${\verb| 0.03|}\\
{\verb|0.6 1.39 0.541 142.5   0.419|}$\pm${\verb| 0.10  -0.184|}$\pm${\verb| 0.11   0.178|}$\pm${\verb| 0.04|}\\
{\verb|0.6 1.41 0.518  67.5   2.856|}$\pm${\verb| 0.47  -0.527|}$\pm${\verb| 0.44   0.574|}$\pm${\verb| 0.26|}\\
{\verb|0.6 1.41 0.518  82.5   2.367|}$\pm${\verb| 0.45   0.044|}$\pm${\verb| 0.46  -0.610|}$\pm${\verb| 0.30|}\\
{\verb|0.6 1.41 0.518  97.5   1.817|}$\pm${\verb| 0.56  -0.631|}$\pm${\verb| 0.54  -2.507|}$\pm${\verb| 0.33|}\\
\end{verse} 
}

%
%

\clearpage

%
%


\begin{thebibliography}{99}

\bibitem{Burkert:04} V.~Burkert, T.-S.H.~Lee, Int. J. Mod. Phys. \textbf{E13}, 1035, 2004.

\bibitem{Foster:83} F.~Foster and G.~Hughes,  Rept. Prog. Phys. \textbf{46}, 1445, 1983.

\bibitem{Arndt:04} R.A. Arndt \textit{et al}., Phys. Rev. C\textbf{69}, 035213, 2004.

\bibitem{Isgur:82} N.~Isgur \textit{et al}., Phys. Rev. D\textbf{25}, 2394, 1982.

\bibitem{Capstick:90} S.~Capstick and G.~Karl,  Phys. Rev. D\textbf{41}, 2767, 1990.

\bibitem{Carlson:86} C.E.~Carlson, Phys. Rev. D\textbf{34}, 2704, 1986.

\bibitem{Buchmann:01} A.J.~Buchmann and E.M.~Henley, Phys. Rev. C\textbf{63}, 015202, 2001.

\bibitem{Isgur:85} N.~Isgur,  Acta Austriaca, Suppl. XXVII, 177, 1985.

\bibitem{Burkert:92}  Zh.~Li, V.~Burkert, Zh.~Li, Phys. Rev. D\textbf{46}, 70, 1992.

\bibitem{Gonzalez:98} F.~Cano and P.~Gonzales, Phys. Lett. \textbf{B431}, 270, 1998.

\bibitem{Arndt:02} R.~Arndt \textit{et al}., Phys. Rev. C\textbf{66}, 055213, 2002.

\bibitem{Gerhard:80} Ch.~Gerhard, Z. Phys \textbf{C4}, 311, 1980.

\bibitem{Hagiwara:02} K.~Hagiwara  \textit{et al}., Phys. Rev. D\textbf{66}, 010001, 2002.

\bibitem{Evangelides:74}  E.~Evangelides \textit{et al}., Nucl. Phys. \textbf{B71}, 381, 1974.

\bibitem{Breuker:78} H.~Breuker \textit{et al}., Nucl. Phys. \textbf{B146}, 285, 1978.

\bibitem{Breuker:82} H.~Breuker \textit{et al}., Z. Phys. \textbf{C13}, 113, 1982. 

\bibitem{Mecking:03} B.~Mecking \textit{et al}., Nucl. Inst. and Meth. \textbf{A503}, 513, 2003. 

\bibitem{Mestayer:00} M.D.~Mestayer \textit{et al}.,  Nucl. Inst. and Meth. \textbf{A449}, 81, 2000.

\bibitem{Adams:01} G.~Adams \textit{et al}.,   Nucl. Inst. and Meth. \textbf{A465}, 414, 2001.

\bibitem{Smith:99} E.S.~Smith \textit{et al}.,  Nucl. Inst. and Meth. \textbf{A432}, 265, 1999.

\bibitem{Amarian:01} M.~Amarian \textit{et al}.,  Nucl. Inst. and Meth. \textbf{A460}, 239, 2001. 

\bibitem{Brun:97} R.~Brun and F.~Rademakers,  Nucl. Inst. and Meth. \textbf{A389},  81, 1997. 

\bibitem{Drechsel:99} D.~Drechsel \textit{et al}., Nucl. Phys. \textbf{A645}, 145, 1999.

\bibitem{Joo:02} K.~Joo  \textit{et al}., Phys. Rev. Lett. \textbf{88}, 122001, 2002.

\bibitem{Afanasev:02} A.~Afanasev \textit{et al}., Phys. Rev. D\textbf{66}, 074004, 2002. 

\bibitem{Mo:69} L.W.~Mo and Y.S.~Tsai, Rev. Mod. Phys. \textbf{41}, 205, 1969. 

\bibitem{Aznauryan:03} I.G.~Aznauryan, Phys. Rev. C\textbf{67}, 015209, 2003.

\bibitem{Bosted:95} P.E.~Bosted,  Phys. Rev. C \textbf{51}, 409, 1995.

\bibitem{Tiator:04} L.~Tiator \textit{et al}., Eur. Phys. J. \textbf{A19}, 25, 2004.

\bibitem{Sato:96} T.~Sato and T.-S.H.~Lee, Phys. Rev.  C \textbf{54}, 2660, 1996.

\bibitem{CLAS_DB} http://clasweb.jlab.org/physicsdb/intro.html 




\end{thebibliography}
\end{document}